\documentclass[aps,prd,10pt,twocolumn,superscriptaddress,floatfix,nofootinbib,amsmath,amssymb,altaffilletter,preprintnumbers]{revtex4-1}

\usepackage{graphicx}
\usepackage{dcolumn}
\usepackage{bm}
\usepackage{tensor}
\usepackage{slashed}
\usepackage{bm}
\usepackage{multirow}
\usepackage{soul}
\usepackage{amsmath}
\usepackage{mathrsfs}
\usepackage{amssymb}
\usepackage{yfonts}
\usepackage{color}
\usepackage{xspace}
\usepackage{url}
\usepackage{verbatim}
\usepackage{mathtools}
\usepackage{upgreek}
\usepackage{amstext}
\usepackage{booktabs}
\usepackage{tabulary}
\usepackage{tabularx}
\usepackage{etoolbox}
\usepackage[utf8]{inputenc}
\usepackage[dvipsnames,table]{xcolor}
\newcolumntype{Y}{>{\centering\arraybackslash}X}
\usepackage[colorlinks=true,pagebackref=true,pdfstartview=FitV,breaklinks=true]{hyperref}
\hypersetup{urlcolor=BlueViolet,
	    citecolor=Plum,
	    linkcolor=PineGreen}
\usepackage[official]{eurosym}
\definecolor{lightgray}{rgb}{0.9,0.9,0.9}	    
\definecolor{green}{rgb}{0,0.5,0}
\definecolor{red}{rgb}{1,0,0}
\definecolor{blue}{rgb}{0,0,0.5}

\newcommand{\dbd}[2]{\ifmmode \frac{\textrm{d}#1}{\textrm{d}#2}\else $\textrm{d}#1/\textrm{d}#2$\fi}
\newcommand{\pbp}[2]{\ifmmode \frac{\partial#1}{\partial#2}\else $\partial#1/\partial#2$\fi}

\newcommand{\drm}{\mathrm{d}}

\DeclareMathAlphabet{\mathpzc}{OT1}{pzc}{m}{it}
\newcommand{\eV}{\text{e\kern-0.15ex V}\xspace}

\newcommand{\TeV}{\text{T\kern-0.1ex \eV}\xspace}

\newcommand{\cevns}{CE$\nu$NS\xspace}

\newcommand{\eves}{E$\nu e$S\xspace}
\newcommand{\keVr}{\text{k\text{e\kern-0.15ex V}$_\mathrm{r}$}\xspace}

\newcommand{\Cygnus}{\textsc{Cygnus}\xspace}

\newcommand{\be}{\begin{equation}}
\newcommand{\ee}{\end{equation}}
\newcommand{\bea}{\begin{eqnarray}}
\newcommand{\eea}{\end{eqnarray}}

\begin{document}

\title{Searching for beyond-Standard-Model solar neutrino interactions using directional detectors}

\author{Anirudh Chandra Shekar}
\email{anirudhcs@tamu.edu}
\affiliation{Department of Physics and Astronomy, Mitchell Institute for Fundamental Physics and Astronomy,
Texas A$\&$M  University, College Station, Texas 77843, USA}

\author{Chiara Lisotti}
\email{maria.lisotti@sydney.edu.au}
\affiliation{ARC Centre of Excellence for Dark Matter Particle Physics, The University of Sydney, School of Physics, NSW 2006, Australia}

\author{Nityasa Mishra}
\email{nityasa$\_$mishra@tamu.edu}
\affiliation{Department of Physics and Astronomy, Mitchell Institute for Fundamental Physics and Astronomy,
Texas A$\&$M  University, College Station, Texas 77843, USA}

\author{Ciaran A.~J.~O'Hare}
\email{ciaran.ohare@sydney.edu.au}
\affiliation{ARC Centre of Excellence for Dark Matter Particle Physics, The University of Sydney, School of Physics, NSW 2006, Australia}

\author{Louis E. Strigari}
\email{strigari@tamu.edu}
\affiliation{Department of Physics and Astronomy, Mitchell Institute for Fundamental Physics and Astronomy,
Texas A$\&$M University, College Station, Texas 77843, USA}

\smallskip
\begin{abstract}
Micro-pattern gaseous detectors (MPGDs) are a class of technologies that enable the full three-dimensional spatial reconstruction of ionisation tracks from nuclear and electron recoils in gas. Anticipating near-future 30~m$^3$-scale time projection chambers with MPGD-based readout, we forecast the sensitivity of such directionally-sensitive low-energy recoil detectors to neutrino interactions beyond the Standard Model. We work in the framework of neutrino non-standard interactions (NSIs), and calculate the combined recoil energy-angle distributions of the electron recoil signal generated by solar neutrinos in atmospheric-pressure He:CF$_4$ gas. We estimate the expected exclusion limits that such an experiment could place on various NSI parameters, as well as the mass and coupling of a new light mediator that interacts with electrons and neutrinos. We find that with an achievable background reduction of around a factor of ten from current estimates for a 30~m$^3$ optical-readout detector using this gas mixture, sensitivity to NSI parameters would already approach Borexino's sensitivity. Directionality also allows for event-by-event neutrino energy reconstruction, which would provide a means to resolve some parameter degeneracies present in the modified neutrino cross section in this formalism. Our results strongly motivate the development of small-scale directionally-sensitive gas detectors for neutrino physics.
\end{abstract}

\maketitle

\section{Introduction}
Almost seventy years since their first experimental detection, the fundamental nature of the neutrino and its connection to the rest of the otherwise well-tested Standard Model (SM) of particle physics is still mysterious. Although neutrinos remain challenging to detect, the pursuit of answers to questions such as the origin of their mass and the precise phenomenology of their oscillations has accelerated the proliferation of many diverse classes of large and highly sensitive detectors, see e.g.~Refs.~\cite{Huber:2022lpm,Klein:2022lrf,Hyper-Kamiokande:2018ofw,Super-Kamiokande:1998kpq,DUNE:2020lwj,SNO:2021xpa,MINOS:2011neo,Borexino:2008dzn,IceCube:2013low,KamLAND:2002uet,GRAND:2018iaj,Baudis:2013qla,Pattavina:2020cqc,MINER:2016igy,NUCLEUS:2017gvo,CONNIE:2019swq,Ricochet:2021rjo,KM3Net:2016zxf,T2K:2023smv,MicroBooNE:2023tzj,Baikal-GVD:2019kwy} for an incomplete sample. Detectors are able to observe and study neutrinos from a range of sources at the precision level, extending from low-energy neutrinos created by human beings, to ultra-high-energy neutrinos generated by extreme astrophysical environments, see e.g.~Ref.~\cite{Vitagliano:2019yzm} for a review.

Although most successive generations of neutrino observatories have \textit{increased} in size over the years, technological developments have also allowed for the size of a detector that is capable of making competitive measurements of a neutrino source to decrease---for example by trading target mass for lower energy thresholds, better signal reconstruction, and enhanced background rejection. In the case of neutrinos produced by the Sun, the first hints of detection by the liquid-xenon-based XENONnT~\cite{XENON:2024ijk} and PandaX~\cite{PandaX:2024muv} experiments are setting new records for the smallest physical size of a detector able to observe neutrinos from an astrophysical source.

These recent detections are possible thanks to the combination of the large cross section for coherent elastic neutrino-nucleus scattering (\cevns)~\cite{Freedman:1973yd,Abdullah:2022zue} as well as the excellent nuclear/electron recoil discrimination and low backgrounds of modern dual-phase xenon time projection chambers---the latter advance having primarily been driven by the (so-far unsuccessful) hunt for galactic dark matter particles interacting with nuclei~\cite{Schumann:2019eaa,Akerib:2022ort,Cooley:2022ufh}. These results, when combined with the experimental measurements of \cevns by COHERENT~\cite{Akimov:2017ade, COHERENT:2020iec, COHERENT:2020ybo} and CONUS+~\cite{Ackermann:2025obx}, are inspiring explorations into the possible manifestations of beyond-SM neutrino interactions at lower energies than have been studied before, see e.g.~Refs.~\cite{Cerdeno:2016sfi,Dutta:2017nht,Abdullah:2018ykz,Denton:2018xmq,Amaral:2020tga,Miranda:2020zji,delaVega:2021wpx,Majumdar:2021vdw,Li:2022jfl,Schwemberger:2022fjl,AtzoriCorona:2022moj,AtzoriCorona:2022jeb,Amaral:2023tbs,Giunti:2023yha,AristizabalSierra:2024nwf,DeRomeri:2024iaw,DeRomeri:2024hvc,DeRomeri:2025csu,Maity:2024aji,Blanco-Mas:2024ale,Xia:2024ytb}. However, at the $\gtrsim$keV recoil energy scale searched for by these low-background detectors, the \cevns process is actually responsible for a subdominant number of neutrino events. In terms of total numbers, elastic neutrino-\textit{electron} scattering (\eves) due to the solar $pp$ flux is a stronger signal by around an order of magnitude. Unfortunately, at present, these events are hidden beneath a large background of electron recoils due to radioactivity in the detector and environment, making it challenging to perform precision measurements of the neutrino-electron cross section. In the long-term, future xenon experiments like XLZD~\cite{DARWIN:2020bnc,deGouvea:2021ymm,Aalbers:2022dzr,XLZD:2024nsu} should be poised to set constraints on new physics manifesting in low-energy E$\nu$eS at an interesting level of sensitivity (given estimates for the sensitivity of the current generation like LZ, XENONnT and PandaX~\cite{Cerdeno:2016sfi,AristizabalSierra:2020edu,A:2022acy,Khan:2020csx,Amaral:2023tbs,DeRomeri:2024dbv}), but this will depend strongly on the level of electron recoil background that is achieved. At present, the best limits still come from Borexino~\cite{Borexino:2008fkj,Bellini:2013lnn,Bellini:2014uqa,Borexino:2017uhp,Borexino:2017rsf,BOREXINO:2018ohr,BOREXINO:2017yyp,Agostini:2020mfq}, whose sensitivity mostly (but not entirely) supersedes~\cite{Khan:2019jvr,Coloma:2022umy} that of experiments using terrestrial neutrino sources like beam-dumps or reactors~\cite{Bauer:2018onh,Ilten:2018crw,Han:2019zkz,Dutta:2020che,Antel:2023hkf}. 

Nevertheless, due to the relatively high energy threshold of Borexino when compared to dark matter detectors, attempting to test beyond-SM physics in the neutrino sector using low-energy electron recoils remains a well-motivated challenge. To meet this challenge, however, we require detectors that are able to accumulate a sufficient number of reliable neutrino events that are not buried beneath backgrounds which would otherwise preclude the robust measurements of their cross section required to discern differences from the SM prediction. A promising idea to achieve this---one which was proposed many years ago in this context~\cite{Seguinot:1992zu,Arzarello:1994jv,Arpesella:1996uc}\footnote{See also Ref.~\cite{Leyton:2017tza} for a similar proposal in the context of geoneutrinos.}, but has only become experimentally feasible recently---is to attempt the full spatial reconstruction of the recoil track \textit{directions}. 

Although neutrino directionality is an important and well-established observable at higher energies~\cite{IceCube:2013low,Super-Kamiokande:1998kpq,KM3Net:2016zxf,Yang:2023rbg}, there are currently only a very limited number of techniques able to reconstruct this information in the regime relevant for the low-energy solar neutrino fluxes~\cite{Theia:2019non,BOREXINO:2021efb, BOREXINO:2021xzc,BOREXINO:2023ygs}. So far, directional detection of low-energy recoil tracks at the necessary $\mathcal{O}(10$--$100)$ keV energies has only been shown to be experimentally feasible for time-projection chambers (TPCs) using gaseous targets~\cite{Battat:2016pap, Vahsen:2021gnb}. This technology has seen significant advances over the last few decades, driven partially by the goal of finding a dark matter detector that is able to circumvent~\cite{Grothaus:2014hja, O'Hare:2015mda, Mayet:2016zxu, OHare:2017rag, OHare:2020lva} the so-called ``neutrino fog'' limit~\cite{Billard:2013qya,Ruppin:2014bra,OHare:2016pjy, Dent:2016iht, Dent:2016wor, Gelmini:2018ogy,OHare:2021utq,Akerib:2022ort,Carew:2023qrj,Maity:2024vkj} imposed upon conventional direct detection experiments. It has been shown that a cost-balanced trade-off is possible using the high-resolution capabilities of highly-segmented micro-pattern gas detectors (MPGDs), which enable sub-mm spatial resolution and high signal-to-noise reconstructions of fully 3-dimensional ionisation tracks~\cite{Vahsen:2020pzb,Vahsen:2021gnb, OHare:2022jnx,Surrow:2022ptn}.\footnote{This type of directional detection is referred to as ``recoil imaging''~\cite{Vahsen:2021gnb, OHare:2022jnx} to distinguish it from other, less powerful types of directional detection that seek to only measure spatial projections of ionisation~\cite{Agafonova:2017ajg,Marshall:2020azl} or to infer directionality indirectly~\cite{Bernabei:2003ct,Belli:2020hit,Drukier:2012hj,Capparelli:2014lua,OHare:2021cgj}, e.g. via daily modulation signals.} 

The reconstruction of ionisation tracks in gas can provide a large amount of event-level information about the recoils that generated them. However, the detrimental effects of diffusion require that the distance over which this ionisation is drifted through the gas to the readout plane must be kept to a minimum, i.e.~25--50 cm at most. This in turn implies that a large-scale experiment of, say, 100~m$^3$ volume or more, could only be achieved in a modular configuration composed of multiple 10~m$^3$-scale detectors. Working towards the ultimate vision of a large-scale ``recoil observatory'' of this kind is the goal of both the \Cygnus Consortium~\cite{Vahsen:2020pzb,Schueler:2022lvr,Ghrear:2024rku,Battat:2016xxe,Ikeda:2021ckk} and the CYGNO collaboration~\cite{Baracchini:2020btb,Amaro:2023dxb,Almeida:2023tgn,CYGNO:2023gud,Amaro:2025pms}. The scale of an experiment necessary for a dark matter search into the neutrino fog would be naturally preceded by several stepping-stone detectors of $\mathcal{O}(10~{\rm m}^3)$ volume. These experiments--- which are in development now~\cite{Amaro:2025pms}---are the focus of this study as they will already have a sufficient target mass and background level to obtain sensitivity to solar neutrinos via \eves.

As for the science case supporting a gas detector of this scale over, say, a non-directional liquid-based detector, there are several notable advantages a directional measurement offers. In particular, monodirectional sources like the Sun or a neutrino beam are interesting because of the fixed kinematic relationship between the neutrino's initial energy and the energy and scattering angle of the subsequent recoil. In principle, a simultaneous measurement of the recoil energy and angle (as enabled by MPGDs) allows for the event-by-event reconstruction of the neutrino energy spectrum. On top of this, a detector able to reconstruct the full spatial distributions of recoil-induced ionisation clouds comes with a powerful intrinsic method for background rejection and nuclear/electron recoil discrimination, both of which have been shown to work down to $\sim$10 keV electron recoil energies~\cite{Deaconu:2017vam,Ghrear:2020pzk,Santos:2011kf,Baracchini:2020btb,Battat:2016xxe,Yakabe:2020rua,Vahsen:2011qx,Jaegle:2019jpx,Tao:2019wfh}.

All of the advantages of directional gas detectors mentioned above inspired Ref.~\cite{Lisotti:2024fco}, which evaluated the detector requirements to perform solar neutrino flux spectroscopy using \eves. Here, we consider the extent to which the science case for this kind of measurement extends beyond this. As well as simply reconstructing the neutrino flux, there will be additional, more subtle information about the neutrino-electron cross section accessible to an experiment measuring angular information that might uncover corrections originating from new physics. So in this paper, we extend Ref.~\cite{Lisotti:2024fco} by evaluating the size and performance of a gas-based directional recoil detector required to test whether or not neutrino-electron scattering kinematics continues to be described by the SM at low energies.

The paper is structured as follows. In Sec.~\ref{sec:scattering} we describe the theoretical calculation of the direction-dependent scattering rate of solar neutrinos with electrons, within and beyond the SM. In Sec.~\ref{sec:detector} we describe our model for the detector technology, and discuss the relevant effects that limit its ability to reconstruct the energy and direction of neutrino events. In Sec.~\ref{sec:statistics} we explain our statistical methodology, and then present results in Sec.~\ref{sec:results}.

\section{Neutrino-electron scattering}\label{sec:scattering}

We begin by describing neutrino-electron elastic scattering, both within the SM and after allowing for the possibility of so-called ``Non-Standard Interactions'' (NSIs) due to the presence of new mediators beyond the SM. We also describe the helpful density matrix formalism, which we adopt to handle the effects of NSIs on neutrino scattering and oscillations in this formalism. 

\subsection{Directional event rate}
For a detector containing $N_T$ scattering targets (e.g.~electrons or nuclei) per unit detector mass, the differential event rate for elastic neutrino scattering on those targets is given by,
\begin{equation}
\begin{split}   
    \frac{\drm R}{\drm E_r} &= N_T \int_{E_\nu^{\rm min}}^{E_\nu^{\rm max}} \drm E_\nu \frac{\drm\Phi}{\drm E_\nu} \sum_\alpha P_{e\alpha}(E_\nu)\frac{\drm\sigma_{\alpha}}{\drm E_r}   \, .
\end{split}
    \label{eq: Recoil Rate}
\end{equation}
Here $\drm \Phi/\drm E_\nu$ is the neutrino flux, $E_r$ is the recoil energy, and $E_\nu$ is the neutrino energy. The quantity $\drm \sigma_\alpha/\drm E_r$ is the differential scattering cross section for the interaction of an $\alpha$-flavoured neutrino with the target particle. The neutrino-energy-dependent function $P_{e\alpha}(E_\nu)$ is the probability that the electron neutrino $\nu_e$ produced in the solar core is detected as a neutrino with flavour $\alpha=\{e,\mu,\tau\}$ after oscillations. It is not possible to identify any flavour information at the level of individual recoil events, so we sum over $\alpha$ when computing $\drm R/\drm E_r$. 

To describe the directional dependence of the recoil event rate, we follow the discussion in Refs.~\cite{O'Hare:2015mda,Abdullah:2020iiv}. For a detector with directional sensitivity, we expand the space of observables to include the recoil direction $\hat{q}_r$, in addition to the recoil energy, $E_r$. To find the distribution of recoil energies and angles we compute the double-differential event rate, $\drm^2 R/\drm E_r \drm \Omega_r$, where $\drm\Omega_r$ is a solid angle element around $\hat{q}_r$. The expression for this is,
\begin{equation}
    \frac{\drm^2R}{\drm E_r\drm\Omega_r} = N_T \int \drm E_\nu \drm\Omega_\nu \sum_\alpha P_{e\alpha}(E_\nu) \frac{\drm^2\Phi}{\drm E_\nu \drm \Omega_\nu} \frac{\drm^2\sigma_\alpha}{\drm E_r \drm\Omega_r} 
    \label{eq: DRS} \, .
\end{equation}
The first quantity inside the integral is the double-differential neutrino flux, which in the case of the Sun can be written straightforwardly as,
\begin{equation}
    \frac{\drm^2\Phi}{\drm E_\nu \drm\Omega_\nu} = \frac{\drm\Phi}{\drm E_\nu} \delta \left(\hat{q}_\nu-\hat{q}_\odot \right) \, ,
    \label{eq: 2-D Flux}
\end{equation}
where $\hat{q}_{\nu}$ is the incoming neutrino direction and $\hat{q}_\odot$ is the unit vector pointing from the Sun to the detector. As for the double-differential cross section, we can assume the scattering is azimuthally symmetric, which means our solid angle element is $\drm\Omega_r = 2\pi \drm\cos\theta_r$, or $\drm\Omega_r = 2\pi \drm\cos\theta_\odot$ if we work in a heliocentric coordinate system. The scattering kinematics relates this angle to the neutrino energy and the energy of the recoil via,
\begin{equation}
    \cos\theta_r = \hat{q}_\nu \cdot \hat{q}_r = \frac{E_\nu + m}{E_\nu} \sqrt{\frac{E_r}{E_r + 2m}} \, ,
    \label{eq: Kinematics-1}
\end{equation}
so that
\begin{equation}
    \frac{\drm^2\sigma_\alpha}{\drm E_r \drm\Omega_r} = \frac{1}{2\pi} \frac{\drm\sigma_\alpha}{\drm E_r} \delta \left(\hat{q}_\nu \cdot \hat{q}_r   - \frac{E_\nu + m}{E_\nu} \sqrt{\frac{E_r}{E_r + 2m}} \right) \, ,
    \label{eq: 2-D CS}
\end{equation}
where $m$ is the mass of the target particle. Using Eqs.\eqref{eq: 2-D Flux} and \eqref{eq: 2-D CS} in \eqref{eq: DRS}, we find,
\begin{equation}
\begin{split}
    \frac{\drm^2R}{\drm E_r \drm\Omega_r} = &\frac{N_T}{2\pi} \int \drm E_\nu \sum_\alpha P_{e\alpha}(E_\nu) \frac{\drm\sigma_\alpha}{\drm E_r} \frac{\drm\Phi}{\drm E_\nu} \times
    \\ 
    &\delta \left( \hat{q}_r \cdot \hat{q}_\odot - \frac{E_\nu + m}{E_\nu} \sqrt{\frac{E_r}{E_r + 2m}} \right)  \, .
\end{split}
    \label{eq: DRS Step1}
\end{equation}
The $\delta$-function in Eq.\eqref{eq: DRS Step1} can be rewritten as $\delta \left( x + \frac{1}{\mathcal{E}} \right)$ where we define:
\begin{equation}
    x=-\frac{1}{E_\nu}, \quad y=\sqrt{\frac{m^2E_r}{E_r + 2m}}, \quad \frac{1}{\mathcal{E}} = \frac{\hat{q}_r \cdot \hat{q}_\odot}{y} - \frac{1}{m} \, ,
    \label{eq: Definitions-1}
\end{equation}
where $\hat{q}_r \cdot \hat{q}_\odot = \cos\theta_\odot$. We can then perform the $\delta$-function integral over this new variable $x$ to get 
\begin{equation}
    \frac{\drm^2R}{\drm E_r \drm\Omega_r} = \frac{N_T}{2\pi} \frac{\mathcal{E}^2}{y} \left. \left( \sum_\alpha P_{e\alpha} \frac{\drm\sigma_\alpha}{\drm E_r} \frac{\drm\Phi}{ \drm E_\nu} \right) \right| _{E_\nu = \mathcal{E}} \, .
    \label{eq: DRS Final}
\end{equation}

\subsubsection{Oscillation probabilities}
The Sun produces electron-flavoured neutrinos $\nu_e$ at its core via nuclear processes~\cite{Haxton:2012wfz}. The neutrino flavour states then evolve during propagation through the solar matter and in the vacuum until they are detected on Earth. The Hamiltonian for neutrino propagation from the core of the Sun to the Earth in the flavour basis is given by,
\begin{align}\label{eq:Hamiltonian}
     H =&  \frac{1}{2E_\nu} U\begin{pmatrix}
                            0 & 0 & 0\\
                            0 & \Delta m^2_{21} & 0 \\
                            0 & 0 & \Delta m^2_{31}
                        \end{pmatrix} U^\dagger \\
                        &+ \sqrt{2}G_F N_e(x) \begin{pmatrix}
                            1 & 0 & 0\\
                            0 & 0 & 0\\
                            0 & 0 & 0\\
                        \end{pmatrix} \, , \nonumber
\end{align}
where $U$ is the Pontecorvo–Maki–Nakagawa–Sakata (PMNS) matrix that transforms flavour eigenstates $\nu_\alpha$ to mass eigenstates $\nu_i$, and the mass-squared differences that govern the mixing are labelled as $\Delta m_{i j}^2 \equiv m_i^2-m_j^2$. The term $\sqrt{2} G_F N_e(x)$ is the matter potential, capturing the coherent forward scattering of neutrinos in the interior of the Sun. This term depends upon the number density of electrons in the Sun, $N_e(x)$, which is a function of position $x$, which is measured relative to the centre of the Sun.

At the detector, the neutrino flavour state $\nu_\alpha$ can be expressed as a superposition of $\nu_i$: 
\begin{equation}
    P_{e\alpha} = \sum_{i=1}^{3} P^m\left(\nu_e \to \nu_i\right) \left\lvert U_{\alpha i}\right\rvert^2,
    \label{eq: Prob sun}
\end{equation}
where $P^m\left(\nu_e \to \nu_i\right)$ is the probability of adiabatic transition of $\nu_e$ to $\nu_i$ in the Sun. For a more detailed discussion and derivation of Eq.\eqref{eq: Prob sun}, we refer to Refs.~\cite{Blennow:2003xw,Mishra:2023jlq,Kelly:2024tvh}.

\subsubsection{Standard Model neutrino cross section}
The expression for $\drm^2R/\drm E_r\drm \Omega_r$ derived above applies to both electron and nuclear scattering. However, we will now focus our attention towards \eves for the remainder of this study. The motivation behind this is simply that the event rate from solar $pp$ and $^7$Be neutrinos is within reach of the $\sim$30 m$^3$-scale gas detectors currently under preparation. Nuclear recoils from the $^8$B neutrino flux could also be measured, but require much larger gas volumes ($\sim 1000$~m$^3$) as well as nuclear-recoil energy thresholds that still remain challenging ($\sim{\rm keV}_{\rm r}$). We relegate a discussion of the prospects for measuring solar NSIs through \cevns to Appendix~\ref{app:cevns}.

The SM differential cross section for \eves ($\nu_\alpha e^- \rightarrow \nu_\alpha e^-$) is,
\begin{equation}
    \frac{\drm \sigma_{\alpha}}{\drm E_r} 
    =\frac{2G_F^2m_e}{\pi}\left[g_{L\alpha}^2 + g_R^2\left(1-\frac{E_r}{E_\nu}\right)^2 - 2g_{L\alpha}g_R\frac{m_eE_r}{2E_\nu^2} \right] \, ,
    \label{eq: ES SM cross section}
\end{equation}
where $m_e$ is the electron mass, $G_F$ is the Fermi constant, and $g_L$ and $g_R$ are the SM couplings. The process $\nu_e e^- \rightarrow \nu_e e^-$ can occur through both a charged current via $W$ exchange, as well as a neutral current via a $Z$. On the other hand the process $\nu_{\mu/\tau}\ e^- \rightarrow \nu_{\mu/\tau}\ e^-$ is purely a neutral current interaction. As a result, the flavour-dependent SM couplings are given by,
\begin{align}
    g_{Le} &= 1 + \left(\sin^2{\theta_w} - \frac{1}{2}\right) \, , \nonumber \\ 
    g_{L\mu} &= g_{L\tau} =\sin^2{\theta_w} - \frac{1}{2} \, ,
    \\
    g_R &= \sin^2{\theta_w} \, , \nonumber
    \label{SM couplings}
\end{align}
where $\sin^2{\theta_w}$ = 0.2387 is the weak mixing angle \cite{ParticleDataGroup:2024cfk}. The differential cross section for $\nu_\mu$ interactions with $e^-$ is the same as $\nu_\tau$ interactions at tree level, but there are percent-level differences between the two with higher-loop corrections (see e.g.~Refs.~\cite{Mishra:2023jlq, Kelly:2024tvh}). Since an experiment of the scale we are interested in here will not be sensitive to these differences, we neglect any radiative corrections to Eq.\eqref{eq: ES SM cross section}.    


\subsection{NSI formalism}
We have now laid out the expected rate of electron recoils due to solar neutrinos under the SM. That calculation will form our null hypothesis, which we will test against a possible extension to the SM. The extension we consider is expressed in terms of the popular non-standard neutrino interactions (NSI) framework. For concreteness, we will consider NSIs due to a new vector mediator and examine two limiting regimes for its mass. A ``light mediator'' refers to one whose mass is comparable to, or smaller than, the typical momentum transfer $q$ involved in the scattering process, while a ``heavy mediator'' is one whose mass is considerably higher than $q$. We expect qualitatively similar results for axial-vector interactions, so for brevity, we do not explore these cases here (compare figures for these cases in e.g.~Refs.~\cite{Cerdeno:2016sfi,Coloma:2022umy,Amaral:2023tbs,AristizabalSierra:2024nwf,DeRomeri:2024iaw,DeRomeri:2024hvc,DeRomeri:2025csu}). We emphasise that our motivation in this study is to determine whether the size, performance and background of a 30~m$^3$-scale directional detector is sufficient to distinguish NSIs from the SM at a competitive level---an exhaustive comparison of phenomenologically similar models is unnecessary to arrive at such a conclusion. 

Let us assume the existence of a new vector mediator $Z'$ of a gauged $U(1)$ symmetry extending the SM. The mediator will couple to fermions $f$ and neutrinos $\nu_\alpha$. The NSI Lagrangian in this case can be written as \cite{Ohlsson_2013,Dev_2019,Coloma:2022umy,DeRomeri:2024dbv},
\begin{equation}
    \mathcal{L}^{\mathrm{NSI}} \subset g_{Z'} \left( \sum_\alpha Q_{Z'}^{\nu_\alpha} \bar{\nu}_{\alpha L} \gamma^\mu \nu_{\alpha L} + \sum_f Q_{Z'}^f\bar{f} \gamma^\mu f \right) Z'_\mu \, .
    \label{eq: Light NSI Lagrangian}
\end{equation}
Here, $g_{Z'}$ is the NSI coupling and $Q_{Z'}^i$ denotes the individual model-dependent vector charges. A few specific examples of anomaly-free realisations of such models are given in Table~\ref{tab:U(1) charges}. In this study, we will show results for only the $U(1)_{B-L}$ case, since the results for the various models are very similar. In the light mediator limit, the free parameters of the model are $(g_{Z^\prime},m_{Z^\prime})$, where $m_{Z^\prime}$ is the mediator mass.

\begin{table}[h]
\centering
\begin{tabular}{|l|c|c|c|c|}
\hline
\textbf{Model} & $Q^{e/\nu_e}_{Z'}$ & $Q^{\mu/\nu_\mu}_{Z'}$ & $Q^{\tau/\nu_\tau}_{Z'}$ & $Q^{u/d}_{Z'}$ \\
\hline
$B - L$ & $-1$ & $-1$ & $-1$ & $1/3$ \\
\hline
$L_e - L_\mu$ & $1$ & $-1$ & $0$ & $0$ \\
\hline
$L_e - L_\tau$ & $1$ & $0$ & $-1$ & $0$ \\
\hline
\end{tabular}
\caption{$Q_{Z'}^i$ values for a few models}
\label{tab:U(1) charges}
\end{table}

When the mass of the mediator significantly exceeds the momentum transfer of the interaction ($m_{Z'}\gg q$), the Lagrangian of Eq.\eqref{eq: Light NSI Lagrangian} can be approximated as an effective four-fermion operator interaction similar to the SM Lagrangian~\cite{Coloma:2022umy,Davidson:2003ha,Lindner:2016wff,Miranda:2015dra},
\begin{equation}
    \mathcal{L}^{\mathrm{eff,NSI}} \subset -2 \sqrt{2} G_F \sum_{f,P,\alpha,\beta} \varepsilon_{\alpha\beta}^{f,P} \left( \bar{\nu}_\alpha \gamma_\mu L \nu_\beta \right) \left( \bar{f} \gamma^\mu P f \right) \, ,
    \label{eq: Heavy NSI Lagrangian}
\end{equation}
where $P=\{L,R\}$ is the projection operator. Here, we have introduced the dimensionless effective coupling strengths $\varepsilon_{\alpha\beta}^{f,P}$, which we take to be real valued.  For vector NSIs in the heavy mediator limit, we take the free parameters of the model to be the following combination of $L$ and $R$ couplings,
\begin{equation}\label{eq:epsilon_def}
    \varepsilon_{\alpha\beta}^{f} \equiv \varepsilon_{\alpha\beta}^{f,L} + \varepsilon_{\alpha\beta}^{f, R} \, .
\end{equation}
In contrast to this, the axial-vector NSI coupling is $\tilde{\varepsilon}_{\alpha\beta}^{f} \equiv \varepsilon_{\alpha\beta}^{f,L} - \varepsilon_{\alpha\beta}^{f, R}$. We will focus on the pure vector case in this work by setting the axial-vector coupling $\tilde{\varepsilon}_{\alpha\beta}^{f}$ to zero, implying in turn that $\varepsilon^{e,L}_{\alpha \beta} = \varepsilon^{e,R}_{\alpha \beta}$.

We label flavours as $\alpha,\beta \in \{e,\mu,\tau\}$. Cases where $\alpha=\beta$ signify ``flavour-conserving'' interactions while $\alpha \neq  \beta$ are ``flavour-violating''. The models quoted in Table~\ref{tab:U(1) charges} are all flavour-conserving, but it is possible to include flavour-violating terms in analogy with Eq.\eqref{eq: Heavy NSI Lagrangian}, see e.g.~Ref.~\cite{Farzan:2015hkd}.

\subsection{Density matrix formalism}
We now move on to describe the density matrix formalism, which we adopt to handle the fact that the introduction of NSIs modifies both the neutrino oscillations and scattering at the same time. For more details on this formalism, see~Refs.~\cite{Coloma:2022umy,Amaral:2023tbs,Gehrlein:2024vwz}. 

We first write down the labelling for the scattering processes, which will be, 
\[\nu_\alpha T\rightarrow\nu_f T \, ,\]
where $T$ represents the scattering target, $\nu_\alpha$ refers to the neutrino flavour state taking part in the scattering, and $\nu_f$ is the final neutrino flavour state. As mentioned above, NSIs open up the possibility of flavour-violating terms where $f \neq \alpha$.

In addition to scattering, neutrino oscillations allow the neutrino flavour at detection ($\nu_\alpha$) to be different from its flavour at production---$\nu_e$ for the case of solar neutrinos---which we label as,
\[\nu_e \rightarrow\nu_\alpha \, .\]
In order to account for the combined effect of neutrino propagation and neutrino detection, one needs to combine the two processes above, summing over the intermediate state $\nu_\alpha$. 

Let $S_{\rm int}$ and $S_{\rm prop}$ be the $S$-matrices for scattering and oscillation, respectively. The square of the amplitude of the combined process can be written as,
\begin{equation}
    |A_{\nu_e\rightarrow\nu_f}|^2 = \bigg|\sum_\alpha \langle \nu_f| S_{\rm int} |\nu_\alpha\rangle \langle \nu_\alpha| S_{\rm prop} |\nu_e\rangle \bigg|^2 \, .
\end{equation}
Any individual detected recoil event could be associated with any neutrino flavour. So we take the sum over amplitudes for all $f$ when computing the rate,
\begin{equation}
    \frac{\drm R}{\drm E_r} \propto \sum_f \bigg|\sum_\alpha \langle \nu_f| S_{\rm int} |\nu_\alpha\rangle \langle \nu_\alpha| S_{\rm prop} |\nu_e\rangle \bigg|^2 \, .
    \label{eq:step1_density_matrix}
\end{equation}
We can manipulate this expression into a more convenient form as follows,
\begin{align}
&\sum_f \bigg|\sum_\alpha \langle \nu_f| S_{\rm int} |\nu_\alpha\rangle \langle \nu_\alpha| S_{\rm prop} |\nu_e\rangle \bigg|^2 \nonumber \\ =& \sum_{\alpha\beta} \sum_{\gamma\lambda} (S_{\rm prop})_{\alpha\gamma}\, \rho^{e}_{\gamma\lambda}(0)\, (S_{\rm prop})^*_{\beta\lambda} 
    \sum_f (S_{\rm int})^*_{f\beta} (S_{\rm int})_{f\alpha} 
    \nonumber \\
    \equiv& \sum_{\alpha\beta} \rho^{e}_{\alpha\beta}(L) \left(\frac{\drm \zeta}{\drm E_r}\right)_{\beta\alpha} = \text{Tr} \left[\boldsymbol{\rho}^e(L) \frac{\drm \boldsymbol{\zeta}}{\drm E_r}\right] \, ,
    \label{eq:step3_density_matrix}
\end{align}
where $(S_{i})_{\alpha\beta} = \langle \nu_\alpha| S_{i}|\nu_\beta\rangle$ and we have used the fact that $\sum_\alpha |\nu_\alpha\rangle \langle\nu_\alpha|=1$. The term $\rho^e_{\gamma\lambda}(0) = \langle \nu_\gamma|\nu_e\rangle \langle \nu_e|\nu_\lambda\rangle$ is the density matrix element describing the initial neutrino system at production. The quantity $\drm \boldsymbol{\zeta}/\drm E_r$ is the generalised scattering cross section, which accounts for the correlation between different neutrino flavour scattering processes. The elements of this quantity can be written as,
\begin{align}
     \left( \frac{\drm\zeta}{\drm E_r} \right)_{\alpha\beta} &= \sum_{f=e,\mu\,\tau} (S_{\rm int})^*_{f\alpha} (S_{\rm int})_{f\beta} \nonumber \\
     &\propto \mathcal{M}^*(\nu_\alpha \rightarrow \nu_f) \mathcal{M}(\nu_\beta \rightarrow \nu_f) \, ,
     \label{eq: NSI CS Matrix Elements}
\end{align}
where the two matrix elements, $\mathcal{M}^*(\nu_\alpha \rightarrow \nu_f)$ and $\mathcal{M}(\nu_\beta \rightarrow \nu_f)$ describe the processes $\nu_\alpha e^- \rightarrow \nu_fe^-$ and $\nu_\beta e^- \rightarrow \nu_f e^-$, respectively.

Finally, the neutrino density matrix $\boldsymbol{\rho}^e(L)$ at the detector after propagation can be written as:
\begin{align}\label{eq: density matrix and H}
     \boldsymbol{\rho}^e(L) &= S_{\rm prop}\, \boldsymbol{\rho}^e (0) \,S^\dagger_{\rm prop}  \\
      &= e^{-iHL} \begin{pmatrix} 
                    1 & 0 & 0\\
                    0 & 0 & 0\\
                    0 & 0 & 0
                    \end{pmatrix} e^{iHL} \, , \nonumber
\end{align}
or, if labelling the matrix elements,
\begin{equation}
        \rho^e_{\alpha\beta}(L) = (S_{\rm prop})_{\alpha e}(S_{\rm prop})^*_{\beta e} \, .
\end{equation}
Here, $H$ represents the Hamiltonian of propagation (see below) while $L$ is the distance travelled by the neutrinos. The resulting density matrix after oscillation is no longer diagonal. Physically, this implies that the incoming neutrino at the detector before scattering is in a superposition of flavour states.

\subsubsection{Oscillation probability with NSIs}\label{sec: NSI Prob.}


The inclusion of NSIs modifies the vector potential influencing neutrino oscillations in matter, cf.~Eq.(\ref{eq:Hamiltonian}). The Hamiltonian for neutrino propagation from the core of the Sun to the Earth in the flavour basis becomes,
\begin{multline}
     H =  \frac{1}{2E_\nu} U\begin{pmatrix}
                            0 & 0 & 0\\
                            0 & \Delta m^2_{21} & 0 \\
                            0 & 0 & \Delta m^2_{31}
                        \end{pmatrix} U^\dagger \\
                        + \sqrt{2}G_F N_e(x) \begin{pmatrix}
                            1+\epsilon_{ee}(x) & \epsilon_{e\mu}(x) & \epsilon_{e\tau}(x)\\
                            \epsilon^*_{e\mu}(x) & \epsilon_{\mu\mu}(x) & \epsilon_{\mu\tau}(x)\\
                            \epsilon^*_{e\tau}(x) & \epsilon^*_{\mu\tau}(x) & \epsilon_{\tau\tau}(x)
                        \end{pmatrix} \, .
    \label{eq: Hamiltonian NSI}
\end{multline}
The $\epsilon_{\alpha\beta}$ are related to the NSI free parameters $\varepsilon_{\alpha\beta}^{f}$ defined in Eq.(\ref{eq:epsilon_def}) as follows\footnote{Note that we take all $\varepsilon$ parameters to be real-valued in this study, although in general they may be complex-valued.},
\begin{align}
    \epsilon_{\alpha\beta}(x) = \varepsilon^e_{\alpha\beta} + (2\varepsilon^u_{\alpha\beta}+ \varepsilon^d_{\alpha\beta}) + \frac{N_n(x)}{N_e(x)} ( \varepsilon^u_{\alpha\beta} + 2\varepsilon^d_{\alpha\beta}) \, ,
    \label{eq: epsilon in hamiltoninan}
\end{align}
where $N_n(x)$ is the neutron number density, and we have implicitly assumed the proton number density $N_p(x) = N_e(x)$ for neutral matter. The above expression is already appropriate for describing NSIs in the heavy mediator limit. However, since the flavour conversion of neutrinos in the presence of matter gets affected by the coherent forward neutrino-matter scattering, we work in the $q^2\rightarrow0$ limit when accounting for the effect of NSIs on oscillations. This makes it possible to use the same expressions also in the light mediator parameterisation---we simply need to translate $(g_{Z^\prime},m_{Z^\prime})$ into an equivalent set of (flavour-conserving in our case) $\varepsilon^f_{\alpha\alpha}$ parameters as follows,
\begin{equation}
    \varepsilon^f_{\alpha\alpha} = \frac{g_{Z'}^2 Q^{f}_{Z'}Q^{\nu_\alpha}_{Z'}}{\sqrt{2}G_F m_{Z'}^2} \, .
    \label{eq: light mediator epsilon}
\end{equation}

Working in the One Mass Dominance (OMD) approximation ($\Delta m^2_{31}\rightarrow \infty$), the density matrix elements can be calculated using an effective 3-flavour oscillation scheme. The $(S_{\rm prop})$ in equation Eq.\eqref{eq: density matrix and H} is given by,
\begin{align}
    (S_{\rm prop})_{\alpha e} &= O_{e1}[O_{\alpha 1}S^{(2)}_{11} + O_{\alpha 2}S^{(2)}_{21} ] \nonumber\\
    &+  O_{\alpha 3} O_{e3} e^{-i\Delta_{33}L} \, .
\end{align}
More details about how $S^{(2)}_{11}$ and $S^{(2)}_{21}$ are calculated can be found in Appendix~\ref{app: prob}.

After taking the average over the distribution of production locations in the Sun (which differs for each neutrino flux), the density matrix elements for general flavour indices \( \alpha, \beta \in \{e,\mu,\tau\} \) become,
\begin{align}
\rho^e_{\alpha\beta}(L) &= (O_{e1})^2 
    [(O_{\alpha 1} O_{\beta 1}) P_{ee}^{2\nu}
    + (O_{\alpha 2} O_{\beta 2}) P_{\text{ext}}\nonumber \\
    &\quad + (O_{\alpha 1} O_{\beta 2}) P_{\text{int}}
    + (O_{\alpha 2} O_{\beta 1}) P_{\text{int}}^*]
 \nonumber \\
&\quad + (O_{\alpha 3} O_{\beta 3}) (O_{e3})^2.
\end{align}
The terms $O \equiv R_{23} R_{13}$, $P_{ee}^{2\nu}, P_{\text{int}}$ and $P_{\text{ext}}$ are defined as:
\begin{equation}
O =
\begin{pmatrix}
    c_{13} & 0 & s_{13} \\
    -s_{13}s_{23} & c_{23} & c_{13}s_{23} \\
    -s_{13}c_{23} & -s_{23} & c_{13}c_{23}
\end{pmatrix} \, ,
\end{equation}
\begin{align}
P_{ee}^{2\nu} &= \frac{1}{2} \left(1 + \cos(2\theta_{12}) \langle\cos(2\theta_m)\rangle\right) \, , \\
P_{\text{int}} &= -\frac{1}{2} \langle\cos(2\theta_m)\rangle\sin(2\theta_{12})\, e^{i\delta_{\text{CP}}} \, , \\
P_{\text{ext}} &= \frac{1}{2} \left(1 - \cos(2\theta_{12}) \langle\cos(2\theta_m)\rangle \right) \, ,
\end{align}
where $s_{ij} = \sin\theta_{ij}, c_{ij} = \cos\theta_{ij} $, $\theta_{ij}$ are the vacuum mixing angles and $\delta_{\rm CP}$ is the CP-violating phase. We use the best fit values given in ref~\cite{Esteban_2024}. $\theta_m$ is the effective mixing angle in matter, calculated by finding the eigenvalues of the Hamiltonian in Eq.\eqref{eq: Hamiltonian NSI} in the OMD approximation, detailed further in Appendix~\ref{app: prob}. The terms containing factors of $e^{i\Delta_{33}L}$ rapidly oscillate as a function of neutrino energy, and the resulting density matrix elements after averaging do not depend on them.

The electron flavour diagonal density matrix component is calculated to be,
\begin{equation}
\rho^e_{ee}(L) = s_{13}^4 + c_{13}^4\, P_{ee}^{2\nu} = P^{3\nu}_{ee} \, .
\end{equation}
Hence, the diagonal terms of the density matrices, $\rho_{ee}$, $\rho_{\mu\mu}$ and $\rho_{\tau\tau}$ correspond to the measurement probabilities of the system, i.e. $P_{ee},P_{e\mu}$ and $P_{e\tau}$ respectively. These will equate to the SM case, when all $\epsilon_{\alpha\beta}$ parameters in Eq.\eqref{eq: Hamiltonian NSI} are set to zero.

\subsubsection{Cross section}\label{sec:NSI Cross section}
For \eves we set $f=e$ in the Lagrangians of Eqs.\eqref{eq: Light NSI Lagrangian} and~\eqref{eq: Heavy NSI Lagrangian}. In this work, we consider purely vector couplings for which $g_L = g_R$. The effects of the light mediator NSI in the cross section can be included by modifying the couplings in Eq.\eqref{eq: ES SM cross section} \cite{Coloma:2022umy,DeRomeri:2024dbv}:
\begin{align}
\textit{Light mediator:}& \\
    g_{L/R}^{\mathrm{NSI}} &= g_{L/R}^{\mathrm{SM}} + \frac{g_{Z'}^2 Q_{Z'}^e Q_{Z'}^{\nu_\alpha}}{2\sqrt{2}G_F \left( 2m_eE_r + m_{Z'}^2 \right)} \, . \nonumber
    \label{eq: ES Light NSI Cross Section}
    \end{align}
where the charges $Q$ are chosen from Table~\ref{tab:U(1) charges}. In this case, it is readily apparent that we recover the SM when we switch off NSIs by setting $g_{Z^\prime} = 0$.
    
For the heavy mediator case, to account for flavour-violating effects, we replace the SM cross section with the generalised scattering cross section~\cite{Amaral:2023tbs,Coloma:2022umy,Gehrlein:2024vwz},
\begin{align}
    &\textit{Heavy mediator:} \\
        &\left(\frac{\drm\zeta}{\drm E_r} \right)^{\mathrm{NSI}}_{\alpha\beta}  =\frac{2G_F^2 m_e}{\pi} \sum_f \bigg[ G_{\alpha f}^L G_{f \beta}^L \nonumber  
        \\ 
        & + G_{\alpha f}^R G_{f \beta}^R \left( 1 - \frac{E_r}{E_v} \right)^2 - \left( G_{\alpha f}^LG_{f \beta}^R + G_{\alpha f}^R G_{f \beta}^L \right) \frac{m_e E_r}{2E_\nu^2} \bigg] \, , \nonumber 
\end{align}
where
\begin{equation}
    G_{\alpha\beta}^L = g_{L\beta} \delta_{\alpha\beta} + \varepsilon_{\alpha\beta}^{e,L}\,, \quad G_{\alpha\beta}^R = g_{R} \delta_{\alpha\beta} + \varepsilon_{\alpha\beta}^{e,R} \, .
    \label{eq: Heavy NSI Couplings}
\end{equation}
If we turn off NSIs by setting $\varepsilon_{\alpha\beta}^{e,L}$ and $\varepsilon_{\alpha\beta}^{e,R}$ to zero, the generalised cross section becomes diagonal, and the diagonal components $\left(\frac{\drm\zeta}{\drm E_r}\right)_{ee}$,$\left(\frac{\drm\zeta}{\drm E_r}\right)_{\mu\mu}$ and $\left(\frac{\drm \zeta}{\drm E_r}\right)_{\tau\tau}$ correspond to the SM cross sections $\frac{\drm\sigma_e}{\drm E_r}$, $\frac{\drm\sigma_{\mu}}{\drm E_r}$ and $\frac{\drm\sigma_{\tau}}{\drm E_r}$ of Eq.\eqref{eq: ES SM cross section} respectively. Note that the label $\alpha\beta$ does not mean that $\alpha$ is the initial and $\beta$ the final neutrino state---they are both initial-state indices, as shown in Eq.\eqref{eq: NSI CS Matrix Elements}.

\subsubsection{Double-differential event rate with NSI}
\begin{figure}[t]
    \centering
    \includegraphics[width=\linewidth]{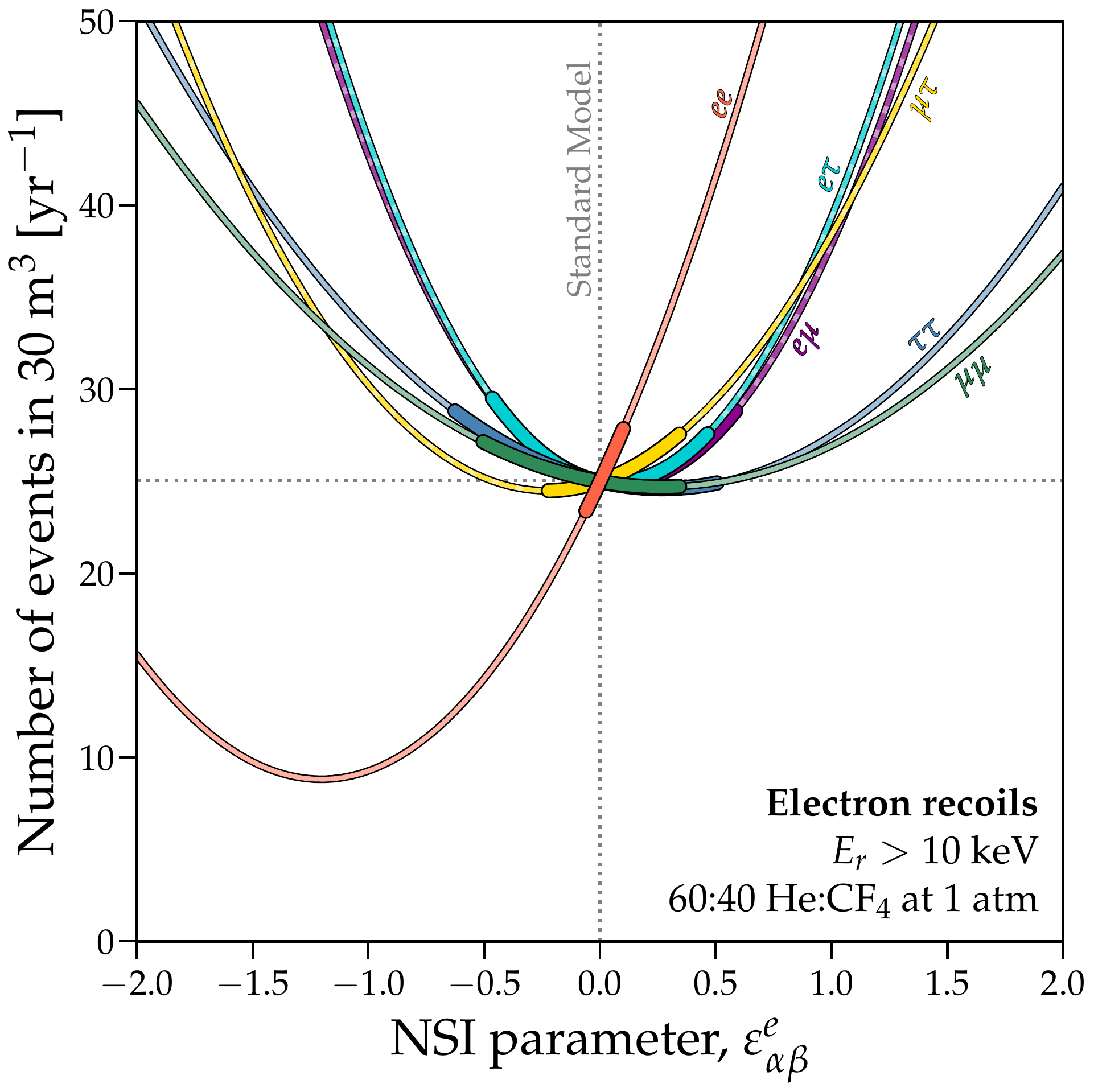}
    \caption{Yearly expected number of electron recoil events with energies $E_r>10$~keV, in a 30 m$^3$ gas detector containing a 60:40 mixture of He and CF$_4$ at atmospheric pressure. Each line shows the total rate as a function of a different $\varepsilon_{\alpha\beta}^e$ where all other values of $\varepsilon_{\alpha\beta}^e$ are kept at zero. The case for the Standard Model is shown with a dotted grey line. The thick part of each line shows the range of $\varepsilon$ allowed by Borexino at 90\% CL. The event numbers shown here correspond to the sum over the $pp$ and $^7$Be fluxes.}
    \label{fig:NumberOfEvents}
\end{figure}
A change of neutrino flavour during a scattering process is disallowed in the SM. In terms of Eq.\eqref{eq: NSI CS Matrix Elements}, this means that the final neutrino state $f$ should be the same as the initial neutrino states $\alpha$ and $\beta$; i.e $f=\alpha=\beta$. This implies that the generalised cross section is diagonal, and as seen in Sec.~\ref{sec:NSI Cross section}, these diagonal elements correspond to the standard model cross sections. We also saw in Sec.~\ref{sec: NSI Prob.} that the diagonal elements of the density matrix $\boldsymbol{\rho}^e$ correspond to the transition probabilities $P_{e\alpha}$. Thus, if we were to express our SM event rate in terms of the density matrix and generalised cross section, it would reduce to a product of transition probability and cross section:
\begin{align}
    \left( \frac{\drm^2R}{\drm E_r \drm\Omega_r} \right)^{\mathrm{SM}} &\propto \mathrm{Tr} \left[ \boldsymbol{\rho}^e \left( \frac{\drm \boldsymbol{\zeta}}{\drm E_r} \right) \right]^{\mathrm{SM}} \nonumber \\ 
    &\propto \sum_{\alpha\beta} \left[ \rho^e_{\alpha\beta} \left( \frac{\drm \zeta}{\drm E_r} \right)_{\beta\alpha} \delta_{\alpha\beta} \right] = \sum_\alpha P_{e\alpha} 
 \frac{\drm\sigma_\alpha}{\drm E_r} \, . \nonumber
\end{align}

However, with NSI effects included, we can have $f \neq \alpha \neq \beta$ which means that the generalised cross section can have off-diagonal terms. To account for all these processes, we must express the NSI-modified event rate as,
\begin{equation}
    \left( \frac{\drm^2R}{\drm E_r \drm\Omega_r} \right)^{\mathrm{NSI}} \propto \mathrm{Tr} \left[ \boldsymbol{\rho}^e \left( \frac{\drm \boldsymbol{\zeta}}{\drm E_r} \right) \right]^{\mathrm{NSI}} \, .
    \label{eq: Solar NSI rate}
\end{equation}
This expression is therefore a more general expression for event rate calculations---it can be used for both SM flavour-conserving interactions, as well as NSI flavour-violating interactions. 

So in summary, we rewrite the full expression for the double-differential event rate that we wrote down in Eq.\eqref{eq: DRS Final} in terms of a density matrix and generalised cross section,
\begin{equation}
    \frac{\drm^2R}{\drm E_r \drm\Omega_r} = \frac{N_T}{2\pi} \frac{\mathcal{E}^2}{y} \left. \left(  \mathrm{Tr} \left[  \boldsymbol{\rho}^e \frac{\drm \boldsymbol{\zeta}}{\drm E_r} \right] \frac{\drm\Phi}{\drm E_\nu} \right) \right|_{E_\nu = \mathcal{E}} \, .
    \label{eq: DRS Density Matrix}
\end{equation}
Finally, we note that we must also calculate this rate for each solar neutrino flux separately, as opposed to defining $\Phi = \Phi_{pp} + \Phi_{^7{\rm Be}} + ...$ as the sum of all fluxes. This is because the density matrix depends on the neutrino production profiles within the Sun, which are different for each flux. For electron recoils, we consider both $pp$ and $^7$Be fluxes, with $pp$ contributing around 75\% of the rate. The remaining solar fluxes contribute less than a handful of events, given the detector volumes we are interested in here. For nuclear recoils (Appendix~\ref{app:cevns}), we consider only the $^8$B flux. For the full electron and nuclear recoil rates across all of the solar neutrino fluxes, as well as prospects for their detection and spectroscopy using directional gas detectors, we refer to our previous study Ref.~\cite{Lisotti:2024fco}.

\subsection{Comparison of SM \& NSI total rates and angular recoil spectra}
In general, the inclusion of nonzero $\varepsilon^e_{\alpha \beta}$ leads to an enhancement in the cross section and so a larger event rate compared to the SM prediction, roughly scaling as $R \propto (\varepsilon^e_{\alpha \beta})^2$. That said, it is possible for interference between positive and negative terms in the cross section, leading to a suppression in the rate relative to the SM. 

In Fig.~\ref{fig:NumberOfEvents}, we display the total expected number of electron recoils per year in a 30~m$^3$ gas volume containing a mixture of He and CF$_4$ at atmospheric pressure. This choice of volume and gas mixture is motivated further in the next section---for now, we focus on the dependence of the total rate on the NSI parameters, where Fig.~\ref{fig:NumberOfEvents} applies to the heavy mediator limit. Here we consider varying one NSI parameter at a time, with all others fixed at zero, including $\varepsilon_{\alpha\beta}^{u,d}$. We see that, in general, turning the NSI parameters on leads to an enhancement in the total rate. The most obvious exception is $\varepsilon^e_{ee} \approx -1.2$ for which there is a partial cancellation in the cross section. Cancellations also occur for the other couplings but are very small in magnitude and occur for small values of $\varepsilon$, so they are not easily visible on the plot. For comparison, we have highlighted the current range of $\varepsilon_{\alpha\beta}^e$ allowed at 90\% CL, from an analysis of Borexino's Phase-II spectral data~\cite{Coloma:2022umy}, under the same assumption that only one NSI coupling is active at a time. 

From Fig.~\ref{fig:NumberOfEvents}, we can already conclude that while the event numbers are at an observable level in an experiment of this scale, obtaining statistics needed to discriminate against the SM for $\varepsilon^e_{\alpha \beta}\lesssim 0.5$ will require either a substantial increase in exposure (volume $\times$ time) or an extremely low electron recoil background. However, as we will motivate now, the ability of MPGD-based experiments to reconstruct both the recoil energy and the recoil direction independently will greatly alleviate the required background level.

\begin{figure*}[t]
    \centering
    \includegraphics[width=0.49\linewidth]{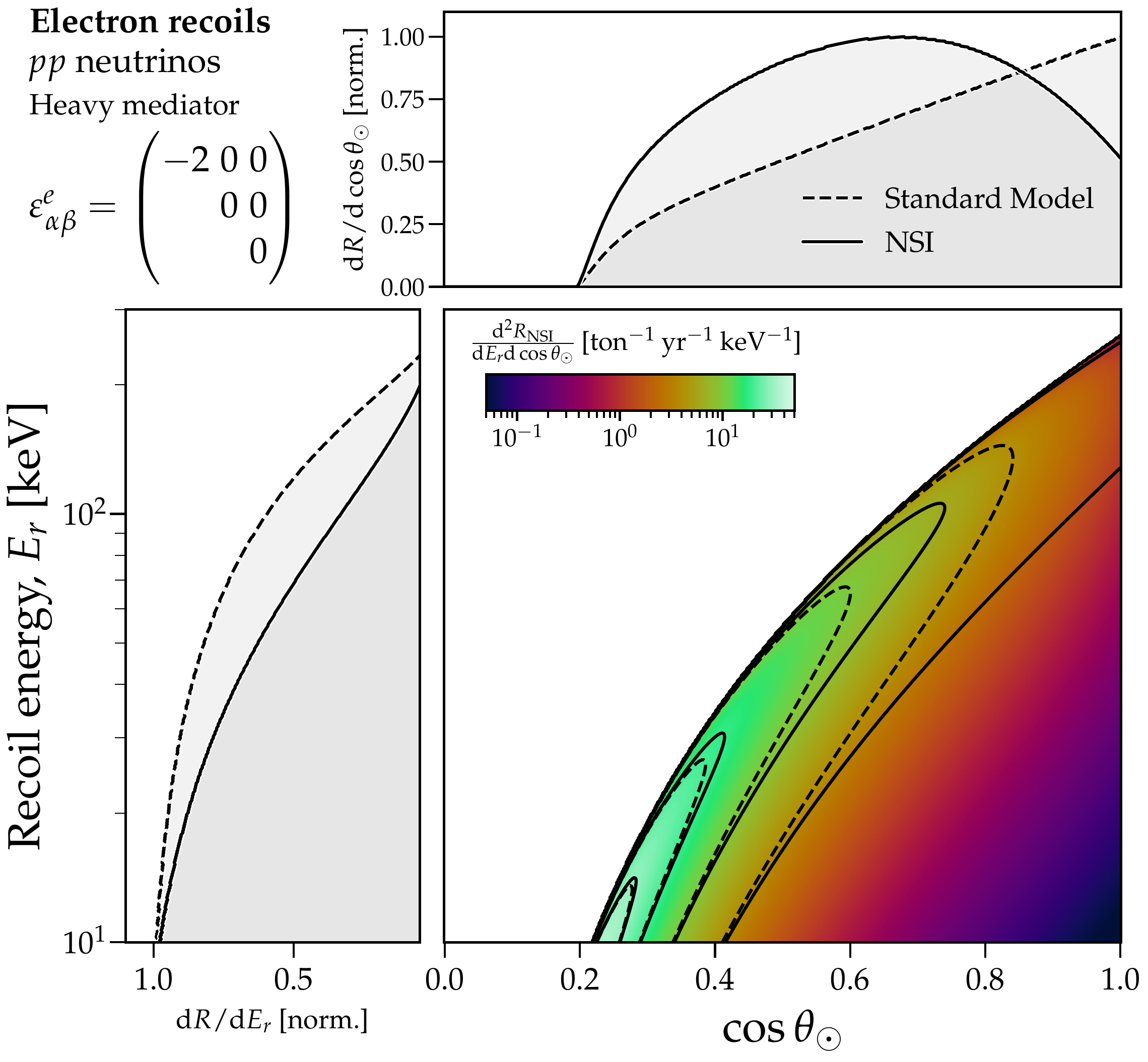}
    \includegraphics[width=0.49\linewidth]{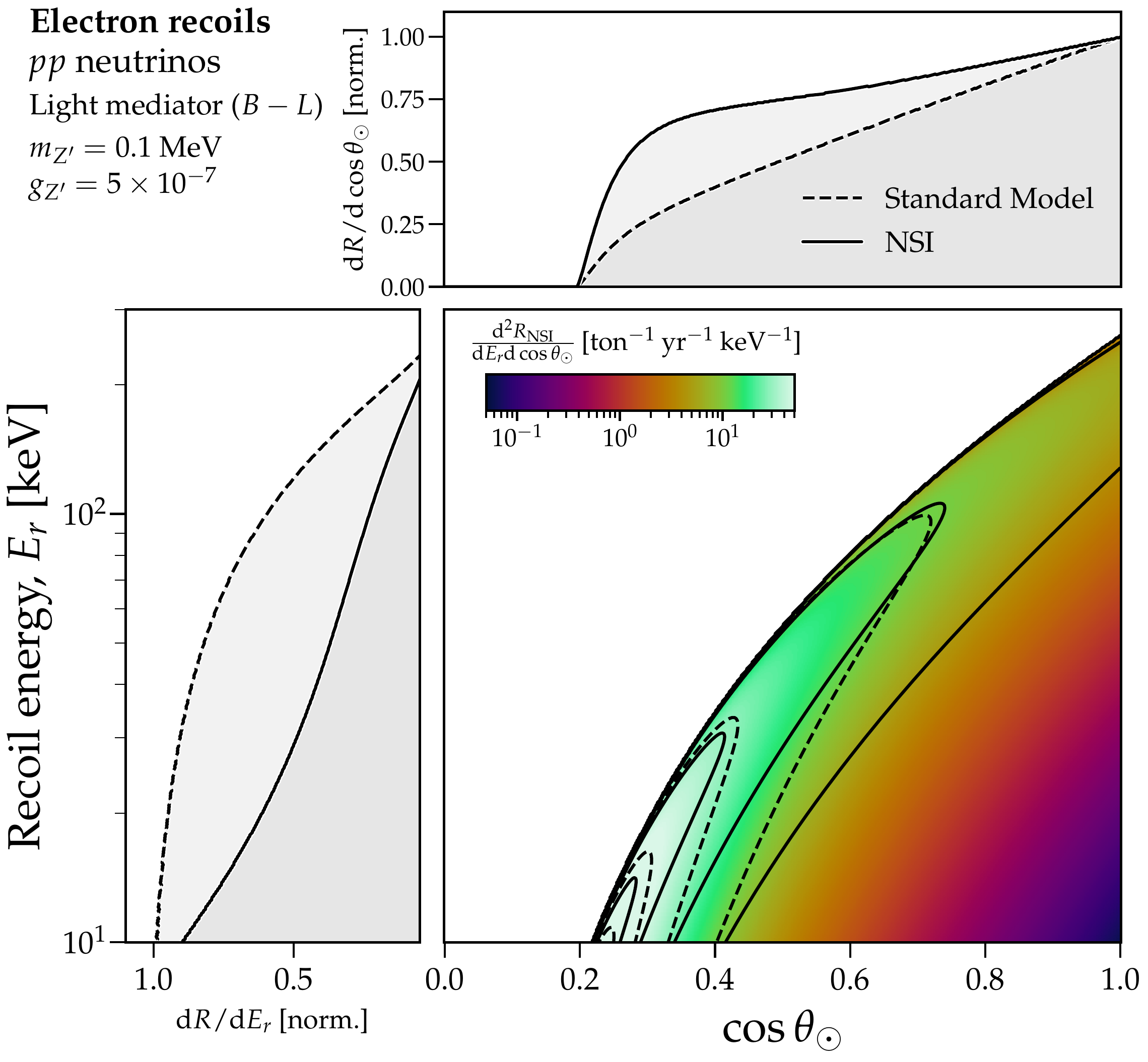}
    \caption{Two illustrative cases of the double-differential event rates as a function of recoil energy, $E_r$, and recoil angle with respect to the Sun, $\cos\theta_\odot$. The central coloured panels display $\textrm{d}^2R/\textrm{d}E_r\textrm{d}\cos\theta_\odot$ for the NSI case. We assign a set of $\varepsilon^e_{\alpha \beta}$ as labelled for the heavy mediator example (left), and ($m_{Z^\prime},g_{Z^\prime})$ as labelled for the $B-L$ light mediator example (right). The smaller panels to the left and above the main panel in each case show the event rate integrated over angle and energy, respectively. In all panels, the solid lines correspond to the NSI case while the dashed lines show the SM ($\forall \varepsilon_{\alpha \beta}^e = 0$ or $g_{Z^\prime} = 0$ for heavy and light mediators, respectively). In all of these cases, the lines correspond to the rate after it has been normalised by the maximum value so as to highlight differences in shape (although the total rate also differs, as shown in Fig.~\ref{fig:NumberOfEvents}). In the main panels, the solid and dashed contours also delineate equal values of the rate normalised by the maximum value---the four levels correspond to 0.75, 0.5, 0.25 and 0.1.}
    \label{fig:RecoilDistribution}
\end{figure*}

\begin{figure}[t]
    \centering
    \includegraphics[width=\linewidth]{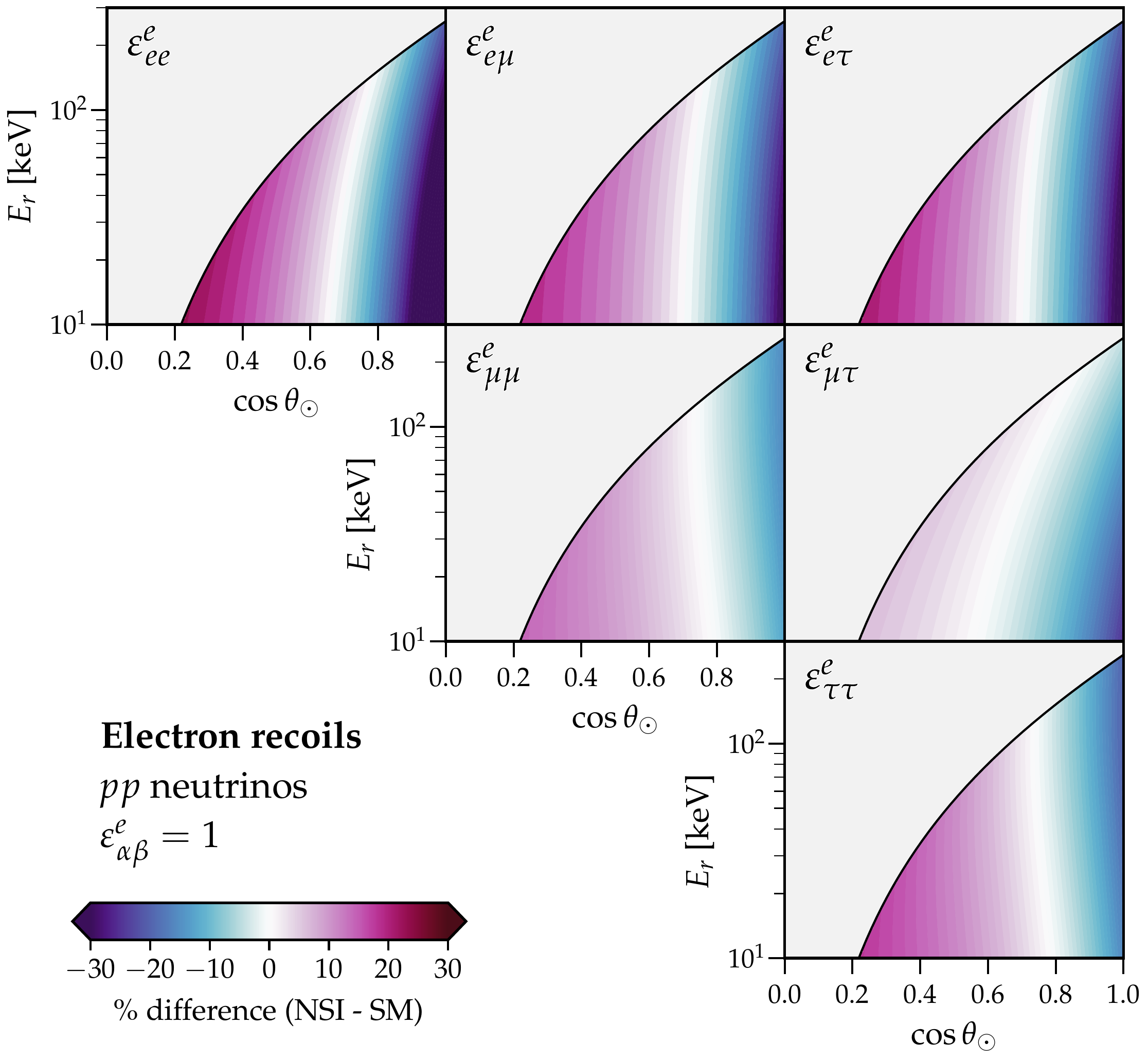}
    \caption{Similar to Fig.~\ref{fig:RecoilDistribution}, only now we show the relative difference in the energy-angle distributions of the recoils, which quantifies the differences compared to the SM for all six of the NSI parameters. More precisely, we are showing $(r_{\rm NSI} - r_{\rm SM})/r_{\rm SM}$ where $r = (1/R)\textrm{d}^2R/\textrm{d}E_r\textrm{d}\cos\theta_\odot$ is the double differential event rate normalised by the total rate. For each panel, we set all other $\varepsilon_{\alpha \beta}^e$ to zero apart from the component labelled, which we set equal to 1. This figure highlights that $\varepsilon^e_{ee}$, $\varepsilon^e_{e\mu}$, and $\varepsilon^e_{e\tau}$ have the largest relative impact on the angular distributions.}
    \label{fig:RelativeDifference}
\end{figure}

To highlight the effect of the NSI parameters on the energy-angle distribution, Figs.~\ref{fig:RecoilDistribution} and~\ref{fig:RelativeDifference} display the double-differential event rates. In this case, we are showing the double-differential event rate in heliocentric coordinates, i.e.~in terms of $\cos{\theta_\odot}$, where $\cos{\theta_\odot} = 1$ corresponds to a recoil pointed in the direction of the solar vector, $\hat{q}_\odot$. The side panels of Fig.~\ref{fig:RecoilDistribution} also show the differential event rate as a function of $\cos{\theta_\odot}$ and $E_r$ alone, i.e.~after integrating $\drm^2 R/\drm E_r\drm \cos{\theta_\odot}$ over energy and angle respectively. Both panels in Fig.~\ref{fig:RecoilDistribution} demonstrate that the recoil distribution remains pointed towards the solar direction. Recall that energy and angle are correlated: the highest energy recoils carry more of the incoming neutrino's forward momentum, so they will have small scattering angles or values of $\cos{\theta_\odot}$ close to 1. When NSIs are turned on, the event rate is generally enhanced more at low energies than at high energies, which means the angular distribution becomes slightly less focused towards the solar direction. We show two illustrative cases, one for the heavy mediator limit and one for the $B-L$ light mediator, but we have checked many different configurations and find that the behaviour seen in Fig.~\ref{fig:RecoilDistribution} is generic.

To illustrate this latter point further, Fig.~\ref{fig:RelativeDifference} shows the relative difference between the SM and the heavy mediator NSI cases, again when each coupling $\varepsilon^e_{\alpha\beta}$ is turned on, while all others are fixed at zero. The blue colours show regions of this space of observables where the NSIs lead to a suppression in the rate compared to the SM, while pink regions are where it receives an enhancement. We note that while the general behaviour is the same, there are subtle differences in the effects of each set of $\alpha\beta$ on $\drm^2 R/\drm E_r\drm\cos\theta_\odot$. These differences are not present at the level of the total event rate (Fig.~\ref{fig:NumberOfEvents}), where turning on different $\varepsilon^e_{\alpha\beta}$ couplings can lead to identical values for the total expected number of events. This result motivates the use of a detector able to measure both $E_r$ and $\cos\theta_\odot$ to attempt to search for the presence of NSIs and identify which interactions may or may not be active.

\section{Directional electron-recoil detection with gas TPCs}\label{sec:detector}
We now describe how we model a directional gas-based detector. The first quantity to fix is the target medium. In line with the promising simulations and sensitivity studies performed in the context of the CYGNO experiment, we will assume the use of atmospheric pressures of He:CF$_4$. This gas mixture is an electron drift gas and can allow high gas gains of $\mathcal{O}(10^6)$ with gas electron multipliers (GEMs), at the minor cost of worse diffusion compared with negative-ion drift gases like SF$_6$. At the relevant energies, electron recoils have much longer tracks than nuclear recoils, but lower ionisation densities, which means enhancing signal-to-noise is preferable over spatial resolution. The CF$_4$ gas also scintillates in the visible range, which as well as enabling the use of an optical readout (i.e.~cameras), also means the light can be timed to use for reconstruction of a track's angle with respect to the drift direction orthogonal to the readout. This is the strategy for full 3-dimensional spatial reconstruction adopted by CYGNO.

The addition of helium to the CF$_4$ has many advantages. In particular, it provides more scattering sites without substantially increasing the gas density, and so degrading the directional performance. It also allows the detector to operate at atmospheric pressure, simplifying the design and engineering of the gas vessel.

In Ref.~\cite{Lisotti:2024fco}, we motivated the use of a 60:40 ratio for He:CF$_4$ at atmospheric pressure for solar neutrino detection. Since preparations are underway by the CYGNO collaboration for a 30~m$^3$ scale detector using this gas mixture, we will fix $30$~m$^3 \times 5$~years as a near-term exposure benchmark, but will entertain further scale-ups if competitive sensitivity to some physics channel requires it. Since ongoing design studies are now focused primarily on materials screening and background reduction, our discussion will primarily focus on the impact of the total background rate $R_{\rm bg}$ as opposed to total volume. Discussion of our assumed background model can be found below in Sec.~\ref{sec:backgrounds}.
 
\subsection{Detector performance}
Having derived the input theoretical double-differential event rate $\textrm{d}^2 R/\textrm{d}E_r\textrm{d}\Omega_r$ in Sec.~\ref{sec:scattering}, we now describe how these event rates can be converted into functions that more closely resemble the space of observables for a gas-based time-projection chamber. 

The two main quantities a direction-sensitive detector aims to reconstruct are the total recoil energy and the initial recoil direction. This is done by reconstructing the three-dimensional ionisation cloud produced when the recoil deposits energy as it slows down in the detector medium. The full three-dimensional ionisation distribution is typically obtained by combining a 2-dimensional projection onto the highly pixelated MPGD readout, with the third axis relative to that plane reconstructed using timing. Depending on the type of readout involved, the arriving charge can either be detected directly (electronic readout) or the scintillation produced during the amplification stage can be imaged with high-resolution cameras (optical readout). A detailed understanding of the expected ionisation distributions from recoils of a given energy and particle type is then required to infer the initial energy and direction from the reconstructed ionisation track. This procedure involves either fitting the distributions using traditional methods, or using dedicated machine-learning algorithms~\cite{Vahsen:2021gnb, Ghrear:2024rku}. Since we are not performing full gas simulations here, our goal is to write down simple functional descriptions of the physical effects that limit the ability to reconstruct the initial recoil direction and energy perfectly, inspired by the performance of existing algorithms tested on experimental data. We do this following the same strategy as in Ref.~\cite{Lisotti:2024fco}, which we recap briefly here.

We emulate the performance of the detector by convolving the underlying theoretical event rate with several performance kernels. These describe the major effects that degrade the ability of the detector to reconstruct the correct energy and $\cos\theta$. These effects are especially important at low energies when the tracks become small and signal-to-noise decreases. The most efficient approach to apply these effects is at the event level, by running a Monte Carlo simulation. The procedure we follow is described further in Ref.~\cite{Lisotti:2024fco}, in short, it amounts to first generating recoil energies and angles from $\drm^2 R/\drm E_r\drm \cos\theta_\odot$, and then resampling those energies and angles from a set of kernels that describe the energy and angular resolutions. The final step is then to randomly assign a vector sense to the recoil direction according to a head-tail efficiency function. We describe these functions below. Then, when those effects are applied, we re-bin the events into a new `observed' double-differential event rate, $\drm^2 R_{\rm obs}/\drm E_r \drm \cos{\theta_\odot}$.

The effect of finite energy resolution is captured by a Gaussian with width $\sigma_E$ that depends on energy. Given a true underlying value of $E_r$ we draw a `measured' recoil energy $E^\prime_r$ from the distribution,
\begin{equation}
    K_E\left(E_{r}, E_{r}^{\prime}\right)=\frac{1}{\sqrt{2 \pi} \sigma_E\left(E_{r}\right)} \exp\left(-\frac{\left(E_{r}-E_{r}^{\prime}\right)^{2}}{2 \sigma_E^{2}\left(E_{r}\right)}\right) \, .
\end{equation}
We use the following simple function for the energy spread,
\begin{equation}\label{eq:energyresolution}
    \frac{\sigma_E}{E_r} \,[\%]=11+\frac{21}{\sqrt{E_r/{\rm keV}}} \, ,
\end{equation}
which is fit to electron recoil data from the CYGNO-LIME detector~\cite{SamueleThesis}. The reference energy of 35~keV is arbitrary, but here we choose it to lie at the 10th percentile of the recoil energy distribution above the threshold of 10 keV, which means we can interpret the value of $p_E$ as the resolution achieved by 90\% of the recoil spectrum. 

Next, the reconstruction of $\cos{\theta}$ is degraded by an angular resolution kernel taken to be a von Mises-Fisher distribution~\cite{Ghrear:2024rku},
\begin{equation}
K_{\theta}(\hat{\mathbf{q}}, \hat{\mathbf{q}}^\prime,E_r)=\frac{\kappa(E_r)}{4 \pi \sinh \kappa(E_r)} \exp \left(\kappa(E_r) \mathbf{q} \cdot \mathbf{q}^\prime \right) \, ,
\end{equation}
where the parameter that specifies the variance of the von Mises-Fisher distribution is $\kappa$. Rather than quoting this parameter, we will express the angular resolution in terms of the more intuitive variable, the root-mean-squared axial angular resolution $\sigma_\theta$, which is a monotonic function of $\kappa$ that can be evaluated numerically. We find this relationship by calculating the root-mean-square axial angle between a sample of recoil vectors generated according to the distribution above, and then repeating this process over a suitable range of $\kappa$. The von Mises-Fisher distribution has the advantage over, for example, a truncated Gaussian, in that it reflects the fact that the distribution exists on the sphere.

The root-mean-squared angle between two axes approaches $\sigma_\theta = 1$~radian for completely uncorrelated directions (i.e.~the limit of no directional sensitivity). Following on from the discussion of $\sigma_E(E_r)$ above, we approximate the energy-dependence of the angular resolution with the following function form,
\begin{equation}\label{eq:angularresolution}
    \sigma_\theta(E_r) = p_\theta \sqrt{\frac{E_{\rm ref}}{E_r}} \, ,
\end{equation}
where $p_\theta = 10^\circ$ for our baseline detector performance benchmark, as inspired by results from gas simulations of electron recoils at the relevant energies~\cite{SamueleThesis, Ghrear:2025iry}. Note that the $E_r$ appearing in this equation is the true recoil energy, not the reconstructed energy, $E_r^\prime$.

One important assumption that we should emphasise here is that, by taking $\cos{\theta_\odot}$ as our derived measurable quantity, we are assuming that the detector has three-dimensional reconstruction, and so the track direction can be rotated into heliocentric coordinates. Although we do this for simplicity (relaxing this assumption would force us to also consider the event time and detector orientation), the ability to perform three-dimensional direction reconstruction is considered to be a major design feature in all MPGD-based detectors being developed right now. CYGNO is able to achieve this by first measuring the two-dimensional projection of the recoil track along the optical readout plane made up of highly sensitive CMOS cameras, which are placed beyond an amplification stage made up of GEMs. This projection is then combined with the one-dimensional projection along the drift axis provided through the timing of scintillation light detected by photomultiplier tubes. The combination of these signals is what also enables the precise reconstruction of the recoil energy.

The final detector performance effect we account for is the limitation in reconstructing the correct $\pm$ sense of the recoil direction. This is a particularly important effect at the lowest energies, where tracks are shortest and the ionisation density is low. It is possible at these energies that the track alignments could be reconstructed even while the head and the tail of the track are not. The head and tail of a recoil are primarily reconstructed by using the differences in the $\drm E/\drm x$ along the track. Unlike nuclear recoils, which are on the stopping side of the Bragg peak, the electrons deposit \textit{more} ionisation as they slow down. So the tail (end) of the track curls up and contains the most ionisation. Since in our simplified description, head-tail assignment is binary, we simply randomly assign a $\pm$ sign to each recoil vector with a probability given by the function,
\begin{equation}
    \varepsilon_{\rm HT}(E_r) = \frac{1}{2}\left(\frac{1}{1+e^{-a_{\rm HT}(E_r-E_{\rm HT}))}}+1 \right) \, .
\end{equation}
This functional form captures the behaviour of head-tail performance well for all recoils, which, by definition, must begin at zero at $E_r = 0$, and plateau at 1 for high-enough energies. For our baseline performance we assume $a_{\rm HT} = 0.2$ and $E_{\rm HT} = 20$~keV. Since the neutrino recoils are restricted to a single hemisphere due to the scattering kinematics (up to the effects of angular resolution), the primary effect of the head-tail performance is to limit the ability to distinguish neutrino events from isotropic backgrounds. 

The final effect we should include in principle is the event-level efficiency of the experiment. This is more difficult to quantify at present, as it will depend on the specifics of a given detector and the way the data is collected and analysed. To keep the discussion in this study as clear as possible for when our results are applied in real detector simulations by the CYGNO collaboration, we simply incorporate a step-function efficiency curve by imposing a hard threshold at 10 keV. Our results are relatively weakly dependent on this threshold since the electron recoil distribution due to $pp$ and $^7$Be neutrinos is not very steeply falling at these energies, and in any case, all of our energy and angular performance curves have already degraded significantly by this energy.

\subsection{Backgrounds}\label{sec:backgrounds}
The rate of non-neutrino electron-recoil backgrounds in experiments of this type is the main quantity that will limit their sensitivity. Fortunately, the directionality will provide an excellent technique to reject backgrounds, which are expected to be close to isotropic. Nevertheless, backgrounds will appear in the signal region defined by $\cos\theta_\odot \in [0,1]$, so we must incorporate some background model that quantifies this. 

We make an educated guess as to the rate and energy spectrum of the background using recent background simulations performed in preparation for a future 30~m$^3$ scale-up of the CYGNO experiment to be located at LNGS. Note that the general design parameters of CYGNO-30, in terms of gas mixture and directional performance, have already inspired most of the choices we have made in this study so far. The differential recoil spectrum for an isotropic background model can be written as follows,
\begin{equation}
    \frac{\textrm{d}^2 R_{\rm bg}}{\textrm{d} E_r \textrm{d}\Omega_r} = \frac{R_{\rm bg}}{4\pi E_{\rm bg}(e^{-E_{\rm th}/E_{\rm bg}} - e^{-E_{\rm max}/E_{\rm bg}})} e^{-E_r/E_{\rm bg}} \, .
\end{equation}
For our chosen model we have $E_{\rm th} = 10$~keV and $E_{\rm max} = 700$~keV. The model is parameterised by one free parameter, the total background rate $R_{\rm bg}$. We fix the scale of the exponential fall off of the background energy spectrum to be $E_{\rm bg} = 67$~keV, which we find to be a good fit to the spectrum of the total internal background from a simulation of CYGNO-30~\cite{SamueleThesis}. This simulation includes all material electron backgrounds in the 0--1000~keV energy window, with the field cage being the dominant contributor to the background budget, followed by the GEMs and camera sensors.

As for the amplitude of the background rate, $R_{\rm bg}$, to make the discussion more intuitive, we will express its value relative to the total event rate from solar neutrinos $R_{\nu}$ under the SM (see the grey dashed line in Fig.~\ref{fig:NumberOfEvents}). Importantly, the simulations for CYGNO~\cite{SamueleThesis} suggest $R_{\rm bg}/R_{\nu}\sim 500$ is \textit{already} achievable without any dedicated background reduction which, as we will discuss, is quite promising. In particular, this study did not account for the possibility of using the ultra-low radioactivity kapton presented in Ref.~\cite{Arnquist:2019fkc}, and so naively a further background reduction is possible already. To see the impact of the size of the background on the scientific reach of the experiment we will explore a wide range of possible background reduction factors, while keeping in mind that $R_{\rm bg} = 50 R_{\nu}$ is a realistic expectation for a fully-deployed experiment of $30~$m$^3$ volume. To demonstrate the power of directionality, in our likelihood analysis described in the next section, we will float the value of $R_{\rm bg}$ as an unknown nuisance parameter.

\section{Likelihood analysis}\label{sec:statistics}
\begin{figure*}[t]
    \centering
    \includegraphics[width=\linewidth]{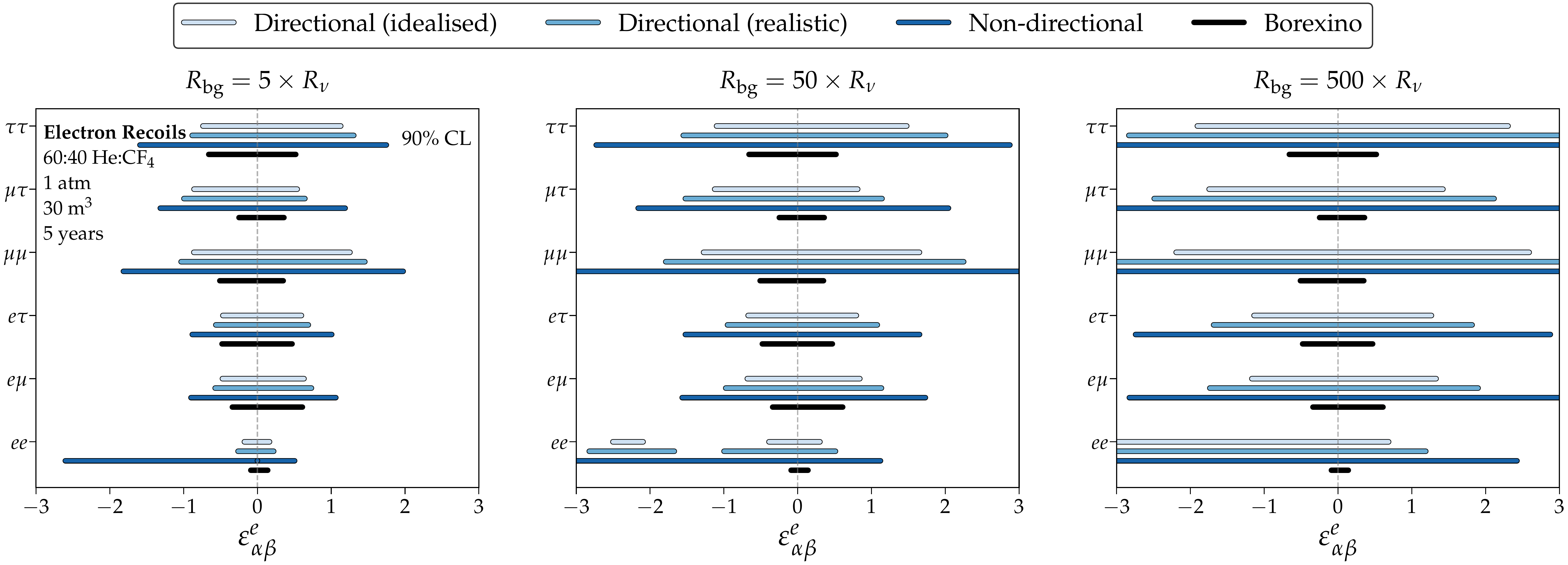}
    \caption{Projected median 90\% CL exclusion limits on each $\varepsilon_{\alpha\beta}^e$ parameter (assuming all other $\varepsilon_{\alpha\beta}^e = 0$). We assume a detector of 30 m$^3$ volume with a 60:40 He:CF$_4$ gas mixture at atmospheric pressure. The three panels from left to right are for different background rates, measured relative to the total rate of neutrino-induced electron recoils under the SM, $R_\nu$. The various projected limits shown in blue are for different levels of directional sensitivity: idealised performance, realistic performance, and no directionality, from lightest to darkest. We emphasise that the ``non-directional'' case has an identical event rate, energy threshold and energy resolution as the ``directional'' cases. For comparison, in black we show the existing constraints from Borexino, demonstrating a competitive level of sensitivity despite the small size of the proposed experiment.}
    \label{fig:Exclusion_ERs_1d}
\end{figure*}

\begin{figure*}[t]
    \centering
    \includegraphics[width=\linewidth]{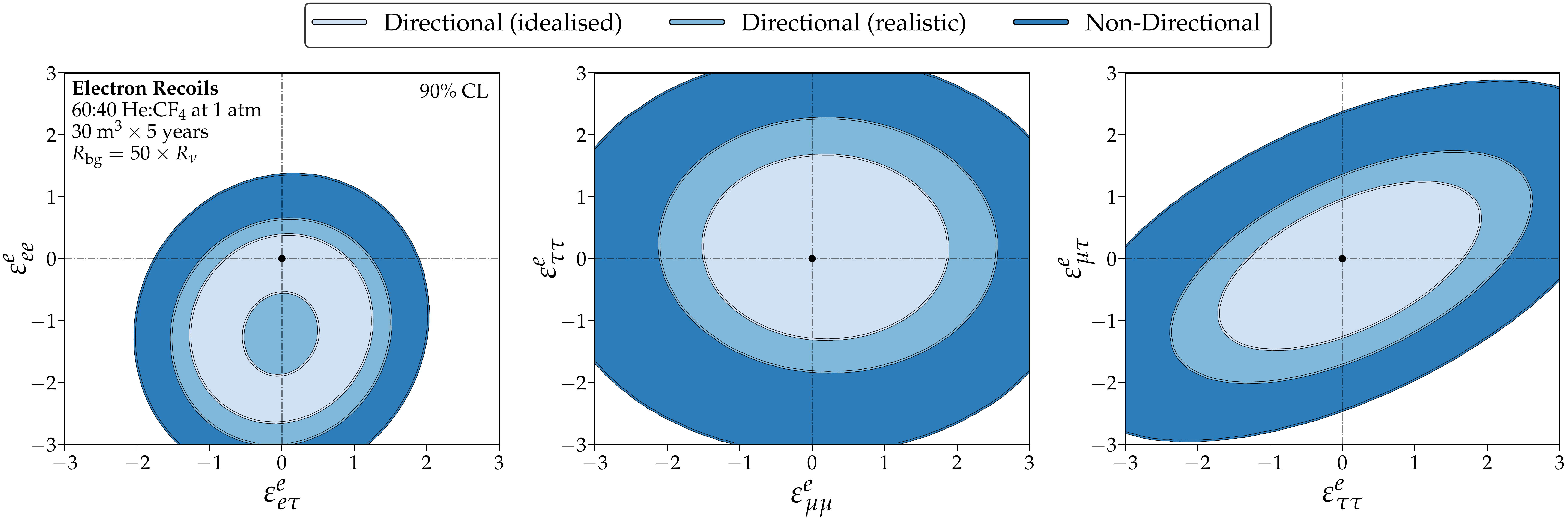}
    \includegraphics[width=\linewidth,trim={0 0 0 130},clip]{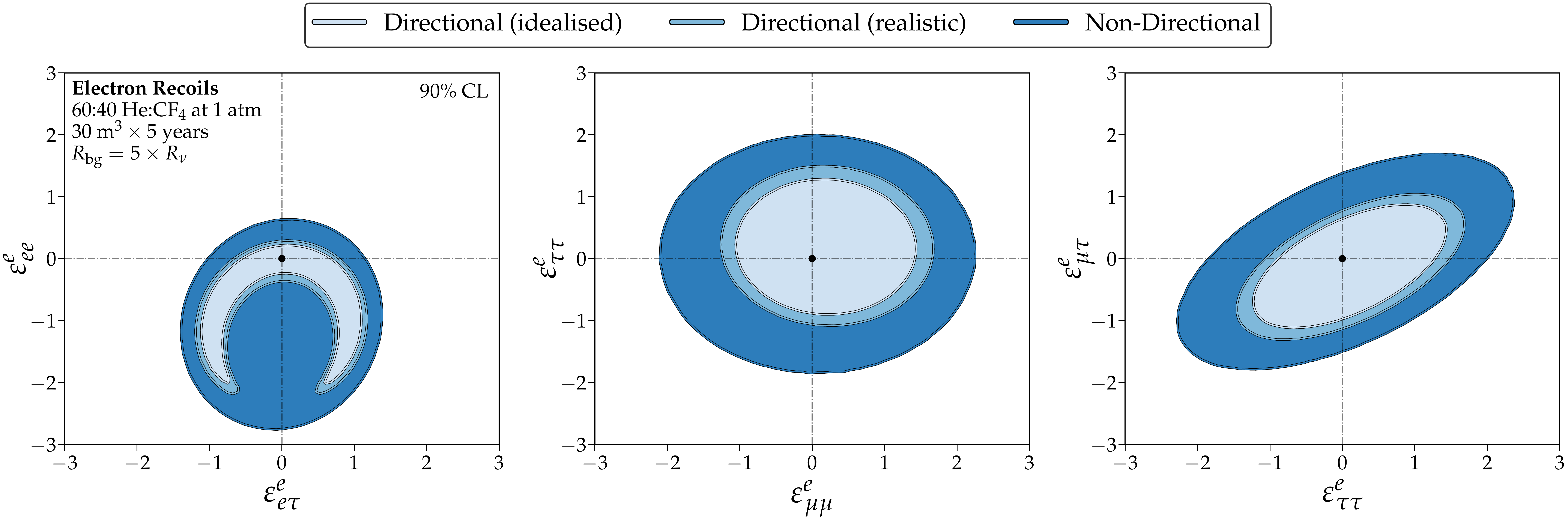}
    \caption{Projected median 90\% CL exclusion limits when two $\varepsilon_{\alpha\beta}^e$ parameters are considered simultaneously (assuming all other $\varepsilon_{\alpha\beta}^e = 0$). We assume a detector of 30 m$^3$ volume with a 60:40 He:CF$_4$ gas mixture at atmospheric pressure. The three columns correspond to three different pairs of NSI parameters. The top set of panels is for a background rate that is fifty times the neutrino electron recoil rate under the SM, while the bottom set of panels is for five times the neutrino rate. The various projected limits shown in blue are for different levels of directional sensitivity: idealised performance, realistic performance, and no directionality, from lightest to darkest. We emphasise that the ``non-directional'' case has an identical event rate, energy threshold and energy resolution as the ``directional'' cases.}
    \label{fig:Exclusion_ERs_2d}
\end{figure*}

\begin{figure}[t]
    \centering
    \includegraphics[width=\linewidth]{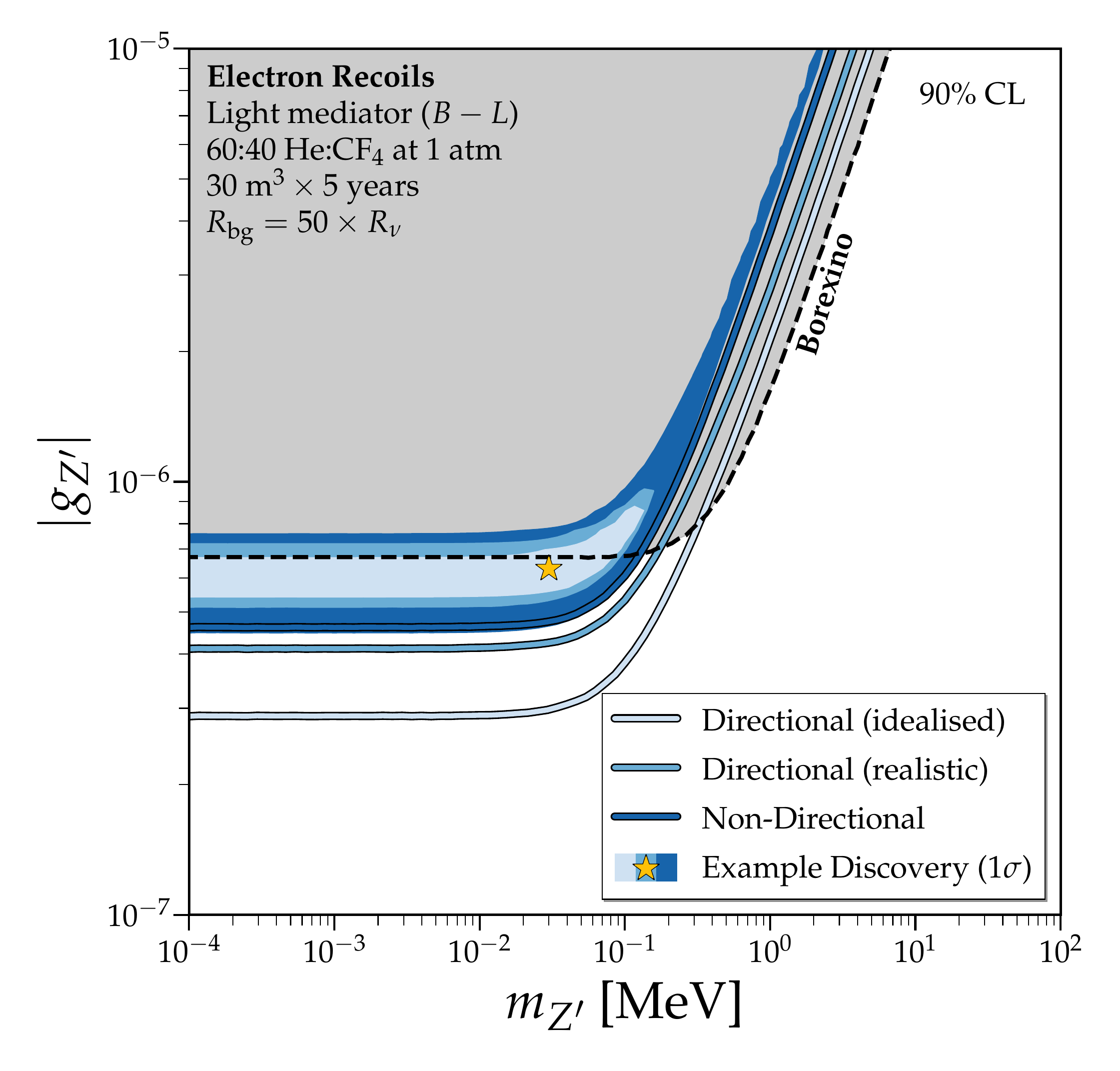}
    \caption{Projected constraints on a light vector mediator under the $U(1)_{B-L}$ model, as a function of the coupling $g_{Z^\prime}$ and mediator mass $m_{Z^\prime}$. The solid lines show median 90\% CL exclusion limits for the three experimental benchmarks used in the previous figures. The constraint from Borexino from Ref.~\cite{Coloma:2022umy}  is shown for comparison. The shaded regions show example 1$\sigma$ \textit{discovery} contours if the true values of the parameters were situated at the gold star. We emphasise that the directional information (and the background rejection that comes with it) is what would allow such a discovery claim to be made. Following conventions, the shaded region encloses the \textit{excluded} parameter space, as opposed to the previous figures, where the shaded regions were the \textit{allowed} parameter space.}
    \label{fig:LightMediator}
\end{figure}
Having described all the necessary ingredients for calculating a reasonable approximation to the neutrino signal detectable by a gas-based directional detector, the final step is to develop a statistical methodology. Our previous analysis in Ref.~\cite{Lisotti:2024fco} focused on the ability of the detector to measure the neutrino flux over some experimental background. Here, our scientific question is slightly different. Given an observed set of neutrino-induced electron recoil energies and directions, is their distribution consistent with the SM or not? Fortunately, the SM event rate can be expressed as a special case of the more general NSI formalism (i.e.~$\forall \varepsilon_{\alpha\beta}^e = 0$, or $g_{Z^\prime} = 0$) which means our null hypothesis that the SM is true is nested within a more general alternative hypothesis that the event rate is described by Eq.(\ref{eq: Solar NSI rate}). This then permits us to use the profile likelihood ratio test and apply Wilks' theorem.

We start by writing down a likelihood. To facilitate the use of Asimov statistics (see below), we will use a binned likelihood based on the event rate having been binned in energy and $\cos{\theta_\odot}$, i.e.,
\begin{align}
    N^{ij}_{\nu}(\varepsilon)& = MT\times\\&\int_{\cos\theta_\odot^i}^{\cos\theta_\odot^i+\Delta \cos\theta} \int_{E_r^j}^{E_r^j+\Delta E_r} \frac{\drm^2 R_{\rm obs}(\varepsilon)}{\drm E_r \drm\cos\theta_\odot} \nonumber \drm E_r  \  \drm\cos\theta_\odot
\end{align}
where $MT$ is the exposure (mass $\times$ duration) of the experiment. Our bins run over the domain $\cos\theta_\odot \in [-1,1]$ and $E_r \in [10,700]$~keV across 50 linearly spaced bins each. The exact binning scheme does not impact our results since the finite energy and angle reconstruction are already taken care of, as discussed in Sec.~\ref{sec:detector}. For compactness, we use the shorthand here $\varepsilon \equiv \{\varepsilon^e_{\alpha\beta}\}$ to denote the set of all $\varepsilon$ values upon which we are performing the statistical test. The background distribution is constructed using the same binning scheme. We parameterise the background using a normalisation $b$ whose correct value is one. This means we can parameterise an expected number of events in bin $ij$ as,
\begin{equation}
    N_{\rm exp}^{ij}(\varepsilon,b) = N_\nu^{ij}(\varepsilon) + b N_{\rm bg}^{ij}
\end{equation}

We now construct a binned likelihood for the parameters $\{\varepsilon,b\}$ given some observed data $N_{\rm obs}^{ij}$ which we assume to be Poisson distributed with mean $N^{ij}_{\rm exp}$,
\begin{equation}
    \mathcal{L}(\varepsilon,b) = \prod_i \prod_j {\rm Poiss}\left[N^{ij}_{\rm obs};N^{ij}_{\rm exp}(\varepsilon,b)\right] \, .
\end{equation}
In practice, we use the log-likelihood, which can be written as,
\begin{align}
    \ln \mathcal{L}(\varepsilon,b) = \sum_{ij} \bigg(N^{ij}_{\rm obs} \ln [N^{ij}_{\nu}(\varepsilon) + bN_{\rm bg}^{ij}]  \\ - (N^{ij}_{\nu}(\varepsilon) + bN_{\rm bg}^{ij})\bigg)\nonumber
\end{align}
we neglect the $N_{\rm obs}!$ term which will cancel when we take likelihood ratios. 

We wish to construct a profile-likelihood ratio test that quantifies the extent to which the null hypothesis of the SM, i.e.~$\varepsilon = 0$, describes the data. To do this, we will assume that the observed data \textit{is} described by the SM and forecast the expected exclusion limits one could then set on $\varepsilon$. We write down the standard profile likelihood ratio test statistic used for deriving exclusion limits as~\cite{Cowan:2010js},
\begin{equation}
    {\rm TS}(\varepsilon) = -2\left( \ln \mathcal{L}(\varepsilon,\hat{\hat{b}}) - \ln \mathcal{L}(\hat{\varepsilon},{\hat{b}}) \right)\nonumber
\end{equation}
where $\hat{\hat{b}}$ is the maximum likelihood estimator (MLE) at some fixed value of $\varepsilon$, whereas $\hat{b}$ and $\hat{\varepsilon}$ are the MLEs for the full likelihood. Wilks' theorem holds in this model setup, which states that the distribution of TS asymptotically follows a $\chi^2_{n}$ distribution where $n$ is the number of parameters in $\varepsilon$.

To make an estimate of the expected exclusion limit that a future experiment could set, we employ the Asimov dataset. The Asimov dataset is a hypothetical situation in which the observed data exactly follows the expectation under some assumed set of ``true'' parameter values. Since we are projecting exclusion limits, where we imagine that the SM hypothesis correctly describes the data, we set $N_{\rm obs} = N_{\rm exp}(\forall \varepsilon = 0,b=1)$. 
The MLEs are then $\hat{b} = 1$ and $\hat{\varepsilon} = 0$, while $\hat{\hat{b}}$ can be found numerically by solving,
\begin{equation}
    \frac{\partial \ln \mathcal{L}(\varepsilon,b)}{\partial b} = \sum_{ij} \left(\frac{N_{\rm obs}^{ij}N_{\rm bg}^{ij}}{N^{ij}_\nu(\varepsilon) + b N_{\rm bg}^{ij}} - N_{\rm bg}^{ij}\right) = 0 \, .
\end{equation}

As described in Ref.~\cite{Cowan:2010js}, the value of TS under the Asimov data set asymptotically approaches the median of its distribution. That is to say, when we forecast an exclusion limit using the Asimov data set, we find a good estimate of the level of sensitivity that at least 50\% of hypothetical experiments could achieve. To forecast a limit at a given CL we then simply need to find values of $\varepsilon$ at which the TS crosses some threshold value. For setting two-sided limits at the $1-p$ level on the $n$ parameters in $\varepsilon$ we use the $1-p$ percentile function of the $\chi^2_n$ distribution. For one-sided limits, e.g.~on the parameter $g_V$ in the light-mediator case, we use the $1-2p$ percentile function.

\section{Results}\label{sec:results}
We will now show estimated median 90\% exclusion limits on the various $\varepsilon_{\alpha \beta}^e$ and $(g_{Z^\prime}, m_{Z^\prime})$ parameters for a range of choices for the directional performance and background conditions. Since we are motivated by the upcoming 30~m$^3$ detector being developed by the CYGNO collaboration, we will present results for this fixed total volume, and provide an analytic scaling to understand how to extrapolate our results to larger volumes. Our primary goal is to motivate small-scale direction-sensitive gas detectors in general, as well as to understand the level of electron recoil background that is required to set relevant constraints on the neutrino cross section using a detector of this type.

To have something to compare our results against, it is useful to define what we call an equivalent ``non-directional experiment''. This is a hypothetical experiment that observes an identical event rate as our directional experiment, but does not have access to any $\cos\theta_\odot$ information. We can forecast limits for this experiment in the same way, but instead of $N_{\rm exp}^{ij}$ appearing in the likelihood, we first take the sum over $\cos\theta_\odot$ bins in the equation above, i.e.
\begin{equation}
    N_{\rm exp,non-dir}^{j}(\varepsilon) = \sum_i N^{ij}_{\rm exp}(\varepsilon) \, .
\end{equation}
As another point of comparison, we also define an ``idealised performance'' benchmark where the energy and angular performances are not included. This amounts to calculating $N_{\rm exp}$ directly from the underlying $\textrm{d}^2R/\textrm{d}E_R \textrm{d}\cos\theta_\odot$ as opposed to $\textrm{d}^2R_{\rm obs}/\textrm{d}E_R \textrm{d}\cos\theta_\odot$ which includes the performance effects. All of the figures in this section will then show three benchmark experiments: non-directional, directional (realistic) and directional (idealised)---with the expected sensitivity improving in that order. We reiterate at this point that our goal is not to claim world-leading sensitivity to NSIs, but to highlight that---thanks to directionality---a gas-based experiment can make an interesting contribution to the global picture of the neutrino's properties, despite being several orders of magnitude smaller and cheaper than its nearest competitors, e.g., Borexino.

First, in Fig.~\ref{fig:Exclusion_ERs_1d} we show a set of forecasted two-sided median 90\% CL exclusion limits (${\rm TS} = 2.71$) on various $\varepsilon_{\alpha \beta}^e$ parameters in the heavy mediator limit. These are one-dimensional constraints where we turn on one NSI coupling from $\alpha\beta = \{ee,e\mu,e\tau,\mu\mu,\mu\tau,\tau\tau\}$, with all others set to zero. As a direct point of comparison, we show the two-sided 90\% CL exclusion limits from an analysis of Borexino Phase-II spectral data~\cite{Coloma:2022umy}. In all cases, the gas mixture, exposure time, volume, and pressure are fixed as labelled in the figure. The three panels from left to right vary only the background rate, $R_{\rm bg}$, from 5 to 500 times the SM neutrino rate. The latter value is the current estimate of the background level from simulations, prior to any dedicated screening or reduction efforts. We see that the constraints for the worst background condition are not at a competitive level, but after only a $\times 10$ reduction, the constraints for certain couplings like $\varepsilon_{e\tau}^e$ reach within a modest factor of Borexino's. For an even more aggressive reduction of a factor of 100, the exclusion limits almost match those of Borexino. The case for directionality in particular is made the most apparent for $\varepsilon_{ee}^e$ where the added information provided through the $\cos{\theta_\odot}$ distribution as well as the background rejection power allows for the region around $\varepsilon_{ee}^e \approx -2$ to be excluded---this point in the parameter space is mildly degenerate with the SM because it leads to the same total number of events. Across all cases, we observe that the exclusion limit on $\varepsilon$ scales as,
\begin{equation}
    \varepsilon \propto (MT)^{-1/4} R_{\rm bg}^{1/4} \, .
\end{equation}

It is also possible for multiple NSI couplings to be active at once and to interplay with each other at the level of the event rate. To capture this possibility, we show three more examples in Fig.~\ref{fig:Exclusion_ERs_2d}. Now we are showing two-dimensional 90\% CL exclusion limits on two degrees of freedom (${\rm TS} = 4.61$), where two $\varepsilon^e_{\alpha\beta}$ are turned on, with all others fixed at zero. The top and bottom sets of panels are again for two background assumptions: 50$\times$ and 5$\times$ the neutrino rate, respectively. Again, we see from comparing the darkest blue contours (non-directional) with the two lighter contours that the power of directionality is apparent. It is clearest again in the $ee$ case, where the degeneracy with the SM for large negative values of $\varepsilon^e_{ee}$ is broken.

Then finally, we come to the light mediator example in Fig.~\ref{fig:LightMediator}. As discussed in Sec.~\ref{sec:scattering} we are showing only the $B-L$ model here with $\forall Q_{Z^\prime}^{e,\nu_{\alpha}} = -1$, but have checked that we obtain almost identical results for other models. We adopt a slightly different definition for the exclusion limits here and following the typical convention---here we have forecasted the median \textit{one-sided} 90\% CL exclusion limits on one parameter, $g_{Z^\prime}$, with the two-dimensional constraint then built up by repeating the test over a range of values for $m_{Z^\prime}$. The Borexino constraint is again from Ref.~\cite{Coloma:2022umy}.\footnote{We note that the Borexino constraint shown here is actually a two-sided 90\% CL exclusion on $g_{Z^\prime}$ rather than the one-sided limit on $|g_{Z^\prime}|$ as we have shown. The differences between the two definitions are almost invisible on the plot at the scale we have shown it, so we choose not to debate this choice.} We note that there are various other constraints from stellar cooling, fixed target experiments, and BBN/$N_{\rm eff}$ that are also present for these light mediator models that partially overlap this region. We choose to remove these for the sake of clutter and because our main message is the comparison between these two solar neutrino experiments. We note that it is always possible to model build around any of these constraints, and not all are present for every instance of an anomaly-free gauged $U(1)^\prime$ model.

As expected, the sensitivity shown in Fig.~\ref{fig:LightMediator} is best at low mediator masses where the deviation in the cross section due to NSIs emerges at energies that are below Borexino's threshold. Borexino, of course, still leads at higher mediator masses. We show the case here for $R_{\rm bg} = 50~R_{\nu}$, but sensitivity scales slowly $g_{Z^\prime}\propto (MT)^{-1/8} R_{\rm bg}^{1/8}$, so the results for $5~R_{\nu}$ are similar.

For this figure, in addition to the forecasted exclusion limits, we show another way to demonstrate the power of directionality by forecasting a possible discovery. The ability to make discoveries is another of the major motivations behind directional detection that has been appreciated for many years (see e.g.~the discussion in the context of dark matter in Ref.~\cite{Vahsen:2021gnb}). Although experiments like XENONnT, LZ and PandaX have the statistics to set \textit{constraints} at a similar level to what is shown here in principle, in practice, it is challenging for them to make positive claims of a signal due to the high electron background preventing the robust identification of neutrino events. This is not the case for a directional experiment, where a much stronger statement about the probability of any given recoil as being due to a neutrino is possible, thanks to the angular domain being split into an (almost) pure background hemisphere and a signal+background hemisphere. To showcase this we show on Fig.~\ref{fig:LightMediator} some example median 1$\sigma$ discovery contours, given some arbitrarily chosen hypothetical ``true'' value for $(g_{Z^\prime},m_{Z^\prime}) =(g^{\rm true}_{Z^\prime},m^{\rm true}_{Z^\prime}) $ shown by the gold star. We calculate this contour using a ``discovery'' test statistic, which essentially amounts (in the Asimov procedure) to fixing $N_{\rm obs} = N_{\nu}(g^{\rm true}_{Z^\prime},m^{\rm true}_{Z^\prime}) + N_{\rm bg}$, whereas previously for exclusion limits we set $g_{Z^\prime}^{\rm true} = 0$. Comparing the resulting contours between the directional and non-directional examples, we again see the benefit that comes with directionality.

\section{Conclusions}\label{sec:conc}
In this paper, we have motivated the continued development of 30~m$^3$-scale gaseous detectors for the directional detection of solar neutrinos through electron recoils. In alignment with the message of Ref.~\cite{Lisotti:2024fco}, we argue that an experiment of this type can perform neutrino physics, trading in cost and overall size for the added signal information and background rejection that comes from directional sensitivity. We illustrated this in the context of the neutrino NSI formalism, where the interaction between neutrinos and electrons is modified by the presence of a new vector $Z^\prime$ mediator beyond the SM. 

Our results are illustrated in Figs.~\ref{fig:Exclusion_ERs_1d} and~\ref{fig:Exclusion_ERs_2d} in the heavy mediator limit ($m_{Z^\prime}\gg q$), where we permit both flavour-conserving and flavour-violating interactions. Then, in Fig.~\ref{fig:LightMediator}, we showed the case for a light mediator, where we allow for only flavour-preserving interactions, as expected in the gauged $U(1)_{B-L}$ model. We have not attempted an exhaustive study of all possible combinations of NSI flavour structures, alternative mediator types, or other $U(1)^\prime$ models, as the results presented here are sufficient to convey our main message. We believe a background reduction of around a factor of ten is needed from preliminary detector background simulations to reach an interesting level of sensitivity to neutrino interactions beyond the SM using solar neutrinos.

Our results remain to be validated in a full Monte Carlo simulation of a detector like this. This work, as well as the background reduction efforts that we advocate for here, is ongoing within the CYGNO collaboration~\cite{SamueleThesis}. Our results suggest a promising future for this technology. The next stage of CYGNO and the detectors of other groups within the \Cygnus consortium~\cite{Vahsen:2020pzb} will be the first in a sequence of gas-based directional detectors to bring a range of complementary capabilities to the fleet of worldwide neutrino observatories.

\acknowledgments
We thank the members of the CYGNO and \Cygnus collaborations for discussions, and in particular Elisabetta Baracchini and Samuele Torelli for providing details about CYGNO's simulation results. CL and CAJO are supported by the Australian Research Council under the grant numbers DE220100225 and CE200100008. LES, NM, ACS are supported by the DOE Grant No. DE-SC0010813. 

\appendix

\section{Effective neutrino oscillations}\label{app: prob}
In Sec.~\ref{sec: NSI Prob.}, when describing the computation of the density matrix in the context of NSIs, we referred to the effective description of the neutrino oscillations in the Sun within the OMD approximation, as well as the averaging over the unknown neutrino production location. In this Appendix, we give more details of this calculation.

Within the OMD approximation, the Hamiltonian in Eq.\eqref{eq: Hamiltonian NSI} can be rotated by $O \equiv R_{23} R_{13}$, such that the rotated Hamiltonian $\Tilde{H} = O^\dagger H O$, and its evolution $\Tilde{S}_{\rm prop} = O^\dagger S_{\rm prop} O$, is given by,
\begin{align}
\Tilde{H} &= \begin{pmatrix}
    H^{(2)} & 0 \\
    0 & \Delta_{33}
\end{pmatrix},
&
\Tilde{S}_{prop} &= \begin{pmatrix}
    S^{(2)} & 0 \\
    0 & e^{-i\Delta_{33}L}
\end{pmatrix} \, .
\end{align}
Here, $H^{(2)}$ is the effective 2$\times$2 Hamiltonian and $S^{(2)}\equiv$ Evol[$H^{(2)}$], describes the evolution in an effective 2-flavour scenario. Following the conventions in Refs.~\cite{Esteban_2018, Coloma:2022umy, Amaral:2023tbs, AristizabalSierra:2024nwf} we get,
\begin{align}
     H^{(2)} &= \frac{\Delta m^2_{21}}{4E_\nu}\begin{pmatrix}
         -\cos{2\theta_{12}} &  \sin{2\theta_{12}} e^{i\delta_{\rm CP}}\\
        \sin{2\theta_{12}}e^{-i\delta_{\rm CP}} &  \cos{2\theta_{12}}
     \end{pmatrix} \nonumber\\
     &+ \sqrt{2}G_FN_e(x)\begin{pmatrix}
         c^2_{13} - \epsilon_D & \epsilon_N\\
         \epsilon_N & \epsilon_D
     \end{pmatrix} \, ,
\end{align}
where the terms $\epsilon_D$ and $\epsilon_N$ are related to the combined NSI parameter $\epsilon_{\alpha\beta}$ defined in Eq.\eqref{eq: epsilon in hamiltoninan} by,
\begin{align}
    \epsilon_D &= c_{13} s_{13} (s_{23} \epsilon_{e\mu} + c_{23} \epsilon_{e\tau}) 
    - (1 + s_{13}^2) c_{23} s_{23} \epsilon_{\mu\tau} \nonumber \\
    & 
    - \frac{1}{2} c_{13}^2 (\epsilon_{ee} - \epsilon_{\mu\mu}) + \frac{1}{2} \left( s_{23}^2 - (s_{13} c_{23})^2 \right) (\epsilon_{\tau\tau} - \epsilon_{\mu\mu}) \, ,\nonumber\\ 
    \epsilon_N &= c_{13} (c_{23} \epsilon_{e\mu} - s_{23} \epsilon_{e\tau}) 
    + s_{13} [ (s_{23}^2 - c_{23}^2) \epsilon_{\mu\tau}\nonumber \\
    & + c_{23}s_{23}(\epsilon_{\tau\tau} - \epsilon_{\mu\mu}) ] \, .
\end{align}

We then define the effective matter potential $A_{\text{CC}}$ and the matter-modified mixing $p$ and $q$,
\begin{align}
    &A_{\text{CC}} = 2\sqrt{2} G_F N_e(x) E_\nu \, , \\
    &p = \bigg|\sin(2\theta_{12}) e^{i\delta_{\text{CP}}} + \frac{2 A_{\text{CC}} \epsilon_n}{\Delta m_{21}^2}\bigg| \, , \\
    &q = \cos(2\theta_{12}) - \frac{2 A_{\text{CC}}}{\Delta m_{21}^2} \left( \frac{c_{13}^2}{2} - \epsilon_d \right) \, ,
\end{align}
to obtain the following relations for the mass-squared difference and mixing angle in matter:
\begin{align}
    \Delta m^2_{21,m} &= \Delta m_{21}^2 \sqrt{p^2 + q^2} \, ,\\
    \sin(2\theta_m) &= \frac{p}{\sqrt{p^2 + q^2}} \, , \\
    \cos(2\theta_m) &= \frac{q}{\sqrt{p^2 + q^2}} \, .
\end{align}
Taking the spatial average of $\cos(2\theta_m) $ over the different production zones within the Sun, weighted by the fraction of neutrinos produced in each zone $f(x)$, we get,
\begin{equation}
    \langle \cos(2\theta_m) \rangle = \int_0^1 \cos(2\theta_m(x)) f(x) \, \drm x \, ,
\end{equation}
where $x=r/R_\odot$ is the normalised solar radial coordinate. The production site distributions, $f(x)$, are different for each flux, and are taken from Ref.~\cite{Vinyoles_2017}.

\section{Coherent elastic neutrino-nucleus scattering}\label{app:cevns}

\begin{figure*}[t]
    \centering
    \includegraphics[width=\linewidth]{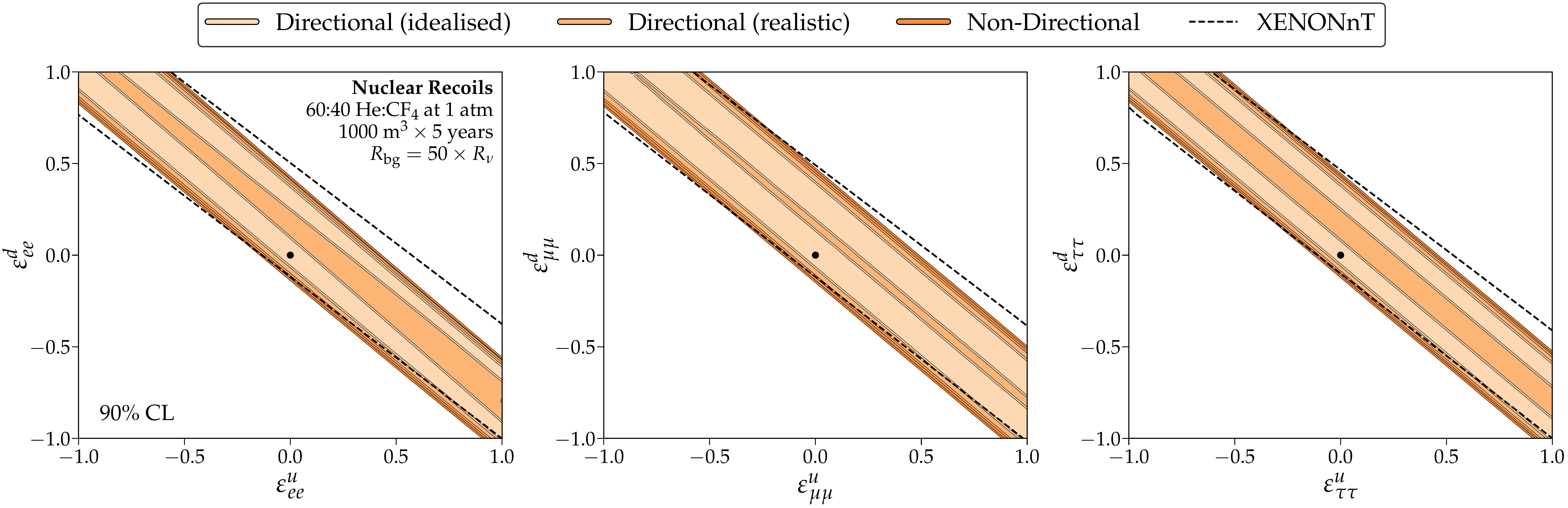}
    \caption{Projected median 90\% CL exclusion limits for a directional gas detector at the required scale needed to measure \cevns with solar neutrinos, i.e.~a volume of 1000 m$^3$ with atmospheric pressure gas. We assume a nuclear recoil threshold of $3$~keV$_{\rm r}$ and performance as described in Appendix~\ref{app:NRs}. We project constraints onto the two-dimensional plane of up and down type NSI couplings across three example flavour-conserving interaction combinations, i.e. $ee$, $\mu\mu$ and $\tau\tau$. For comparison, we show the constraints from Ref.~\cite{AristizabalSierra:2024nwf}, for XENONnT's recent measurement of $^8$B neutrinos~\cite{XENON:2024ijk}. This represents a further future and less immediate goal for a potential large-scale recoil observatory, as designed for sensitivity to dark matter~\cite{Vahsen:2020pzb}.}
    \label{fig:CEvNS}
\end{figure*}

\subsection{Standard Model}
Given a nucleus $N$ of mass $m_N$, under the SM the CE$\nu$NS process $\nu N \rightarrow \nu N$ occurs only through neutral current interactions via the $Z$ exchange, which couples to all neutrino flavours equally. As a result, this process is flavour-blind at tree-level, and the cross section is given by,
\begin{equation}
    \left( \frac{\drm \sigma}{\drm E_r} \right)^{\mathrm{SM}} = \frac{G_F^2 m_N F^2(q) g_V^2}{\pi} \left( 1 - \frac{m_N E_r}{2E_\nu^2} \right) \, .
    \label{eq: SM CEVNS CS}
\end{equation}
For a given nucleus with atomic number $Z$, mass number $A$ and the number of neutrons $N=A-Z$, the SM coupling $g_V = Z g_p^V + N g_n^V$. Here, $g_p^V = 2 g_u^V + g_d^V = \frac{1}{2} - 2\sin^2{\theta_w}$ and $g_n^V = g_u^V + 2g_d^V = -\frac{1}{2} $ which signify the couplings with protons and neutrons respectively, in terms of the $u$ and $d$ couplings. The form factor, $F(q)$, is used to capture the effects of nuclear structure, which become important at large momentum transfer, $q=\sqrt{2m_N E_r}$. The commonly-used Helm ansatz for the form factor is~\cite{Lewin:1995rx}:
\begin{equation}
    F(q) = \frac{3j_1(q r_n)}{q r_n} \exp{\left(\frac{-q^2s^2}{2}\right)}
    \label{eq: Form Factor}
\end{equation}
where $j_1(q)$ is the spherical Bessel function of first kind, $r_n = \sqrt{\frac{5}{3}\left(R_{\rm min}^2 - 3s^2\right)}$ where $s=0.9$ fm is a measure of the nuclear skin thickness and $R_{min}$ is the root-mean-square radius of the proton distribution of the nucleus. The form factor plays an essentially negligible role for recoil energies and nuclear targets we are interested in here, so we do not require any more sophisticated models than this---see e.g.~Ref.~\cite{AbdelKhaleq:2024hir}.

\subsection{NSIs for neutrino-nucleus scattering}
Considering again the pure vector NSI case, we now take the Lagrangian in Eq.\eqref{eq: Light NSI Lagrangian} and set $f=\{u,d\}$ to build the modified CE$\nu$NS process. Assuming that the new light mediator has the same coupling with both $u$ and $d$ quarks, the new interactions can be accounted for in the cross section in Eq.\eqref{eq: SM CEVNS CS} by modifying,
\begin{equation}
    g^{\text{NSI}}_V = g^{\text{SM}}_V + \frac{3 g_{Z'}^2 Q_{Z'}^f Q_{Z'}^{\nu_\alpha} (Z+N)}{\sqrt{2} G_F (2m_NE_r + m_{Z'}^2)} \, .
\end{equation}
For a heavy mediator, $m_{Z'} \gg q$, we consider the Lagrangian in Eq.\eqref{eq: Heavy NSI Lagrangian} with $f=\{u,d\}$. Following the same procedure as described in the main text in the context of \eves, the generalised cross section for CE$\nu$NS can be written as~\cite{Amaral:2023tbs},
\begin{equation}
\begin{split}
    \left( \frac{\drm \zeta}{\drm E_r} \right)^{\mathrm{NSI}}_{\alpha\beta} &= \frac{G_F^2 m_N F^2(q)}{\pi} \left( 1 - \frac{m_N E_r}{2E_\nu^2} \right) \quad \times \\
    & \left[ g_V^2 \delta_{\alpha\beta} + 2g_V G_{\alpha\beta} + \sum_f G_{\alpha f} G_{f\beta} \right] \, ,
\end{split}
\end{equation}
where,
\begin{equation}
    G_{\alpha\beta} = Z (2\varepsilon_{\alpha\beta}^u + \varepsilon_{\alpha\beta}^d) + N (\varepsilon_{\alpha\beta}^u + 2\varepsilon_{\alpha\beta}^d) 
\end{equation}
The double-differential event rate for CE$\nu$NS can also be written in exactly the same way as for \eves---cf.~Eq.\eqref{eq: DRS Density Matrix}---up to replacing the electron mass $m_e$ by the nucleus mass $m_N$, and using $N_T$ to count the number of nuclei in the target.

\subsection{Nuclear recoils in directional gas detectors}\label{app:NRs}
The directions and energies of nuclear recoils are also independently measurable in a gas detector through essentially the same procedure as with electron recoils. In the case of nuclear recoils, however, the sizes of the tracks at the relevant energies ($ 1- 15$~keV$_{\rm r}$ for the nuclei in our chosen gas mixture) are much shorter, at $\sim$cm in length. This makes the effects of diffusion as the charge is transported to the readout plane a major obstacle for achieving good directionality. While directionality can still be used for background rejection, the near-disappearance of any details of the recoil tracks at low energies makes this challenging. This is coupled with the fact that we need to extend to much larger volumes to acquire sufficient statistics in the case of CE$\nu$NS, since the only detectable flux, $^8$B, is significantly weaker than that of $pp$ neutrinos. This is why we demote this discussion to an appendix, but nevertheless, since this channel may be of interest in the longer term, we show some example results here.

Largely inspired by the extensive study of nuclear recoil directionality performed in the context of dark matter and solar CE$\nu$NS in Ref.~\cite{Vahsen:2020pzb}, we adopt an ambitious but not unrealistic energy threshold of 3~keV$_r$, taking all nuclear recoil energies to be those before quenching. We set the following performance parameters: 
\begin{equation}
\left\{a_{\rm HT},\frac{E_{\rm HT}}{{\rm keV}},\,\frac{p_\theta}{{\rm deg.}},p_E\right\}  = 
      \{0.1,10.0,25,0.06\}
\end{equation}
using the same functional forms for energy resolution, head-tail efficiency and angular resolution. Unlike in the electron recoil case, the relatively straight nature of the nuclear recoil tracks, combined with the washing-out of the ionisation distribution due to the diffusion, makes head-tail much more challenging to measure.

In Fig.~\ref{fig:CEvNS} we show forecasted 90\% CL exclusion limits on the two-dimensional space of $u$ and $d$-type couplings, i.e. $\varepsilon_{\alpha\alpha}^u$, $\varepsilon_{\alpha\alpha}^d$ across several possible flavour-conserving combinations. Results for other parameters are qualitatively similar. As can be readily observed from the plots, an ambitious volume of 1000~m$^3$ and an aggressive background reduction in the nuclear recoil channel down to 50 times the $^8$B neutrino rate is required to obtain a level of sensitivity already met by XENONnT and PandaX via their detection of this solar flux. This conclusion is in general alignment with the sensitivity projections made in Ref.~\cite{Vahsen:2020pzb}. Although detecting solar CE$\nu$NS using this kind of experiment may be somewhat impractical in the near term, we expect the situation to be much more promising for detecting CE$\nu$NS with a higher-energy neutrino source, which we will explore in a future study.

\maketitle
\flushbottom

\bibliographystyle{bibi}
\bibliography{biblio}

@article{Cowan:2010js,
      author         = "Cowan, Glen and Cranmer, Kyle and Gross, Eilam and
                        Vitells, Ofer",
      title          = "{Asymptotic formulae for likelihood-based tests of new
                        physics}",
      journal        = "Eur. Phys. J. C",
      volume         = "71",
      year           = "2011",
      pages          = "1554",
      doi            = "10.1140/epjc/s10052-011-1554-0,
                        10.1140/epjc/s10052-013-2501-z",
      note           = "[Erratum: Eur. Phys. J.C73,2501(2013)]",
      eprint         = "1007.1727",
      archivePrefix  = "arXiv",
      primaryClass   = "physics.data-an",
      SLACcitation   = "%%CITATION = ARXIV:1007.1727;%%"
}

@article{BOREXINO:2021xzc,
    author = "Agostini, M. and others",
    collaboration = "BOREXINO",
    title = "{Correlated and integrated directionality for sub-MeV solar neutrinos in Borexino}",
    eprint = "2109.04770",
    archivePrefix = "arXiv",
    primaryClass = "hep-ex",
    doi = "10.1103/PhysRevD.105.052002",
    journal = "Phys. Rev. D",
    volume = "105",
    number = "5",
    pages = "052002",
    year = "2022"
}

@article{BOREXINO:2021efb,
    author = "Agostini, M. and others",
    collaboration = "BOREXINO",
    title = "{First Directional Measurement of Sub-MeV Solar Neutrinos with Borexino}",
    eprint = "2112.11816",
    archivePrefix = "arXiv",
    primaryClass = "hep-ex",
    doi = "10.1103/PhysRevLett.128.091803",
    journal = "Phys. Rev. Lett.",
    volume = "128",
    number = "9",
    pages = "091803",
    year = "2022"
}

@article{OHare:2021utq,
    author = "O'Hare, Ciaran A. J.",
    title = "{New Definition of the Neutrino Floor for Direct Dark Matter Searches}",
    eprint = "2109.03116",
    archivePrefix = "arXiv",
    primaryClass = "hep-ph",
    doi = "10.1103/PhysRevLett.127.251802",
    journal = "Phys. Rev. Lett.",
    volume = "127",
    number = "25",
    pages = "251802",
    year = "2021"
}

@article{COHERENT:2020iec,
    author = "Akimov, D. and others",
    collaboration = "COHERENT",
    title = "{First Measurement of Coherent Elastic Neutrino-Nucleus Scattering on Argon}",
    eprint = "2003.10630",
    archivePrefix = "arXiv",
    primaryClass = "nucl-ex",
    doi = "10.1103/PhysRevLett.126.012002",
    journal = "Phys. Rev. Lett.",
    volume = "126",
    number = "1",
    pages = "012002",
    year = "2021"
}

@article{COHERENT:2020ybo,
    author = "Akimov, D. and others",
    collaboration = "COHERENT",
    title = "{COHERENT Collaboration data release from the first detection of coherent elastic neutrino-nucleus scattering on argon}",
    eprint = "2006.12659",
    archivePrefix = "arXiv",
    primaryClass = "nucl-ex",
    doi = "10.5281/zenodo.3903810",
    month = "6",
    year = "2020"
}

@article{Vitagliano:2019yzm,
    author = "Vitagliano, Edoardo and Tamborra, Irene and Raffelt, Georg",
    title = "{Grand Unified Neutrino Spectrum at Earth: Sources and Spectral Components}",
    eprint = "1910.11878",
    archivePrefix = "arXiv",
    primaryClass = "astro-ph.HE",
    reportNumber = "MPP-2019-205",
    doi = "10.1103/RevModPhys.92.045006",
    journal = "Rev. Mod. Phys.",
    volume = "92",
    pages = "45006",
    year = "2020"
}

@article{Abdullah:2020iiv,
    author = "Abdullah, M. and Aristizabal Sierra, D. and Dutta, Bhaskar and Strigari, Louis E.",
    title = "{Coherent Elastic Neutrino-Nucleus Scattering with directional detectors}",
    eprint = "2003.11510",
    archivePrefix = "arXiv",
    primaryClass = "hep-ph",
    doi = "10.1103/PhysRevD.102.015009",
    journal = "Phys. Rev. D",
    volume = "102",
    pages = "015009",
    year = "2020"
}

@inproceedings{Arzarello:1994jv,
    author = "Arzarello, F. and Ciralli, F. and Frasconi, F. and Bonvicini, G. and Laurenti, G. and Zichichi, A. and Seguinot, J. and Ypsilantis, T. and Tzamarias, S.",
    title = "{HELLAZ: A High rate solar neutrino detector with neutrino energy determination}",
    reportNumber = "LPC-94-28, CERN-LAA-94-19",
    pages = "205--228",
    month = "7",
    year = "1994",
    booktitle = "{International School on Cosmological Dark Matter, Valencia, Spain}"
}

@article{Xia:2024ytb,
    author = "Xia, Shuo-yu",
    title = "{Measuring solar neutrino fluxes in direct detection experiments in the presence of light mediators}",
    eprint = "2410.01167",
    archivePrefix = "arXiv",
    primaryClass = "hep-ph",
    doi = "10.1016/j.nuclphysb.2024.116738",
    journal = "Nucl. Phys. B",
    volume = "1009",
    pages = "116738",
    year = "2024"
}

@article{Blanco-Mas:2024ale,
    author = "Blanco-Mas, Pablo and Coloma, Pilar and Herrera, Gonzalo and Huber, Patrick and Kopp, Joachim and Shoemaker, Ian M. and Tabrizi, Zahra",
    title = "{Clarity through the Neutrino Fog: Constraining New Forces in Dark Matter Detectors}",
    eprint = "2411.14206",
    archivePrefix = "arXiv",
    primaryClass = "hep-ph",
    reportNumber = "IFT-UAM/CSIC-24-164",
    month = "11",
    year = "2024"
}

@article{Maity:2024aji,
    author = "Maity, Tarak Nath and Boehm, Celine",
    title = "{First measurement of the weak mixing angle in direct detection experiments}",
    eprint = "2409.04385",
    archivePrefix = "arXiv",
    primaryClass = "hep-ph",
    month = "9",
    year = "2024"
}

@article{Seguinot:1992zu,
    author = "Seguinot, J. and Ypsilantis, T. and Zichichi, A.",
    title = "{A High rate solar neutrino detector with energy determination}",
    reportNumber = "LPC-92-31",
    journal = "Conf. Proc. C",
    volume = "920310",
    pages = "289--313",
    year = "1992"
}

@article{Agostini:2020mfq,
    author = "Agostini, M. and others",
    collaboration = "BOREXINO",
    title = "{Experimental evidence of neutrinos produced in the CNO fusion cycle in the Sun}",
    eprint = "2006.15115",
    archivePrefix = "arXiv",
    primaryClass = "hep-ex",
    doi = "10.1038/s41586-020-2934-0",
    journal = "Nature",
    volume = "587",
    pages = "577--582",
    year = "2020"
}

@inproceedings{Akerib:2022ort,
    author = "Akerib, D. S. and others",
    title = "{Snowmass2021 Cosmic Frontier Dark Matter Direct Detection to the Neutrino Fog}",
    booktitle = "{Snowmass 2021}",
    eprint = "2203.08084",
    archivePrefix = "arXiv",
    primaryClass = "hep-ex",
    reportNumber = "FERMILAB-CONF-22-180-V",
    month = "3",
    year = "2022"
}

@article{Mishra:2023jlq,
    author = "Mishra, Nityasa and Strigari, Louis E.",
    title = "{Solar neutrinos with CE\ensuremath{\nu}NS and flavor-dependent radiative corrections}",
    eprint = "2305.17827",
    archivePrefix = "arXiv",
    primaryClass = "hep-ph",
    reportNumber = "MI-HET-812",
    doi = "10.1103/PhysRevD.108.063023",
    journal = "Phys. Rev. D",
    volume = "108",
    number = "6",
    pages = "063023",
    year = "2023"
}

@article{Abdullah:2022zue,
    author = "Abdullah, M. and others",
    title = "{Coherent elastic neutrino-nucleus scattering: Terrestrial and astrophysical applications}",
    eprint = "2203.07361",
    archivePrefix = "arXiv",
    primaryClass = "hep-ph",
    month = "3",
    year = "2022"
}

@article{Carew:2023qrj,
    author = "Carew, Ben and Caddell, Ashlee R. and Maity, Tarak Nath and O'Hare, Ciaran A. J.",
    title = "{Neutrino fog for dark matter-electron scattering experiments}",
    eprint = "2312.04303",
    archivePrefix = "arXiv",
    primaryClass = "hep-ph",
    doi = "10.1103/PhysRevD.109.083016",
    journal = "Phys. Rev. D",
    volume = "109",
    number = "8",
    pages = "083016",
    year = "2024"
}

@article{Yakabe:2020rua,
    author = "Yakabe, Ryota and others",
    title = "{First limits from a 3D-vector directional dark matter search with the NEWAGE-0.3b\textquoteright{} detector}",
    eprint = "2005.05157",
    archivePrefix = "arXiv",
    primaryClass = "hep-ex",
    doi = "10.1093/ptep/ptaa147",
    journal = "PTEP",
    volume = "2020",
    number = "11",
    pages = "113F01",
    year = "2020"
}

@inproceedings{OHare:2022jnx,
    author = "O'Hare, C. A. J. and others",
    title = "{Recoil imaging for dark matter, neutrinos, and physics beyond the Standard Model}",
    booktitle = "{Snowmass 2021}",
    eprint = "2203.05914",
    archivePrefix = "arXiv",
    primaryClass = "physics.ins-det",
    reportNumber = "FERMILAB-CONF-22-334-ND",
    month = "3",
    year = "2022"
}

@article{Jaegle:2019jpx,
	title = "Compact, directional neutron detectors capable of high-resolution nuclear recoil imaging",
	journal        = "Nucl. Instrum. Meth. A",
	volume = "945",
	pages = "162296",
	year = "2019",
	issn = "0168-9002",
	doi = "https://doi.org/10.1016/j.nima.2019.06.037",
	url = "http://www.sciencedirect.com/science/article/pii/S0168900219308770",
	author = "I. Jaegle and P.M. Lewis and M. Garcia-Sciveres and M.T. Hedges and T. Hemperek and J. Janssen and Q. Ji and D.-L. Pohl and S. Ross and J. Schueler and I. Seong and T.N. Thorpe and S.E. Vahsen"
}

@inproceedings{Surrow:2022ptn,
    author = "Surrow, B. and others",
    title = "{Micro-Pattern Gaseous Detectors}",
    booktitle = "{2022 Snowmass Summer Study}",
    eprint = "2209.05202",
    archivePrefix = "arXiv",
    primaryClass = "physics.ins-det",
    month = "9",
    year = "2022"
}

@article{Mayet:2016zxu,
      author         = "Mayet, F. and others",
      title          = "{A review of the discovery reach of directional Dark
                        Matter detection}",
      journal        = "Phys. Rept.",
      volume         = "627",
      year           = "2016",
      pages          = "1-49",
      doi            = "10.1016/j.physrep.2016.02.007",
      eprint         = "1602.03781",
      archivePrefix  = "arXiv",
      primaryClass   = "astro-ph.CO",
      SLACcitation   = "%%CITATION = ARXIV:1602.03781;%%"
}

@article{Deaconu:2017vam,
      author         = "Deaconu, Cosmin and others",
      title          = "{Measurement of the directional sensitivity of Dark
                        Matter Time Projection Chamber detectors}",
      journal        = "Phys. Rev. D",
      volume         = "95",
      year           = "2017",
      number         = "12",
      pages          = "122002",
      doi            = "10.1103/PhysRevD.95.122002",
      eprint         = "1705.05965",
      archivePrefix  = "arXiv",
      primaryClass   = "astro-ph.IM",
      SLACcitation   = "%%CITATION = ARXIV:1705.05965;%%"
}

@article{Battat:2016xxe,
      author         = "Battat, J. B. R. and others",
      title          = "{Low Threshold Results and Limits from the DRIFT
                        Directional Dark Matter Detector}",
      collaboration  = "DRIFT",
      journal        = "Astropart. Phys.",
      volume         = "91",
      year           = "2017",
      pages          = "65-74",
      doi            = "10.1016/j.astropartphys.2017.03.007",
      eprint         = "1701.00171",
      archivePrefix  = "arXiv",
      primaryClass   = "astro-ph.IM",
      SLACcitation   = "%%CITATION = ARXIV:1701.00171;%%"
}

@article{Marshall:2020azl,
    author = "Marshall, Mason C. and Turner, Matthew J. and Ku, Mark J. H. and Phillips, David F. and Walsworth, Ronald L.",
    title = "{Directional detection of dark matter with diamond}",
    eprint = "2009.01028",
    archivePrefix = "arXiv",
    primaryClass = "physics.ins-det",
    doi = "10.1088/2058-9565/abe5ed",
    journal = "Quantum Sci. Technol.",
    volume = "6",
    number = "2",
    pages = "024011",
    year = "2021"
}

@article{OHare:2021cgj,
    author = "O'Hare, Ciaran A. J. and others",
    title = "{Particle detection and tracking with DNA}",
    eprint = "2105.11949",
    archivePrefix = "arXiv",
    primaryClass = "physics.ins-det",
    doi = "10.1140/epjc/s10052-022-10264-6",
    journal = "Eur. Phys. J. C",
    volume = "82",
    number = "4",
    pages = "306",
    year = "2022"
}

@article{OHare:2017rag,
      author         = "O'Hare, Ciaran A. J. and Kavanagh, Bradley J. and Green,
                        Anne M.",
      title          = "{Time-integrated directional detection of dark matter}",
      journal        = "Phys. Rev. D",
      volume         = "96",
      year           = "2017",
      number         = "8",
      pages          = "083011",
      doi            = "10.1103/PhysRevD.96.083011",
      eprint         = "1708.02959",
      archivePrefix  = "arXiv",
      primaryClass   = "astro-ph.CO",
      SLACcitation   = "%%CITATION = ARXIV:1708.02959;%%"
}

@article{Drukier:2012hj,
      author         = "Drukier, Andrzej and Freese, Katherine and Lopez,
                        Alejandro and Spergel, David and Cantor, Charles and
                        Church, George and Sano, Takeshi",
      title          = "{New Dark Matter Detectors using DNA or RNA for Nanometer
                        Tracking}",
      year           = "2012",
      eprint         = "1206.6809",
      archivePrefix  = "arXiv",
      primaryClass   = "astro-ph.IM",
      reportNumber   = "MCTP-12-14",
      SLACcitation   = "%%CITATION = ARXIV:1206.6809;%%"
}

@article{Belli:2020hit,
    author = "Belli, P. and others",
    title = "{Measurements of $\hbox {ZnWO}_4$ anisotropic response to nuclear recoils for the ADAMO project}",
    eprint = "2002.09482",
    archivePrefix = "arXiv",
    primaryClass = "physics.ins-det",
    doi = "10.1140/epja/s10050-020-00094-z",
    journal = "Eur. Phys. J. A",
    volume = "56",
    number = "3",
    pages = "83",
    year = "2020"
}

@article{Baracchini:2020btb,
    author = "Baracchini, E. and others",
    title = "{CYGNO: a gaseous TPC with optical readout for dark matter directional search}",
    eprint = "2007.12627",
    archivePrefix = "arXiv",
    primaryClass = "physics.ins-det",
    doi = "10.1088/1748-0221/15/07/C07036",
    journal = "JINST",
    volume = "15",
    number = "07",
    pages = "C07036",
    year = "2020"
}

@article{Billard:2013qya,
      author         = "Billard, J. and Strigari, L. and Figueroa-Feliciano, E.",
      title          = "{Implication of neutrino backgrounds on the reach of next
                        generation dark matter direct detection experiments}",
      journal        = "Phys. Rev.~D",
      volume         = "89",
      year           = "2014",
      number         = "2",
      pages          = "023524",
      doi            = "10.1103/PhysRevD.89.023524",
      eprint         = "1307.5458",
      archivePrefix  = "arXiv",
      primaryClass   = "hep-ph",
      SLACcitation   = "%%CITATION = ARXIV:1307.5458;%%"
}

@article{Ruppin:2014bra,
      author         = "Ruppin, F. and Billard, J. and Figueroa-Feliciano, E. and
                        Strigari, L.",
      title          = "{Complementarity of dark matter detectors in light of the
                        neutrino background}",
      journal        = "Phys. Rev.~D",
      volume         = "90",
      year           = "2014",
      number         = "8",
      pages          = "083510",
      doi            = "10.1103/PhysRevD.90.083510",
      eprint         = "1408.3581",
      archivePrefix  = "arXiv",
      primaryClass   = "hep-ph",
      SLACcitation   = "%%CITATION = ARXIV:1408.3581;%%"
}

@article{Vahsen:2020pzb,
    author = "Vahsen, S.E. and others",
    title = "{CYGNUS: Feasibility of a nuclear recoil observatory with directional sensitivity to dark matter and neutrinos}",
    eprint = "2008.12587",
    archivePrefix = "arXiv",
    primaryClass = "physics.ins-det",
    month = "8",
    year = "2020"
}

@article{Vahsen:2011qx,
      author         = "Vahsen, S.E. and Feng, H. and Garcia-Sciveres, M. and
                        Jaegle, I. and Kadyk, J. and others",
      title          = "{The Directional Dark Matter Detector ($D^{3}$)}",
      journal        = "\EASPUB",
      volume         = "53",
      pages          = "43-50",
      doi            = "10.1051/eas/1253006",
      year           = "2012",
      eprint         = "1110.3401",
      archivePrefix  = "arXiv",
      primaryClass   = "astro-ph.IM",
      SLACcitation   = "%%CITATION = ARXIV:1110.3401;%%",
}

@article{Santos:2011kf,
    author = "Santos, D. and others",
    editor = "Mayet, F. and Santos, D.",
    title = "{MIMAC : A micro-tpc matrix for directional detection of dark matter}",
    eprint = "1111.1566",
    archivePrefix = "arXiv",
    primaryClass = "astro-ph.IM",
    doi = "10.1051/eas/1253004",
    journal = "EAS Publ. Ser.",
    volume = "53",
    pages = "25--31",
    year = "2012"
}

@article{OHare:2020lva,
    author = "O'Hare, Ciaran A. J.",
    title = "{Can we overcome the neutrino floor at high masses?}",
    eprint = "2002.07499",
    archivePrefix = "arXiv",
    primaryClass = "astro-ph.CO",
    reportNumber = "CPPC-2020-13",
    doi = "10.1103/PhysRevD.102.063024",
    journal = "Phys. Rev. D",
    volume = "102",
    number = "6",
    pages = "063024",
    year = "2020"
}

@article{Grothaus:2014hja,
      author         = "Grothaus, Philipp and Fairbairn, Malcolm and Monroe,
                        Jocelyn",
      title          = "{Directional Dark Matter Detection Beyond the Neutrino
                        Bound}",
      journal        = "Phys. Rev.~D",
      volume         = "90",
      year           = "2014",
      number         = "5",
      pages          = "055018",
      doi            = "10.1103/PhysRevD.90.055018",
      eprint         = "1406.5047",
      archivePrefix  = "arXiv",
      primaryClass   = "hep-ph",
      reportNumber   = "KCL-PH-TH-2014-28, LCTS-2014-27",
      SLACcitation   = "%%CITATION = ARXIV:1406.5047;%%"
}

@article{Haxton:2012wfz,
    author = "Haxton, W. C. and Hamish Robertson, R. G. and Serenelli, Aldo M.",
    title = "{Solar Neutrinos: Status and Prospects}",
    eprint = "1208.5723",
    archivePrefix = "arXiv",
    primaryClass = "astro-ph.SR",
    reportNumber = "UCB-NPAT-12-012, NT-LBL-12-013",
    doi = "10.1146/annurev-astro-081811-125539",
    journal = "Ann. Rev. Astron. Astrophys.",
    volume = "51",
    pages = "21--61",
    year = "2013"
}

@Article{Lewin:1995rx,
     author    = "Lewin, J. D. and Smith, P. F.",
     title     = "{Review of mathematics, numerical factors, and corrections
                  for dark  matter experiments based on elastic nuclear
                  recoil}",
     journal   = "Astropart. Phys.",
     volume    = "6",
     year      = "1996",
     pages     = "87-112",
     doi       = "10.1016/S0927-6505(96)00047-3",
}

@article{Arpesella:1996uc,
    author = "Arpesella, C. and Broggini, C. and Cattadori, C.",
    title = "{A possible gas for solar neutrino spectroscopy}",
    doi = "10.1016/0927-6505(95)00051-8",
    journal = "Astropart. Phys.",
    volume = "4",
    pages = "333--341",
    year = "1996"
}

@article{Battat:2016pap,
      author         = "Battat, J. B. R. and others",
      title          = "{Readout technologies for directional WIMP Dark Matter
                        detection}",
      journal        = "Phys. Rept.",
      volume         = "662",
      year           = "2016",
      pages          = "1-46",
      doi            = "10.1016/j.physrep.2016.10.001",
      eprint         = "1610.02396",
      archivePrefix  = "arXiv",
      primaryClass   = "physics.ins-det",
      SLACcitation   = "%%CITATION = ARXIV:1610.02396;%%"
}

@article{Schumann:2019eaa,
    author = "Schumann, Marc",
    title = "{Direct Detection of WIMP Dark Matter: Concepts and Status}",
    eprint = "1903.03026",
    archivePrefix = "arXiv",
    primaryClass = "astro-ph.CO",
    doi = "10.1088/1361-6471/ab2ea5",
    journal = "J. Phys. G",
    volume = "46",
    number = "10",
    pages = "103003",
    year = "2019"
}

@article{DARWIN:2020bnc,
    author = "Aalbers, J. and others",
    collaboration = "DARWIN",
    title = "{Solar neutrino detection sensitivity in DARWIN via electron scattering}",
    eprint = "2006.03114",
    archivePrefix = "arXiv",
    primaryClass = "physics.ins-det",
    doi = "10.1140/epjc/s10052-020-08602-7",
    journal = "Eur. Phys. J. C",
    volume = "80",
    number = "12",
    pages = "1133",
    year = "2020"
}

@article{Coloma:2022umy,
    author = "Coloma, Pilar and Gonzalez-Garcia, M. C. and Maltoni, Michele and Pinheiro, Jo\~ao Paulo and Urrea, Salvador",
    title = "{Constraining new physics with Borexino Phase-II spectral data}",
    eprint = "2204.03011",
    archivePrefix = "arXiv",
    primaryClass = "hep-ph",
    reportNumber = "IFT-UAM/CSIC-22-14, IFIC/22-15, FTUV-22-0404.7998, YITP-SB-2022-05",
    doi = "10.1007/JHEP07(2022)138",
    journal = "JHEP",
    volume = "07",
    pages = "138",
    year = "2022",
    note = "[Erratum: JHEP 11, 138 (2022)]"
}

@article{Miranda:2020zji,
    author = "Miranda, O. G. and Papoulias, D. K. and T\'ortola, M. and Valle, J. W. F.",
    title = "{Probing new neutral gauge bosons with $CE\nu NS$ and neutrino-electron scattering}",
    eprint = "2002.01482",
    archivePrefix = "arXiv",
    primaryClass = "hep-ph",
    reportNumber = "IFIC/20-XXX",
    doi = "10.1103/PhysRevD.101.073005",
    journal = "Phys. Rev. D",
    volume = "101",
    number = "7",
    pages = "073005",
    year = "2020"
}

@article{Abdullah:2018ykz,
    author = "Abdullah, Mohammad and Dent, James B. and Dutta, Bhaskar and Kane, Gordon L. and Liao, Shu and Strigari, Louis E.",
    title = "{Coherent elastic neutrino nucleus scattering as a probe of a Z' through kinetic and mass mixing effects}",
    eprint = "1803.01224",
    archivePrefix = "arXiv",
    primaryClass = "hep-ph",
    reportNumber = "MI-TH-1878",
    doi = "10.1103/PhysRevD.98.015005",
    journal = "Phys. Rev. D",
    volume = "98",
    number = "1",
    pages = "015005",
    year = "2018"
}

@article{Amaral:2023tbs,
    author = "Amaral, Dorian W. P. and Cerdeno, David and Cheek, Andrew and Foldenauer, Patrick",
    title = "{A direct detection view of the neutrino NSI landscape}",
    eprint = "2302.12846",
    archivePrefix = "arXiv",
    primaryClass = "hep-ph",
    reportNumber = "IPPP/23/08; IFT-UAM/CSIC-23-19; FT-UAM-23-1",
    doi = "10.1007/JHEP07(2023)071",
    journal = "JHEP",
    volume = "07",
    pages = "071",
    year = "2023"
}

@article{Ackermann:2025obx,
    author = "Ackermann, N. and others",
    title = "{First observation of reactor antineutrinos by coherent scattering}",
    eprint = "2501.05206",
    archivePrefix = "arXiv",
    primaryClass = "hep-ex",
    month = "1",
    year = "2025"
}

@article{Yang:2023rbg,
    author = "Yang, Zekun and others",
    title = "{First attempt of directionality reconstruction for atmospheric neutrinos in a large homogeneous liquid scintillator detector}",
    eprint = "2310.06281",
    archivePrefix = "arXiv",
    primaryClass = "hep-ex",
    doi = "10.1103/PhysRevD.109.052005",
    journal = "Phys. Rev. D",
    volume = "109",
    number = "5",
    pages = "052005",
    year = "2024"
}

@article{DeRomeri:2024iaw,
    author = "De Romeri, Valentina and Papoulias, Dimitrios K. and Ternes, Christoph A.",
    title = "{Bounds on new neutrino interactions from the first CE\ensuremath{\nu}NS data at direct detection experiments}",
    eprint = "2411.11749",
    archivePrefix = "arXiv",
    primaryClass = "hep-ph",
    doi = "10.1088/1475-7516/2025/05/012",
    journal = "JCAP",
    volume = "05",
    pages = "012",
    year = "2025"
}

@article{DeRomeri:2025csu,
    author = "De Romeri, Valentina and Papoulias, Dimitrios K. and Sanchez Garcia, Gonzalo",
    title = "{Implications of the first CONUS+ measurement of coherent elastic neutrino-nucleus scattering}",
    eprint = "2501.17843",
    archivePrefix = "arXiv",
    primaryClass = "hep-ph",
    doi = "10.1103/PhysRevD.111.075025",
    journal = "Phys. Rev. D",
    volume = "111",
    number = "7",
    pages = "075025",
    year = "2025"
}

@article{DeRomeri:2024hvc,
    author = "De Romeri, Valentina and Papoulias, Dimitrios K. and Sanchez Garcia, Gonzalo and Ternes, Christoph A. and T\'ortola, Mariam",
    title = "{Neutrino electromagnetic properties and sterile dipole portal in light of the first solar CE$\nu$NS data}",
    eprint = "2412.14991",
    archivePrefix = "arXiv",
    primaryClass = "hep-ph",
    month = "12",
    year = "2024"
}

@article{Khan:2020csx,
    author = "Khan, Amir N.",
    title = "{Constraints on general light mediators from PandaX-II electron recoil data}",
    eprint = "2008.10279",
    archivePrefix = "arXiv",
    primaryClass = "hep-ph",
    doi = "10.1016/j.physletb.2021.136415",
    journal = "Phys. Lett. B",
    volume = "819",
    pages = "136415",
    year = "2021"
}

@article{Khan:2019jvr,
    author = "Khan, Amir N. and Rodejohann, Werner and Xu, Xun-Jie",
    title = "{Borexino and general neutrino interactions}",
    eprint = "1906.12102",
    archivePrefix = "arXiv",
    primaryClass = "hep-ph",
    reportNumber = "FERMILAB-PUB-19-348-T",
    doi = "10.1103/PhysRevD.101.055047",
    journal = "Phys. Rev. D",
    volume = "101",
    number = "5",
    pages = "055047",
    year = "2020"
}

@article{Maity:2024vkj,
    author = "Maity, Tarak Nath",
    title = "{Neutrinos as background and signal in Migdal search}",
    eprint = "2412.17649",
    archivePrefix = "arXiv",
    primaryClass = "hep-ph",
    month = "12",
    year = "2024"
}

@article{AristizabalSierra:2024nwf,
    author = "Aristizabal Sierra, D. and Mishra, N. and Strigari, L.",
    title = "{Implications of first neutrino-induced nuclear recoil measurements in direct detection experiments: Probing nonstandard interaction via CE\ensuremath{\nu}NS}",
    eprint = "2409.02003",
    archivePrefix = "arXiv",
    primaryClass = "hep-ph",
    doi = "10.1103/PhysRevD.111.055007",
    journal = "Phys. Rev. D",
    volume = "111",
    number = "5",
    pages = "055007",
    year = "2025"
}

@article{XLZD:2024nsu,
    author = "Aalbers, J. and others",
    collaboration = "XLZD",
    title = "{The XLZD Design Book: Towards the Next-Generation Liquid Xenon Observatory for Dark Matter and Neutrino Physics}",
    eprint = "2410.17137",
    archivePrefix = "arXiv",
    primaryClass = "hep-ex",
    month = "10",
    year = "2024"
}

@article{Freedman:1973yd,
	Author = {Freedman, Daniel Z.},
	Doi = {10.1103/PhysRevD.9.1389},
	Journal = {Phys. Rev. D},
	Pages = {1389-1392},
	Slaccitation = {%%CITATION = PHRVA,D9,1389;%%},
	Title = {{Coherent neutrino nucleus scattering as a probe of the weak neutral current}},
	Volume = {9},
	Year = {1974},
}

@article{OHare:2016pjy,
      author         = "O'Hare, Ciaran A. J.",
      title          = "{Dark matter astrophysical uncertainties and the neutrino
                        floor}",
      journal        = "Phys. Rev. D",
      volume         = "94",
      year           = "2016",
      number         = "6",
      pages          = "063527",
      doi            = "10.1103/PhysRevD.94.063527",
      eprint         = "1604.03858",
      archivePrefix  = "arXiv",
      primaryClass   = "astro-ph.CO",
      SLACcitation   = "%%CITATION = ARXIV:1604.03858;%%"
}

@article{Agafonova:2017ajg,
      author         = "Agafonova, N. and others",
      title          = "{Discovery potential for directional Dark Matter
                        detection with nuclear emulsions}",
      collaboration  = "NEWSdm",
      journal        = "Eur. Phys. J. C",
      volume         = "78",
      year           = "2018",
      number         = "7",
      pages          = "578",
      doi            = "10.1140/epjc/s10052-018-6060-1",
      eprint         = "1705.00613",
      archivePrefix  = "arXiv",
      primaryClass   = "astro-ph.CO",
      SLACcitation   = "%%CITATION = ARXIV:1705.00613;%%"
}

@article{Amaral:2020tga,
    author = "Amaral, Dorian Warren Praia do and Cerdeno, David G. and Foldenauer, Patrick and Reid, Elliott",
    title = "{Solar neutrino probes of the muon anomalous magnetic moment in the gauged $ \mathrm{U}{(1)}_{L_{\mu }-{L}_{\tau }} $}",
    eprint = "2006.11225",
    archivePrefix = "arXiv",
    primaryClass = "hep-ph",
    reportNumber = "IPPP/20/24, IFT-UAM/CSIC-20-70",
    doi = "10.1007/JHEP12(2020)155",
    journal = "JHEP",
    volume = "12",
    pages = "155",
    year = "2020"
}

@article{AtzoriCorona:2022moj,
    author = "Atzori Corona, M. and Cadeddu, M. and Cargioli, N. and Dordei, F. and Giunti, C. and Li, Y. F. and Picciau, E. and Ternes, C. A. and Zhang, Y. Y.",
    title = "{Probing light mediators and (g \ensuremath{-} 2)$_{\mu}$ through detection of coherent elastic neutrino nucleus scattering at COHERENT}",
    eprint = "2202.11002",
    archivePrefix = "arXiv",
    primaryClass = "hep-ph",
    doi = "10.1007/JHEP05(2022)109",
    journal = "JHEP",
    volume = "05",
    pages = "109",
    year = "2022"
}

@article{delaVega:2021wpx,
    author = "de la Vega, Leon M. G. and Flores, L. J. and Nath, Newton and Peinado, Eduardo",
    title = "{Complementarity between dark matter direct searches and CE\ensuremath{\nu}NS experiments in U(1)' models}",
    eprint = "2107.04037",
    archivePrefix = "arXiv",
    primaryClass = "hep-ph",
    doi = "10.1007/JHEP09(2021)146",
    journal = "JHEP",
    volume = "09",
    pages = "146",
    year = "2021"
}

@article{A:2022acy,
    author = "A., ShivaSankar K. and Majumdar, Anirban and Papoulias, Dimitrios K. and Prajapati, Hemant and Srivastava, Rahul",
    title = "{Implications of first LZ and XENONnT results: A comparative study of neutrino properties and light mediators}",
    eprint = "2208.06415",
    archivePrefix = "arXiv",
    primaryClass = "hep-ph",
    doi = "10.1016/j.physletb.2023.137742",
    journal = "Phys. Lett. B",
    volume = "839",
    pages = "137742",
    year = "2023"
}

@article{Dutta:2017nht,
    author = "Dutta, Bhaskar and Liao, Shu and Strigari, Louis E. and Walker, Joel W.",
    title = "{Non-standard interactions of solar neutrinos in dark matter experiments}",
    eprint = "1705.00661",
    archivePrefix = "arXiv",
    primaryClass = "hep-ph",
    reportNumber = "MI-TH-1724",
    doi = "10.1016/j.physletb.2017.08.031",
    journal = "Phys. Lett. B",
    volume = "773",
    pages = "242--246",
    year = "2017"
}

@article{Cerdeno:2016sfi,
      author         = "Cerdeño, David G. and Fairbairn, Malcolm and Jubb,
                        Thomas and Machado, Pedro A. N. and Vincent, Aaron C. and
                        Bœhm, Céline",
      title          = "{Physics from solar neutrinos in dark matter direct
                        detection experiments}",
      journal        = "JHEP",
      volume         = "05",
      year           = "2016",
      pages          = "118",
      doi            = "10.1007/JHEP09(2016)048, 10.1007/JHEP05(2016)118",
      note           = "[Erratum: JHEP09,048(2016)]",
      eprint         = "1604.01025",
      archivePrefix  = "arXiv",
      primaryClass   = "hep-ph",
      reportNumber   = "IFT-UAM-CSIC-16-031, FTUAM-16-12, IPPP-16-27, DCTP-16-54,
                        KCL-PH-TH-2016-19, --KCL-PH-TH-2016-19",
      SLACcitation   = "%%CITATION = ARXIV:1604.01025;%%"
}

@article{Akimov:2017ade,
      author         = "Akimov, D. and others",
      title          = "{Observation of Coherent Elastic Neutrino-Nucleus
                        Scattering}",
      collaboration  = "COHERENT",
      journal        = "Science",
      volume         = "357",
      year           = "2017",
      number         = "6356",
      pages          = "1123-1126",
      doi            = "10.1126/science.aao0990",
      eprint         = "1708.01294",
      archivePrefix  = "arXiv",
      primaryClass   = "nucl-ex",
      SLACcitation   = "%%CITATION = ARXIV:1708.01294;%%"
}

@article{Borexino:2017rsf,
    author = "Agostini, M. and others",
    collaboration = "Borexino",
    title = "{First Simultaneous Precision Spectroscopy of $pp$, $^7$Be, and $pep$ Solar Neutrinos with Borexino Phase-II}",
    eprint = "1707.09279",
    archivePrefix = "arXiv",
    primaryClass = "hep-ex",
    doi = "10.1103/PhysRevD.100.082004",
    journal = "Phys. Rev. D",
    volume = "100",
    number = "8",
    pages = "082004",
    year = "2019"
}

@article{Theia:2019non,
    author = "Askins, M. and others",
    collaboration = "Theia",
    title = "{THEIA: an advanced optical neutrino detector}",
    eprint = "1911.03501",
    archivePrefix = "arXiv",
    primaryClass = "physics.ins-det",
    doi = "10.1140/epjc/s10052-020-7977-8",
    journal = "Eur. Phys. J. C",
    volume = "80",
    number = "5",
    pages = "416",
    year = "2020"
}

@article{Amaro:2023dxb,
    author = "Amaro, Fernando Domingues and others",
    title = "{The CYGNO experiment, a directional detector for direct Dark Matter searches}",
    eprint = "2306.04568",
    archivePrefix = "arXiv",
    primaryClass = "physics.ins-det",
    doi = "10.1016/j.nima.2023.168325",
    journal = "Nucl. Instrum. Meth. A",
    volume = "1054",
    pages = "168325",
    year = "2023"
}

@article{Almeida:2023tgn,
    author = "Almeida, B. D. and others",
    title = "{Noise assessment of CMOS active pixel sensors for the CYGNO Experiment}",
    doi = "10.1088/1361-6501/acf7e1",
    journal = "Measur. Sci. Tech.",
    volume = "34",
    number = "12",
    pages = "125145",
    year = "2023"
}

@article{CYGNO:2023gud,
    author = "Amaro, F. D. and others",
    collaboration = "CYGNO",
    title = "{The CYGNO experiment: a directional Dark Matter detector with optical readout}",
    doi = "10.1088/1748-0221/18/09/C09010",
    journal = "JINST",
    volume = "18",
    number = "09",
    pages = "C09010",
    year = "2023"
}

@article{Ghrear:2024rku,
    author = "Ghrear, Majd and Sadowski, Peter and Vahsen, Sven Einar",
    title = "{Deep Probabilistic Direction Prediction in 3D with Applications to Directional Dark Matter Detectors}",
    eprint = "2403.15949",
    archivePrefix = "arXiv",
    primaryClass = "hep-ex",
    month = "3",
    year = "2024"
}

@article{AbdelKhaleq:2024hir,
    author = "Abdel Khaleq, Raghda and Newstead, Jayden L. and Simenel, Cedric and Stuchbery, Andrew E.",
    title = "{Detailed nuclear structure calculations for coherent elastic neutrino-nucleus scattering}",
    eprint = "2405.20060",
    archivePrefix = "arXiv",
    primaryClass = "hep-ph",
    doi = "10.1103/PhysRevD.111.033003",
    journal = "Phys. Rev. D",
    volume = "111",
    number = "3",
    pages = "033003",
    year = "2025"
}

@article{Ghrear:2025iry,
    author = "Ghrear, Majd and Vahsen, Sven E.",
    title = "{Angular Resolution of Electrons in Gaseous Targets}",
    eprint = "2502.12387",
    archivePrefix = "arXiv",
    primaryClass = "physics.ins-det",
    month = "2",
    year = "2025"
}

@article{BOREXINO:2023ygs,
    author = "Basilico, D. and others",
    collaboration = "BOREXINO",
    title = "{Final results of Borexino on CNO solar neutrinos}",
    eprint = "2307.14636",
    archivePrefix = "arXiv",
    primaryClass = "hep-ex",
    doi = "10.1103/PhysRevD.108.102005",
    journal = "Phys. Rev. D",
    volume = "108",
    number = "10",
    pages = "102005",
    year = "2023"
}

@article{Bellini:2013lnn,
      author         = "Bellini, G. and others",
      title          = "{Final results of Borexino Phase-I on low energy solar
                        neutrino spectroscopy}",
      collaboration  = "Borexino",
      journal        = "Phys. Rev. D",
      volume         = "89",
      year           = "2014",
      number         = "11",
      pages          = "112007",
      doi            = "10.1103/PhysRevD.89.112007",
      eprint         = "1308.0443",
      archivePrefix  = "arXiv",
      primaryClass   = "hep-ex",
      SLACcitation   = "%%CITATION = ARXIV:1308.0443;%%"
}

@article{Ghrear:2020pzk,
    author = "Ghrear, Majd and Vahsen, Sven E. and Deaconu, Cosmin",
    title = "{Observables for recoil identification in high-definition Gas Time Projection Chambers}",
    eprint = "2012.13649",
    archivePrefix = "arXiv",
    primaryClass = "physics.ins-det",
    doi = "10.1088/1475-7516/2021/10/005",
    journal = "JCAP",
    volume = "10",
    pages = "005",
    year = "2021"
}

@article{Schueler:2022lvr,
    author = "Schueler, J. and Ghrear, M. and Vahsen, S. E. and Sadowski, P. and Deaconu, C.",
    title = "{Deep learning for improved keV-scale recoil identification in high resolution gas time projection chambers}",
    eprint = "2206.10822",
    archivePrefix = "arXiv",
    primaryClass = "physics.ins-det",
    month = "6",
    year = "2022"
}

@article{ParticleDataGroup:2024cfk,
    author = "Navas, S. and others",
    collaboration = "Particle Data Group",
    title = "{Review of particle physics}",
    doi = "10.1103/PhysRevD.110.030001",
    journal = "Phys. Rev. D",
    volume = "110",
    number = "3",
    pages = "030001",
    year = "2024"
}

@article{Arnquist:2019fkc,
    author = "Arnquist, Isaac J. and Beck, Chelsie and di Vacri, Maria Laura and Harouaka, Khadouja and Saldanha, Richard",
    title = "{Ultra-low radioactivity Kapton and copper-Kapton laminates}",
    eprint = "1910.04317",
    archivePrefix = "arXiv",
    primaryClass = "physics.ins-det",
    doi = "10.1016/j.nima.2020.163573",
    journal = "Nucl. Instrum. Meth. A",
    volume = "959",
    pages = "163573",
    year = "2020"
}

@article{SamueleThesis,
    author = "Torelli, Samuele",
    title = "{Feasibility of a directional solar neutrino measurement with the CYGNO/INITIUM experiment}",
    eprint = "2408.03760",
    archivePrefix = "arXiv",
    primaryClass = "physics.ins-det",
    month = "8",
    year = "2024"
}

@article{Dutta:2020che,
    author = "Dutta, Bhaskar and Lang, Rafael F. and Liao, Shu and Sinha, Samiran and Strigari, Louis and Thompson, Adrian",
    title = "{A global analysis strategy to resolve neutrino NSI degeneracies with scattering and oscillation data}",
    eprint = "2002.03066",
    archivePrefix = "arXiv",
    primaryClass = "hep-ph",
    reportNumber = "MI-TH-204",
    doi = "10.1007/JHEP09(2020)106",
    journal = "JHEP",
    volume = "09",
    pages = "106",
    year = "2020"
}

@article{Borexino:2017uhp,
    author = "Agostini, M. and others",
    collaboration = "Borexino",
    title = "{Improved measurement of $^{8}\mathrm{B}$ solar neutrinos with $1.5\text{ }\text{ }\mathrm{kt}\ifmmode\cdot\else\textperiodcentered\fi{}\mathrm{y}$ of Borexino exposure}",
    eprint = "1709.00756",
    archivePrefix = "arXiv",
    primaryClass = "hep-ex",
    doi = "10.1103/PhysRevD.101.062001",
    journal = "Phys. Rev. D",
    volume = "101",
    number = "6",
    pages = "062001",
    year = "2020"
}

@article{Borexino:2008fkj,
    author = "Bellini, G. and others",
    collaboration = "Borexino",
    title = "{Measurement of the solar $^8$B neutrino rate with a liquid scintillator target and 3 MeV energy threshold in the Borexino detector}",
    eprint = "0808.2868",
    archivePrefix = "arXiv",
    primaryClass = "astro-ph",
    doi = "10.1103/PhysRevD.82.033006",
    journal = "Phys. Rev. D",
    volume = "82",
    pages = "033006",
    year = "2010"
}

@article{Gelmini:2018ogy,
      author         = "Gelmini, Graciela B. and Takhistov, Volodymyr and Witte,
                        Samuel J.",
      title          = "{Casting a Wide Signal Net with Future Direct Dark Matter
                        Detection Experiments}",
      journal        = "JCAP",
      volume         = "1807",
      year           = "2018",
      number         = "07",
      pages          = "009",
      doi            = "10.1088/1475-7516/2018/07/009",
      eprint         = "1804.01638",
      archivePrefix  = "arXiv",
      primaryClass   = "hep-ph",
      SLACcitation   = "%%CITATION = ARXIV:1804.01638;%%"
}

@article{Han:2019zkz,
    author = "Han, Tao and Liao, Jiajun and Liu, Hongkai and Marfatia, Danny",
    title = "{Nonstandard neutrino interactions at COHERENT, DUNE, T2HK and LHC}",
    eprint = "1910.03272",
    archivePrefix = "arXiv",
    primaryClass = "hep-ph",
    reportNumber = "PITT-PACC-1815",
    doi = "10.1007/JHEP11(2019)028",
    journal = "JHEP",
    volume = "11",
    pages = "028",
    year = "2019"
}

@article{Ilten:2018crw,
    author = "Ilten, Philip and Soreq, Yotam and Williams, Mike and Xue, Wei",
    title = "{Serendipity in dark photon searches}",
    eprint = "1801.04847",
    archivePrefix = "arXiv",
    primaryClass = "hep-ph",
    reportNumber = "MIT-CTP/4976, CERN-TH-2017-282, MIT-CTP-4976",
    doi = "10.1007/JHEP06(2018)004",
    journal = "JHEP",
    volume = "06",
    pages = "004",
    year = "2018"
}

@article{Bauer:2018onh,
    author = "Bauer, Martin and Foldenauer, Patrick and Jaeckel, Joerg",
    title = "{Hunting All the Hidden Photons}",
    eprint = "1803.05466",
    archivePrefix = "arXiv",
    primaryClass = "hep-ph",
    doi = "10.1007/JHEP07(2018)094",
    journal = "JHEP",
    volume = "07",
    pages = "094",
    year = "2018"
}

@article{AristizabalSierra:2020edu,
    author = "Aristizabal Sierra, D. and De Romeri, V. and Flores, L. J. and Papoulias, D. K.",
    title = "{Light vector mediators facing XENON1T data}",
    eprint = "2006.12457",
    archivePrefix = "arXiv",
    primaryClass = "hep-ph",
    doi = "10.1016/j.physletb.2020.135681",
    journal = "Phys. Lett. B",
    volume = "809",
    pages = "135681",
    year = "2020"
}

@article{Antel:2023hkf,
    author = "Antel, C. and others",
    title = "{Feebly-interacting particles: FIPs 2022 Workshop Report}",
    eprint = "2305.01715",
    archivePrefix = "arXiv",
    primaryClass = "hep-ph",
    reportNumber = "CERN-TH-2023-061, DESY-23-050, FERMILAB-PUB-23-149-PPD, INFN-23-14-LNF, JLAB-PHY-23-3789, LA-UR-23-21432, MITP-23-015",
    doi = "10.1140/epjc/s10052-023-12168-5",
    journal = "Eur. Phys. J. C",
    volume = "83",
    number = "12",
    pages = "1122",
    year = "2023"
}

@article{Leyton:2017tza,
      author         = "Leyton, Michael and Dye, Stephen and Monroe, Jocelyn",
      title          = "{Exploring the hidden interior of the Earth with
                        directional neutrino measurements}",
      journal        = "Nature Commun.",
      volume         = "8",
      year           = "2017",
      pages          = "15989",
      doi            = "10.1038/ncomms15989",
      eprint         = "1710.06724",
      archivePrefix  = "arXiv",
      primaryClass   = "physics.geo-ph",
      SLACcitation   = "%%CITATION = ARXIV:1710.06724;%%"
}

@article{CYGNUS,
      author         = "Baracchini, E. and others",
      title          = "{CYGNUS: Feasibility of a Nuclear Recoil Observatory with Directional Sensitivity to Dark Matter and Solar Neutrinos}",
      year = 2019,
      journal        = "[In preparation]"
}

@article{Bernabei:2003ct,
      author         = "Bernabei, R. and Belli, P. and Nozzoli, F. and
                        Incicchitti, A.",
      title          = "{Anisotropic scintillators for WIMP direct detection:
                        Revisited}",
      journal        = "Eur. Phys. J. C",
      volume         = "28",
      year           = "2003",
      pages          = "203-209",
      doi            = "10.1140/epjc/s2003-01190-8",
      SLACcitation   = "%%CITATION = EPHJA,C28,203;%%"
}

@article{BOREXINO:2018ohr,
    author = "Agostini, M. and others",
    collaboration = "BOREXINO",
    title = "{Comprehensive measurement of $pp$-chain solar neutrinos}",
    reportNumber = "FERMILAB-PUB-18-592-ND",
    doi = "10.1038/s41586-018-0624-y",
    journal = "Nature",
    volume = "562",
    number = "7728",
    pages = "505--510",
    year = "2018"
}

@article{Denton:2018xmq,
    author = "Denton, Peter B. and Farzan, Yasaman and Shoemaker, Ian M.",
    title = "{Testing large non-standard neutrino interactions with arbitrary mediator mass after COHERENT data}",
    eprint = "1804.03660",
    archivePrefix = "arXiv",
    primaryClass = "hep-ph",
    doi = "10.1007/JHEP07(2018)037",
    journal = "JHEP",
    volume = "07",
    pages = "037",
    year = "2018"
}

@article{Giunti:2023yha,
    author = "Giunti, Carlo and Ternes, Christoph A.",
    title = "{Testing neutrino electromagnetic properties at current and future dark matter experiments}",
    eprint = "2309.17380",
    archivePrefix = "arXiv",
    primaryClass = "hep-ph",
    doi = "10.1103/PhysRevD.108.095044",
    journal = "Phys. Rev. D",
    volume = "108",
    number = "9",
    pages = "095044",
    year = "2023"
}

@article{deGouvea:2021ymm,
    author = "de Gouv\^ea, Andr\'e and McGinness, Emma and Martinez-Soler, Ivan and Perez-Gonzalez, Yuber F.",
    title = "{pp solar neutrinos at DARWIN}",
    eprint = "2111.02421",
    archivePrefix = "arXiv",
    primaryClass = "hep-ph",
    reportNumber = "NUHEP-TH/21-17, FERMILAB-PUB-21-560-T, IPPP/21/46",
    doi = "10.1103/PhysRevD.106.096017",
    journal = "Phys. Rev. D",
    volume = "106",
    number = "9",
    pages = "096017",
    year = "2022"
}

@article{AtzoriCorona:2022jeb,
    author = "Atzori Corona, M. and Bonivento, W. M. and Cadeddu, M. and Cargioli, N. and Dordei, F.",
    title = "{New constraint on neutrino magnetic moment and neutrino millicharge from LUX-ZEPLIN dark matter search results}",
    eprint = "2207.05036",
    archivePrefix = "arXiv",
    primaryClass = "hep-ph",
    doi = "10.1103/PhysRevD.107.053001",
    journal = "Phys. Rev. D",
    volume = "107",
    number = "5",
    pages = "053001",
    year = "2023"
}

@article{Li:2022jfl,
    author = "Li, Yu-Feng and Xia, Shuo-yu",
    title = "{Constraining light mediators via detection of coherent elastic solar neutrino nucleus scattering}",
    eprint = "2201.05015",
    archivePrefix = "arXiv",
    primaryClass = "hep-ph",
    doi = "10.1016/j.nuclphysb.2022.115737",
    journal = "Nucl. Phys. B",
    volume = "977",
    pages = "115737",
    year = "2022"
}

@article{Majumdar:2021vdw,
    author = "Majumdar, Anirban and Papoulias, D. K. and Srivastava, Rahul",
    title = "{Dark matter detectors as a novel probe for light new physics}",
    eprint = "2112.03309",
    archivePrefix = "arXiv",
    primaryClass = "hep-ph",
    doi = "10.1103/PhysRevD.106.013001",
    journal = "Phys. Rev. D",
    volume = "106",
    number = "1",
    pages = "013001",
    year = "2022"
}

@article{Schwemberger:2022fjl,
    author = "Schwemberger, Thomas and Yu, Tien-Tien",
    title = "{Detecting beyond the standard model interactions of solar neutrinos in low-threshold dark matter detectors}",
    eprint = "2202.01254",
    archivePrefix = "arXiv",
    primaryClass = "hep-ph",
    doi = "10.1103/PhysRevD.106.015002",
    journal = "Phys. Rev. D",
    volume = "106",
    number = "1",
    pages = "015002",
    year = "2022"
}

@article{Ikeda:2021ckk,
    author = "Ikeda, Tomonori and others",
    title = "{Direction-sensitive dark matter search with the low-background gaseous detector NEWAGE-0.3b\textquotedblright{}}",
    eprint = "2101.09921",
    archivePrefix = "arXiv",
    primaryClass = "hep-ex",
    doi = "10.1093/ptep/ptab053",
    journal = "PTEP",
    volume = "2021",
    number = "6",
    pages = "063F01",
    year = "2021"
}

@article{Tao:2019wfh,
    author = "Tao, Y. and others",
    title = "{Track length measurement of $^{19}$F$^+$ ions with the MIMAC directional Dark Matter detector prototype}",
    eprint = "1903.02159",
    archivePrefix = "arXiv",
    primaryClass = "physics.ins-det",
    doi = "10.1016/j.nima.2020.164569",
    journal = "Nucl. Instrum. Meth. A",
    volume = "985",
    pages = "164569",
    year = "2021"
}

@article{Dent:2016iht,
      author         = "Dent, James B. and Dutta, Bhaskar and Newstead, Jayden L.
                        and Strigari, Louis E.",
      title          = "{Effective field theory treatment of the neutrino
                        background in direct dark matter detection experiments}",
      journal        = "Phys. Rev. D",
      volume         = "93",
      year           = "2016",
      number         = "7",
      pages          = "075018",
      doi            = "10.1103/PhysRevD.93.075018",
      eprint         = "1602.05300",
      archivePrefix  = "arXiv",
      primaryClass   = "hep-ph",
      SLACcitation   = "%%CITATION = ARXIV:1602.05300;%%"
}

@article{BOREXINO:2017yyp,
    author = "Agostini, M. and others",
    collaboration = "BOREXINO",
    title = "{Seasonal Modulation of the $^7$Be Solar Neutrino Rate in Borexino}",
    eprint = "1701.07970",
    archivePrefix = "arXiv",
    primaryClass = "hep-ex",
    doi = "10.1016/j.astropartphys.2017.04.004",
    journal = "Astropart. Phys.",
    volume = "92",
    pages = "21--29",
    year = "2017"
}

@article{Capparelli:2014lua,
      author         = "Capparelli, L. M. and Cavoto, G. and Mazzilli, D. and
                        Polosa, A. D.",
      title          = "{Directional Dark Matter Searches with Carbon Nanotubes}",
      journal        = "Phys. Dark Univ.",
      volume         = "9-10",
      year           = "2015",
      pages          = "24-30",
      doi            = "10.1016/j.dark.2015.12.004, 10.1016/j.dark.2015.08.002",
      note           = "[Erratum: Phys. Dark Univ.11,79(2016)]",
      eprint         = "1412.8213",
      archivePrefix  = "arXiv",
      primaryClass   = "physics.ins-det",
      SLACcitation   = "%%CITATION = ARXIV:1412.8213;%%"
}

@article{Dent:2016wor,
      author         = "Dent, James B. and Dutta, Bhaskar and Newstead, Jayden L.
                        and Strigari, Louis E.",
      title          = "{Dark matter, light mediators, and the neutrino floor}",
      journal        = "Phys. Rev. D",
      volume         = "95",
      year           = "2017",
      number         = "5",
      pages          = "051701",
      doi            = "10.1103/PhysRevD.95.051701",
      eprint         = "1607.01468",
      archivePrefix  = "arXiv",
      primaryClass   = "hep-ph",
      reportNumber   = "MI-TH-1618, CETUP2016-003",
      SLACcitation   = "%%CITATION = ARXIV:1607.01468;%%"
}

@article{Cooley:2022ufh,
    author = "Cooley, Jodi and others",
    title = "{Report of the Topical Group on Particle Dark Matter for Snowmass 2021}",
    eprint = "2209.07426",
    archivePrefix = "arXiv",
    primaryClass = "hep-ph",
    reportNumber = "FERMILAB-PUB-22-718-PPD",
    month = "9",
    year = "2022"
}

@article{Aalbers:2022dzr,
    author = "Aalbers, J. and others",
    title = "{A next-generation liquid xenon observatory for dark matter and neutrino physics}",
    eprint = "2203.02309",
    archivePrefix = "arXiv",
    primaryClass = "physics.ins-det",
    reportNumber = "INT-PUB-22-003, FERMILAB-PUB-22-112-PPD-QIS-T",
    doi = "10.1088/1361-6471/ac841a",
    journal = "J. Phys. G",
    volume = "50",
    number = "1",
    pages = "013001",
    year = "2023"
}

@article{Bellini:2014uqa,
      author         = "Bellini, G. and others",
      title          = "{Neutrinos from the primary proton–proton fusion
                        process in the Sun}",
      collaboration  = "BOREXINO Collaboration",
      journal        = "Nature",
      volume         = "512",
      year           = "2014",
      number         = "7515",
      pages          = "383-386",
      doi            = "10.1038/nature13702",
      SLACcitation   = "%%CITATION = NATUA,512,383;%%"
}

@article{Vahsen:2021gnb,
    author = "Vahsen, Sven E. and O'Hare, Ciaran A. J. and Loomba, Dinesh",
    title = "{Directional recoil detection}",
    eprint = "2102.04596",
    archivePrefix = "arXiv",
    primaryClass = "physics.ins-det",
    doi = "10.1146/annurev-nucl-020821-035016",
    journal = "Ann. Rev. Nucl. Part. Sci.",
    volume = "71",
    pages = "189--224",
    year = "2021"
}

@article{PandaX:2024muv,
    author = "Bo, Zihao and others",
    collaboration = "PandaX",
    title = "{First Indication of Solar B8 Neutrinos through Coherent Elastic Neutrino-Nucleus Scattering in PandaX-4T}",
    eprint = "2407.10892",
    archivePrefix = "arXiv",
    primaryClass = "hep-ex",
    doi = "10.1103/PhysRevLett.133.191001",
    journal = "Phys. Rev. Lett.",
    volume = "133",
    number = "19",
    pages = "191001",
    year = "2024"
}

@article{XENON:2024ijk,
    author = "Aprile, Elena and others",
    collaboration = "XENON",
    title = "{First Indication of Solar B8 Neutrinos via Coherent Elastic Neutrino-Nucleus Scattering with XENONnT}",
    eprint = "2408.02877",
    archivePrefix = "arXiv",
    primaryClass = "nucl-ex",
    doi = "10.1103/PhysRevLett.133.191002",
    journal = "Phys. Rev. Lett.",
    volume = "133",
    number = "19",
    pages = "191002",
    year = "2024"
}

@article{Amaro:2025pms,
    author = "Amaro, Fernando Dominques and others",
    title = "{Modeling the light response of an optically readout GEM based TPC for the CYGNO experiment}",
    eprint = "2505.06362",
    archivePrefix = "arXiv",
    primaryClass = "physics.ins-det",
    month = "5",
    year = "2025"
}

@article{Lisotti:2024fco,
    author = "Lisotti, Chiara and others",
    title = "{CYG$\nu $S: detecting solar neutrinos with directional gas time projection chambers}",
    eprint = "2404.03690",
    archivePrefix = "arXiv",
    primaryClass = "hep-ph",
    doi = "10.1140/epjc/s10052-024-13392-3",
    journal = "Eur. Phys. J. C",
    volume = "84",
    number = "10",
    pages = "1021",
    year = "2024"
}

@article{Kelly:2024tvh,
    author = "Kelly, Kevin J. and Mishra, Nityasa and Rai, Mudit and Strigari, Louis E.",
    title = "{\ensuremath{\nu}\ensuremath{\mu} and \ensuremath{\nu}\ensuremath{\tau} elastic scattering in Borexino}",
    eprint = "2407.03174",
    archivePrefix = "arXiv",
    primaryClass = "hep-ph",
    reportNumber = "MI-HET-836",
    doi = "10.1103/PhysRevD.110.113004",
    journal = "Phys. Rev. D",
    volume = "110",
    number = "11",
    pages = "113004",
    year = "2024"
}

@article{DeRomeri:2024dbv,
    author = "De Romeri, Valentina and Papoulias, Dimitrios K. and Ternes, Christoph A.",
    title = "{Light vector mediators at direct detection experiments}",
    eprint = "2402.05506",
    archivePrefix = "arXiv",
    primaryClass = "hep-ph",
    doi = "10.1007/JHEP05(2024)165",
    journal = "JHEP",
    volume = "05",
    pages = "165",
    year = "2024"
}

@article{Davidson:2003ha,
    author = "Davidson, S. and Pena-Garay, C. and Rius, N. and Santamaria, A.",
    title = "{Present and future bounds on nonstandard neutrino interactions}",
    eprint = "hep-ph/0302093",
    archivePrefix = "arXiv",
    reportNumber = "IFIC-02-36, FTUV-03-0212, IPPP-02-49, DCPT-02-98",
    doi = "10.1088/1126-6708/2003/03/011",
    journal = "JHEP",
    volume = "03",
    pages = "011",
    year = "2003"
}

@article{Lindner:2016wff,
    author = "Lindner, Manfred and Rodejohann, Werner and Xu, Xun-Jie",
    title = "{Coherent Neutrino-Nucleus Scattering and new Neutrino Interactions}",
    eprint = "1612.04150",
    archivePrefix = "arXiv",
    primaryClass = "hep-ph",
    doi = "10.1007/JHEP03(2017)097",
    journal = "JHEP",
    volume = "03",
    pages = "097",
    year = "2017"
}

@article{Miranda:2015dra,
    author = "Miranda, O. G. and Nunokawa, H.",
    title = "{Non standard neutrino interactions: current status and future prospects}",
    eprint = "1505.06254",
    archivePrefix = "arXiv",
    primaryClass = "hep-ph",
    doi = "10.1088/1367-2630/17/9/095002",
    journal = "New J. Phys.",
    volume = "17",
    number = "9",
    pages = "095002",
    year = "2015"
}

@article{Super-Kamiokande:1998kpq,
    author = "Fukuda, Y. and others",
    collaboration = "Super-Kamiokande",
    title = "{Evidence for oscillation of atmospheric neutrinos}",
    eprint = "hep-ex/9807003",
    archivePrefix = "arXiv",
    reportNumber = "BU-98-17, ICRR-REPORT-422-98-18, UCI-98-8, KEK-PREPRINT-98-95, LSU-HEPA-5-98, UMD-98-003, SBHEP-98-5, TKU-PAP-98-06, TIT-HPE-98-09",
    doi = "10.1103/PhysRevLett.81.1562",
    journal = "Phys. Rev. Lett.",
    volume = "81",
    pages = "1562--1567",
    year = "1998"
}

@article{Baikal-GVD:2019kwy,
    author = "Avrorin, A. D. and others",
    collaboration = "Baikal-GVD",
    title = "{Neutrino Telescope in Lake Baikal: Present and Future}",
    eprint = "1908.05427",
    archivePrefix = "arXiv",
    primaryClass = "astro-ph.HE",
    doi = "10.22323/1.358.1011",
    journal = "PoS",
    volume = "ICRC2019",
    pages = "1011",
    year = "2020"
}

@article{MINOS:2011neo,
    author = "Adamson, P. and others",
    collaboration = "MINOS",
    title = "{Measurement of the Neutrino Mass Splitting and Flavor Mixing by MINOS}",
    eprint = "1103.0340",
    archivePrefix = "arXiv",
    primaryClass = "hep-ex",
    reportNumber = "FERMILAB-PUB-11-040-PPD, BNL-94498-2010-JA",
    doi = "10.1103/PhysRevLett.106.181801",
    journal = "Phys. Rev. Lett.",
    volume = "106",
    pages = "181801",
    year = "2011"
}

@article{MicroBooNE:2023tzj,
    author = "Abratenko, P. and others",
    collaboration = "MicroBooNE",
    title = "{First Double-Differential Measurement of Kinematic Imbalance in Neutrino Interactions with the MicroBooNE Detector}",
    eprint = "2301.03706",
    archivePrefix = "arXiv",
    primaryClass = "hep-ex",
    reportNumber = "FERMILAB-PUB-23-004-ND",
    doi = "10.1103/PhysRevLett.131.101802",
    journal = "Phys. Rev. Lett.",
    volume = "131",
    number = "10",
    pages = "101802",
    year = "2023"
}

@article{Baudis:2013qla,
    author = "Baudis, L. and Ferella, A. and Kish, A. and Manalaysay, A. and Marrodan Undagoitia, T. and Schumann, M.",
    title = "{Neutrino physics with multi-ton scale liquid xenon detectors}",
    eprint = "1309.7024",
    archivePrefix = "arXiv",
    primaryClass = "physics.ins-det",
    doi = "10.1088/1475-7516/2014/01/044",
    journal = "JCAP",
    volume = "01",
    pages = "044",
    year = "2014"
}

@article{Borexino:2008dzn,
    author = "Arpesella, C. and others",
    collaboration = "Borexino",
    title = "{Direct Measurement of the Be-7 Solar Neutrino Flux with 192 Days of Borexino Data}",
    eprint = "0805.3843",
    archivePrefix = "arXiv",
    primaryClass = "astro-ph",
    doi = "10.1103/PhysRevLett.101.091302",
    journal = "Phys. Rev. Lett.",
    volume = "101",
    pages = "091302",
    year = "2008"
}

@article{Gehrlein:2024vwz,
    author = "Gehrlein, Julia and Machado, Pedro A. N. and Pinheiro, Jo\~ao Paulo",
    title = "{Constraining non-standard neutrino interactions with neutral current events at long-baseline oscillation experiments}",
    eprint = "2412.08712",
    archivePrefix = "arXiv",
    primaryClass = "hep-ph",
    reportNumber = "FERMILAB-PUB-24-0776-T, FERMILAB-PUB-24-0776-T",
    doi = "10.1007/JHEP05(2025)065",
    journal = "JHEP",
    volume = "05",
    pages = "065",
    year = "2025"
}

@article{Hyper-Kamiokande:2018ofw,
    author = "Abe, K. and others",
    collaboration = "Hyper-Kamiokande",
    title = "{Hyper-Kamiokande Design Report}",
    eprint = "1805.04163",
    archivePrefix = "arXiv",
    primaryClass = "physics.ins-det",
    month = "5",
    year = "2018"
}

@article{T2K:2023smv,
    author = "Abe, K. and others",
    collaboration = "T2K",
    title = "{Measurements of neutrino oscillation parameters from the T2K experiment using $3.6\times 10^{21}$ protons on target}",
    eprint = "2303.03222",
    archivePrefix = "arXiv",
    primaryClass = "hep-ex",
    doi = "10.1140/epjc/s10052-023-11819-x",
    journal = "Eur. Phys. J. C",
    volume = "83",
    number = "9",
    pages = "782",
    year = "2023"
}

@article{KM3Net:2016zxf,
    author = "Adrian-Martinez, S. and others",
    collaboration = "KM3Net",
    title = "{Letter of intent for KM3NeT 2.0}",
    eprint = "1601.07459",
    archivePrefix = "arXiv",
    primaryClass = "astro-ph.IM",
    doi = "10.1088/0954-3899/43/8/084001",
    journal = "J. Phys. G",
    volume = "43",
    number = "8",
    pages = "084001",
    year = "2016"
}

@article{Ricochet:2021rjo,
    author = "Augier, C. and others",
    collaboration = "Ricochet",
    title = "{Ricochet Progress and Status}",
    eprint = "2111.06745",
    archivePrefix = "arXiv",
    primaryClass = "physics.ins-det",
    doi = "10.1007/s10909-023-02971-5",
    journal = "J. Low Temp. Phys.",
    volume = "212",
    pages = "127--137",
    year = "2023"
}

@article{CONNIE:2019swq,
    author = "Aguilar-Arevalo, Alexis and others",
    collaboration = "CONNIE",
    title = "{Exploring low-energy neutrino physics with the Coherent Neutrino Nucleus Interaction Experiment}",
    eprint = "1906.02200",
    archivePrefix = "arXiv",
    primaryClass = "physics.ins-det",
    reportNumber = "FERMILAB-PUB-19-302-CD-PPD",
    doi = "10.1103/PhysRevD.100.092005",
    journal = "Phys. Rev. D",
    volume = "100",
    number = "9",
    pages = "092005",
    year = "2019"
}

@article{NUCLEUS:2017gvo,
    author = "Strauss, R. and others",
    collaboration = "NUCLEUS",
    title = "{Gram-scale cryogenic calorimeters for rare-event searches}",
    eprint = "1704.04317",
    archivePrefix = "arXiv",
    primaryClass = "physics.ins-det",
    doi = "10.1103/PhysRevD.96.022009",
    journal = "Phys. Rev. D",
    volume = "96",
    number = "2",
    pages = "022009",
    year = "2017"
}

@article{MINER:2016igy,
    author = "Agnolet, G. and others",
    collaboration = "MINER",
    title = "{Background Studies for the MINER Coherent Neutrino Scattering Reactor Experiment}",
    eprint = "1609.02066",
    archivePrefix = "arXiv",
    primaryClass = "physics.ins-det",
    doi = "10.1016/j.nima.2017.02.024",
    journal = "Nucl. Instrum. Meth. A",
    volume = "853",
    pages = "53--60",
    year = "2017"
}

@article{Pattavina:2020cqc,
    author = "Pattavina, Luca and Ferreiro Iachellini, Nahuel and Tamborra, Irene",
    title = "{Neutrino observatory based on archaeological lead}",
    eprint = "2004.06936",
    archivePrefix = "arXiv",
    primaryClass = "astro-ph.HE",
    doi = "10.1103/PhysRevD.102.063001",
    journal = "Phys. Rev. D",
    volume = "102",
    number = "6",
    pages = "063001",
    year = "2020"
}

@article{GRAND:2018iaj,
    author = "\'Alvarez-Mu\~niz, Jaime and others",
    collaboration = "GRAND",
    title = "{The Giant Radio Array for Neutrino Detection (GRAND): Science and Design}",
    eprint = "1810.09994",
    archivePrefix = "arXiv",
    primaryClass = "astro-ph.HE",
    doi = "10.1007/s11433-018-9385-7",
    journal = "Sci. China Phys. Mech. Astron.",
    volume = "63",
    number = "1",
    pages = "219501",
    year = "2020"
}

@article{KamLAND:2002uet,
    author = "Eguchi, K. and others",
    collaboration = "KamLAND",
    title = "{First results from KamLAND: Evidence for reactor anti-neutrino disappearance}",
    eprint = "hep-ex/0212021",
    archivePrefix = "arXiv",
    doi = "10.1103/PhysRevLett.90.021802",
    journal = "Phys. Rev. Lett.",
    volume = "90",
    pages = "021802",
    year = "2003"
}

@article{IceCube:2013low,
    author = "Aartsen, M. G. and others",
    collaboration = "IceCube",
    title = "{Evidence for High-Energy Extraterrestrial Neutrinos at the IceCube Detector}",
    eprint = "1311.5238",
    archivePrefix = "arXiv",
    primaryClass = "astro-ph.HE",
    doi = "10.1126/science.1242856",
    journal = "Science",
    volume = "342",
    pages = "1242856",
    year = "2013"
}

@article{SNO:2021xpa,
    author = "Albanese, V. and others",
    collaboration = "SNO+",
    title = "{The SNO+ experiment}",
    eprint = "2104.11687",
    archivePrefix = "arXiv",
    primaryClass = "physics.ins-det",
    doi = "10.1088/1748-0221/16/08/P08059",
    journal = "JINST",
    volume = "16",
    number = "08",
    pages = "P08059",
    year = "2021"
}

@article{DUNE:2020lwj,
    author = "Abi, Babak and others",
    collaboration = "DUNE",
    title = "{Deep Underground Neutrino Experiment (DUNE), Far Detector Technical Design Report, Volume I Introduction to DUNE}",
    eprint = "2002.02967",
    archivePrefix = "arXiv",
    primaryClass = "physics.ins-det",
    reportNumber = "FERMILAB-PUB-20-024-ND, FERMILAB-DESIGN-2020-01",
    doi = "10.1088/1748-0221/15/08/T08008",
    journal = "JINST",
    volume = "15",
    number = "08",
    pages = "T08008",
    year = "2020"
}

@article{Klein:2022lrf,
    author = "Klein, Joshua R. and others",
    title = "{SNOWMASS Neutrino Frontier NF10 Topical Group Report: Neutrino Detectors}",
    eprint = "2211.09669",
    archivePrefix = "arXiv",
    primaryClass = "hep-ex",
    reportNumber = "FERMILAB-FN-1216-ND",
    month = "11",
    year = "2022"
}

@inproceedings{Huber:2022lpm,
    author = "Huber, Patrick and others",
    title = "{Snowmass Neutrino Frontier Report}",
    booktitle = "{Snowmass 2021}",
    eprint = "2211.08641",
    archivePrefix = "arXiv",
    primaryClass = "hep-ex",
    reportNumber = "FERMILAB-FN-1215-ND-PPD-SCD",
    month = "11",
    year = "2022"
}

@article{Blennow:2003xw,
    author = "Blennow, Mattias and Ohlsson, Tommy and Snellman, Hakan",
    title = "{Day-night effect in solar neutrino oscillations with three flavors}",
    eprint = "hep-ph/0311098",
    archivePrefix = "arXiv",
    doi = "10.1103/PhysRevD.69.073006",
    journal = "Phys. Rev. D",
    volume = "69",
    pages = "073006",
    year = "2004"
}

@article{Farzan:2015hkd,
    author = "Farzan, Yasaman and Shoemaker, Ian M.",
    title = "{Lepton Flavor Violating Non-Standard Interactions via Light Mediators}",
    eprint = "1512.09147",
    archivePrefix = "arXiv",
    primaryClass = "hep-ph",
    doi = "10.1007/JHEP07(2016)033",
    journal = "JHEP",
    volume = "07",
    pages = "033",
    year = "2016"
}

@article{Ohlsson_2013,
   title={Status of non-standard neutrino interactions},
   volume={76},
   ISSN={1361-6633},
   url={http://dx.doi.org/10.1088/0034-4885/76/4/044201},
   DOI={10.1088/0034-4885/76/4/044201},
   number={4},
   journal={Reports on Progress in Physics},
   publisher={IOP Publishing},
   author={Ohlsson, Tommy},
   year={2013},
   month=mar, pages={044201} }

@article{Dev_2019,
   title={Neutrino non-standard interactions: A status report},
   ISSN={2666-4003},
   url={http://dx.doi.org/10.21468/SciPostPhysProc.2.001},
   DOI={10.21468/scipostphysproc.2.001},
   number={2},
   journal={SciPost Physics Proceedings},
   publisher={Stichting SciPost},
   author={Dev, Bhupal and Babu, K. S. and Denton, Peter and Machado, Pedro and Argüelles, Carlos A. and Barrow, Joshua L. and Chatterjee, Sabya Sachi and Chen, Mu-Chun and de Gouvêa, André and Dutta, Bhaskar and Gonçalves, Dorival and Han, Tao and Hostert, Matheus and Jana, Sudip and Kelly, Kevin J. and Li, Shirley Weishi and Martinez-Soler, Ivan and Mehta, Poonam and Mocioiu, Irina and Perez-Gonzalez, Yuber F. and Salvado, Jordi and Shoemaker, Ian and Tammaro, Michele and Thapa, Anil and Turner, Jessica and Xu, Xun-Jie},
   year={2019},
   month=dec }

@article{Esteban_2018,
   title={Updated constraints on non-standard interactions from global analysis of oscillation data},
   volume={2018},
   ISSN={1029-8479},
   url={http://dx.doi.org/10.1007/JHEP08(2018)180},
   DOI={10.1007/jhep08(2018)180},
   number={8},
   journal={Journal of High Energy Physics},
   publisher={Springer Science and Business Media LLC},
   author={Esteban, Ivan and Gonzalez-Garcia, M. C. and Maltoni, Michele and Martinez-Soler, Ivan and Salvado, Jordi},
   year={2018},
   month=aug }

@article{Vinyoles_2017,
   title={A New Generation of Standard Solar Models},
   volume={835},
   ISSN={1538-4357},
   url={http://dx.doi.org/10.3847/1538-4357/835/2/202},
   DOI={10.3847/1538-4357/835/2/202},
   number={2},
   journal={The Astrophysical Journal},
   publisher={American Astronomical Society},
   author={Vinyoles, Núria and Serenelli, Aldo M. and Villante, Francesco L. and Basu, Sarbani and Bergström, Johannes and Gonzalez-Garcia, M. C. and Maltoni, Michele and Peña-Garay, Carlos and Song, Ningqiang},
   year={2017},
   month=jan, pages={202} }

@article{Esteban_2024,
   title={NuFit-6.0: updated global analysis of three-flavor neutrino oscillations},
   volume={2024},
   ISSN={1029-8479},
   url={http://dx.doi.org/10.1007/JHEP12(2024)216},
   DOI={10.1007/jhep12(2024)216},
   number={12},
   journal={Journal of High Energy Physics},
   publisher={Springer Science and Business Media LLC},
   author={Esteban, Ivan and Gonzalez-Garcia, M. C. and Maltoni, Michele and Martinez-Soler, Ivan and Pinheiro, João Paulo and Schwetz, Thomas},
   year={2024},
   month=dec }

\end{document}